\documentclass[aps,prd,reprint,showpacs,twocolumn,amsmath,amssymb]{revtex4}
\usepackage{epsfig}
\usepackage{ifpdf}
\usepackage{graphicx}
\usepackage{amssymb,amsmath,lineno,mathbbol,upgreek}
\usepackage[tight]{subfigure}
\usepackage{mathrsfs}
\usepackage{bbm}
\usepackage{slashed}
\usepackage{natbib}

\usepackage{calligra}
\DeclareMathAlphabet{\mathcalligra}{T1}{calligra}{m}{n}
\DeclareFontShape{T1}{calligra}{m}{n}{<->s*[2.2]callig15}{}

\usepackage{float}
\usepackage{afterpage}
\usepackage{float}
\usepackage{lipsum}

\usepackage{calligra}
\DeclareMathAlphabet{\mathcalligra}{T1}{calligra}{m}{n}
\DeclareFontShape{T1}{calligra}{m}{n}{<->s*[2.2]callig15}{}

\pacs{03.75.Lm, 67.85.-d, 05.45.-a, 03.65.Pm}

\begin{document}

\bibliographystyle{prsty}

\title{Superconductivity and quantum phase transitions in dense QCD$_3$}

\author{Laith H. Haddad}
\affiliation{Department of Physics, Colorado School of Mines, Golden, CO 80401,USA}
\date{\today}

\begin{abstract}
We investigate thermal and quantum phase transitions (QPT) in the $(2+1)$-dimensional two-index antisymmetric formulation of large $N_c$ quantum chromodynamics. This formulation allows for color superconductivity, hence a more realistic model of QCD than standard 't Hooft large $N_c$ extrapolations for quarks in the fundamental representation. Baryon ground state topology, symmetries, and quantum dynamics are examined. Intermediate and high density regimes are modeled through effective four-fermion interactions comprised of attractive scalar and diquark terms and a term to account for vector-meson repulsion, the latter relevant at high densities. Physics beyond mean-field is addressed through the full quantum field theory for Madelung-decomposed fluctuations of the baryon field. A key QPT occurs at the meson-diquark transition driven by the ratio of baryon chemical potential to quark mass, $\mu_B/m$, or to an external applied magnetic field, $\mu_B/|eB|$. Critical behavior at ${\mu^c_B} \equiv m$ (or $|eB|$) is governed by diverging fluctuations of a field $\gamma$ that modulates the angular position of baryon potential minima in spin-space, identified with discrete chiral $\mathbb{Z}_2^\sigma$, $\mathbb{Z}_2^{LR} \times \mathbb{Z}_2^{LR}$, and $\mathbb{Z}_4^\chi$ symmetries for meson, diquark, and asymptotically free regimes, respectively. The topological sector is found to consist of baryonic solitons that map space-time to chiral $U(1)$ circles normal to these discrete symmetries. Remarkably, competition between $\mu_B$ and $m$ (or $B$), destroys superconductivity via a \emph{quantum} Berezinskii-Kosterlitz-Thouless (BKT) phase transition at ${\mu^c_B}$, with signature gap scaling $\Delta_\mathrm{BCS}\sim \exp\!\left(- c/\sqrt{\mu_B - \mu^c_B}\right)$. This peculiar breakdown of quasi-long-range order originates in the bosonic vortex sector of spin-charge-separated current induced by large $\gamma$ fluctuations between paired baryonic solitons. Significantly, the large $\gamma$ fluctuations at ${\mu^c_B }$ behave as an additional momentum scale, resulting in an effective $(3+1)$d conformal critical theory. We use these insights to elucidate the nature of \emph{holographic} BKT transitions, from the field theory side, a result which has remained elusive to date. We derive the QCD phase diagram for $\mu_B$ above the baryon mass and find good agreement with results obtained by other methods.
\end{abstract}

\maketitle

\section{Introduction}

Presently, a thorough knowledge of the phases of quantum chromodynamics (QCD) is lacking and constitutes a significant hurdle towards our understanding of the rich phenomenology encoded within the standard model of particle physics~\cite{Rajagopal2001,Alford2008,Fukushima2011,Weise2012}. Although QCD is notoriously difficult to solve directly, much insight has been gained through a combination of laboratory experimentation, observational cosmology, and effective theoretical models. For instance, data from heavy ion collision experiments tells us that for temperatures above the QCD scale, i.e., $\Lambda_\mathrm{QCD} \sim 200 \,  \mathrm{MeV}$, and at low quark densities the physics is described by a quark-gluon plasma (QGP)~\cite{Siemens2018}. In contrast, we know considerably less about the physics at low temperatures and increasingly higher densities beginning near the baryon density, $\Lambda_\mathrm{QCD}^3 \sim 1 \, \mathrm{fm}^{-3}$~\cite{McLerran2007}. Although, it is true that at extremely high densities perturbative asymptotic freedom applies, as hadrons ultimately dissolve into degenerate quark matter, asymptotic freedom does not apply for the intermediate regimes. Adding to the problem, first principle lattice simulations are impractical here due to the infamous sign problem and empirical data is constrained by the fact that the required quark densities are unreachable in the laboratory, occurring only in the deep interior of dense stellar objects such as neutron stars~\cite{Sauls1989,Alford2004,Alford2009}. Nevertheless, various phenomenological methods such as the Nambu-Jona-Lasinio (NJL)~\cite{1nambu61,2nambu61,Ferrer2015} and random matrix models~\cite{Stephanov1996,Halasz1997}, among others, have lead to conjectured exotic phases including color superconductivity (CSC)~\cite{Alford2001_2,Alford2001,Alford2008}, color-flavor locking (CFL), and mixed phases characterized by the simultaneous coexistence of hadrons and quarks~\cite{Fukushima2011}.

 In QCD, as well as various QCD-like theories, quarks are expected to propagate freely for baryon densities much greater than $1 \,  \mathrm{fm}^{-3}$, which leads to a Fermi surface and potential instabilities towards a number of possible pairing states: Cooper, particle-hole, density wave, etc. This high density state is basically a quark liquid with no chiral symmetry breaking, dynamically generated quark mass, or confinement. The dissolution of baryons into diquarks as an intermediate stage towards free quarks, and the precise mechanism by which this occurs constitutes the main focus of the present work.
 
 In this article, we investigate finite density phase transitions at low temperature, focusing particularly on the interplay between chiral symmetry breaking (CSB), color superconductivity, and vector-meson repulsion near the hadron-quark matter phase transition in (2+1)-dimensional QCD, a fundamental subject that continues to draw interest~$^{\footnotemark[1]}$\footnotetext[1]{See for instance the series of works~\cite{Klimenko2012,Klimenko2013,Klimenko2015,Klimenko2016}, among others.}. Besides a generally rich context, the reduced dimensionality provides a more tractable platform than the (3+1)d case while still retaining key features of the full theory such as confinement and a discrete chiral symmetry~\cite{Frenkel1986,Ferretti1992}. We model the ground state at low-temperatures for the intermediate-to-large quark density regime within the large $N_c$ semiclassical limit for two-index antisymmetric quarks in the Gross-Neveu-Thirring model~(see for example~\cite{Andrianov1992} and references therein), augmented by the inclusion of a diquark term to allow for superconductivity. This form couples $N_c$ color species of fermions through scalar (Gross-Neveu), vector (Thirring), and diquark (BCS) interactions. Inclusion of the vector term reflects conventional wisdom, which argues for significant vector-meson repulsion at high densities.

Elaborating further on the motivation for and character of our chosen interactions. The scalar and diquark terms are taken in the usual way to be attractive in order to allow for chiral and diquark condensation through the quark bilinears $\langle \bar{q} q \rangle$ and $\langle q q \rangle$. Each of these dominates over different density ranges, being more or less mutually exclusive within the mean-field spin landscape: diquarks are favored when minima of the ground state effective potential are localized along the left-right chiral diagonals ($\mu_B \gg  m$), mesons are favored along a single vertical spin-up direction ($\mu_B \ll m $). The choice of a repulsive vector term is justified from multiple lines of reasoning~\cite{Kitazawa2002,Buballa2005,Abuki2009,Coelho2010,Weissenborn2012,Masuda2013,Orsaria2014,Menezes2014,Ferrer2015}. For example, both the instanton-anti-instanton molecule model and the renormalization group approach result in repulsive vector meson corrections in their respective effective actions. From a phenomenological standpoint, the equation of state for neutron stars requires a degree of stiffness that seems to be accounted for only by adding a repulsive vector term to standard NJL-type theories. 
 
 This article is organized as follows. In section~\ref{Preliminary}, we begin with a general discussion and overview of some of the central features of our model and results of our analysis. Section~\ref{Model} establishes the mathematical foundation of our model by specifying the Lagrangian density for attractive and repulsive four-fermion interactions. In Sec.~\ref{CondensateFormation}, we present the various quark bilinear condensates that occur in our system and discuss the low-temperature action at the mean field level. In Sec.~\ref{FiniteTPT}, we investigate finite-temperature phase transitions and associated symmetries. In Sec.~\ref{QPT}, we study quantum phase transitions, order parameters, bound states, and Fermi surface fluctuations at finite baryon chemical potential. Section~\ref{MadForm} presents mathematical details to go beyond mean field theory, calculating quark excitations of bilinear condensates using a Madelung decomposition for the quark degrees of freedom. In Sec.~\ref{QBKTPT}, we present the full derivation of the quantum Berezinskii-Kosterlitz-Thouless phase transition, focusing on both topological and renormalization group aspects. In Sec.~\ref{Connection}, we derive the QCD phase diagram, connecting to established as well as conjectured insights. In addition, we derive, in this section, the relation to holographic theories. In Sec.~\ref{Conclusion}, we conclude.

 \section{Preliminary discussion and overview} 
 \label{Preliminary}

 In this section we wish to provide some general discussion to hopefully contextualize the physics of the results that follow in the main part of this paper.

 \subsection{Color superconductivity and large $N_c$ limits }

 Of particular interest to us is the Cooper instability that occurs through the one-gluon exchange, attractive in the anti-triplet of the $\bar{3}$ channel ($3 \times 3 = 6_S + 3_A$). For one-gluon exchange, the quark-quark amplitude is proportional to the color tensor
 \begin{eqnarray}
  - C_{A}^2 \left( \delta_{a a'} \delta_{b' b} - \delta_{a b'} \delta_{a' b}   \right) +  C_{S}^2 \left( \delta_{a a'} \delta_{b' b} + \delta_{a b'} \delta_{a' b}   \right) \label{colortensor}
 \end{eqnarray}
 where $C_A$ and $C_S$ are constants that depend on the number of colors, $N_c$. The first anti-triplet term (antisymmetric in the color indices $a$ and $b$) captures the attractive channel of the interaction. The second sextet term (symmetric in $a, b$) does not interest us as it provides repulsion between quarks (refs: ``Two lectures on color superconductivity'', others). Though we will look at the limit in which the number of colors is taken to infinity, which does present challenges that must be addressed, antisymmetry in the color indices of the attractive term in Eq.~(\ref{colortensor}) is central to our analysis, as we would like to allow for color superconductivity in our model.

A crucial issue must be addressed regarding competing instabilities of the large $N_c$ Fermi surface. It has been suggested, with strong justification, that particle-hole pairing (Overhauser effect) overtakes particle-particle or hole-hole pairing (BCS effect), in the traditional large $N_c$ expansion ('t Hooft limit)~\cite{Park2000}. In~\cite{Park2000}, the authors studied finite density QCD by constructing a Wilsonian effective action for various scalar-isoscalar excitations around the Fermi surface, and found the Overhauser effect to be the generically dominant one, BCS being exponentially suppressed. Other investigations at finite density address a related process which also suppresses BCS, the Deryagin, Grigoriev, and Rubakov (DGR) instability associated with the formation of a chiral density wave~\cite{Deryagin1992}. The operative issue here lies in the fact the DGR condensate is a global color singlet; the BCS condensate is not.

A viable alternative to the usual 't Hooft extrapolation from $N_c=3$ --- which treats quarks in the fundamental representation --- is the two-index antisymmetric representation \cite{Armoni2003,Armoni2005,DeGrand2015}. We employ this latter method in our investigations. It is interesting that both formulations coincide with standard QCD for $N_c= 3$, but differ radically at large $N_c$ and finite density. One significant difference between these two large $N_c$ limits is that the superconducting state is suppressed in the fundamental representation, whereas it is not in the two-index formulation \cite{Buchoff2010}. This is namely due to active quark loop effects to leading order in the latter case, suppressed in the former. It should also be stressed that the usual large $N_c$ limit for fundamental quarks looks very different from standard $N_c =3$ QCD. These observations simply highlight the fact that large $N_c$ methods should be implemented while heeding the caveats which stress their limits and individual peculiarities. It should not be forgotten, though, that one is ultimately interested in modeling real QCD and that, there, one should expect to observe the formation of a superconducting ground state. Our particular choice of large $N_c$ extrapolation is motivated primarily by its success in achieving that aim.

\subsection{Finite and zero-temperature phase structure}

At the mean field level, much of the general zero-temperature properties of our system can be gleaned from the thermodynamic relationships between various low-temperature order parameters displayed in Table~\ref{table1}. Here, we have defined the key parameters for our system which include extensive quantities such as the quark mass and chemical potential as well as intensive order parameters. Key among the latter are quark bilinears such as the scalar condensate, diquark pseudogap, and BCS gap. The regime where a particular bilinear dominates is determined by the the relative size of the defining parameters, e.g., scalar condensation is disfavored when the quark chemical potential exceeds its mass. More detailed assessment of critical behavior for each quantity characterized by proximity of the quark mass to criticality, with $m_c \equiv |\tilde{\mu}_B|$, at the far right of the table, requires a full quantum treatment beyond the mean field picture. There, the nature of quasi-long-range order on either side of the QCP is traced to phase decoherence induced by diverging fluctuations in the parametric angle $\gamma$, associated with the discrete generator of left $\leftrightarrow$ right chiral symmetry. Though we will exploit mean field arguments unabashedly throughout our work, the wealth of insights offered by quantum path-integral methods cannot be overstated.

 The large $N_c$ mean field formalism we use is enhanced by assuming a quadratic temperature dependence for one of the condensates, a condition which propagates to all other order parameters through a fundamental symmetry of our model. This modification is motivated by first principles and proves effective at predicting the temperature-versus-chemical potential/density phase structure of QCD around the superconducting region of interest to us. The results match predictions obtained through other methods that probe well beyond the large $N_c$ paradigm~\cite{Buballa2005,Alford2008,Fukushima2011}. Indeed, it is remarkable that a rich phase structure consisting of hadronic, superconducting, and asymptotically free quark regimes, emerges out of purely mean-field considerations: critical features naturally emerge such as a first-order phase transition curve separating the confined hadronic region from the deconfined phase at large temperatures (the QGP) and from the superconducting phase towards higher densities, including the expected finite-temperature BEC-BCS crossover characterizing the amorphous superconducting-to-deconfined region above the CSC phase. An additional crossover at large densities appears in the full quantum picture completing the QGP circular swath that sweeps from upper left to lower right, connecting the physics at high temperatures with that of high densities. It is interesting that the crossover in our model at higher density occurs where the ordinary CSC-CFL transition appears in more sophisticated models that include the full flavor spectrum~\cite{Alford2008}.

 In this paper we emphasize heavily the zero-temperature properties of our system, incorporating quantum effects through the full partition function constructed from the ground state degrees of freedom. At zero temperature, we find that increasing the baryon chemical potential from below drives the system first along the lower edge of the hadronic region, where the quark mass is still substantially larger than the effective chemical potential, $m >  \tilde{\mu}_B$. Fluctuations along this edge are described by massive Gross-Neveu (GN) theory with broken chiral symmetry distinguished by a discrete $\mathbb{Z}_2^\sigma$ internal symmetry group of the scalar meson. The physics in this region is essentially covariant but has a small Lorentz violating density perturbation that introduces finite $\tilde{\mu}_B/m$ corrections. It is these small perturbations that eventually destroy coherence of the scalar condensate through chiral symmetry restoring baryonic fluctuations~$^{\footnotemark[2]}$\footnotetext[2]{Note the distinction between the chiral symmetry group of the full theory, which here is partially restored, in contrast to that of the ground state which gets reduced.}, becoming increasingly pervasive as one nears the QCP at $\tilde{\mu}_B = m$. At the QCP, scalar meson physics is overwhelmed by these fluctuations and the GN model no longer provides a valid description. Above the QCP, $\tilde{\mu}_B >  m$, we find the system to be described by classical BCS theory that couples states with the same chirality, but now with a small CSB term proportional to $m$ that mixes left and right chiral states. The discrete part of the (ground state) symmetry group transforms as $\mathbb{Z}_2^\sigma \to \mathbb{Z}_2^{L,R} \times  \mathbb{Z}_2^{L,R}$ across the QCP. Thus, in a reverse manner, approaching the QCP from above one encounters increasingly sharp fluctuations in the meson field, breaking chiral symmetry and subjecting the diquark condensate to a similar fate as that of the scalar condensate when approaching the QCP from the opposite direction.

  \subsection{Spin-charge separation and quantum fluctuations}

   The two-index antisymmetric formulation decomposes fundamental quarks through what amounts to a basis transformation $\psi^{(i)} \to \epsilon^{i j k} \psi_{ j k}$. Note that antisymmetry is often incorporated concisely by writing two-index states with bracketed indices, $\psi_{[ j k]}$, so that, in terms of the fundamental representation, using Dirac notation we have
  \begin{eqnarray}
 | \psi_{[ j k]} \rangle = \frac{1}{\sqrt{2}} \left(  | \psi^{(j )} \rangle \,  |  \psi^{(k )} \rangle  - |  \psi^{(k)} \rangle  \,  |  \psi^{(j)} \rangle    \right) \, , 
  \end{eqnarray}
  where $1 \leq j \leq k \leq N_c$. Thus, there are $N_c ( N_c -1 )/2$ color components in the two-index antisymmetric representation contrasted with $N_c$ for fundamental quarks. For a single quark flavor in the fundamental representation, antisymmetry in the color index requires the remaining product $\psi_\mathrm{space} \times  \psi_\mathrm{spin}$ to be symmetric under exchange. Thus, in typical large $N_c$ treatments of QCD the combined spatial and spin part of the quark many-body wavefunction is bosonic (spin-1/2 boson), satisfying either a relativistic interacting Klein-Gordon equation, for light quarks, or a non-relativistic Schr\"odinger type equation, for massive baryons~\cite{Cohen2011}.

  In the present context, the additional antisymmetry acquired when shifting from the fundamental to the two-index representation forces antisymmetry onto the product of spin and space. Hence, in our work we will retain the full Dirac structure for two-index quarks. Moreover, in the presence of background condensates the effective potential for the quark field is minimized by allowing the fermionic structure to reside only in the Dirac spin index. This allows the amplitude and overall phase of the quark field to occupy a single energy state associated with some macroscopic expectation value at one of the potential minima. This perspective amounts to squeezing the quark fluctuations into the spin of the quark many-body wavefunction such that the internal spin-phase factor associated with the direction of Dirac current coherently samples all directions over the Fermi sphere.

 This picture may at first seem exotic, but upon further scrutiny one can see that it simply provides an alternative description of the underlying microscopic quark-quark or quark-antiquark bound states in terms of spin-charge separated bosonic current and fermionic spin. One cannot overemphasize the significance of this point. That the quark field develops a finite expectation value at a minimum of the effective potential away from the location of zero-quark amplitude, may seem a strange notion to those unfamiliar with large $N_c$ fermion statistics --- we are essentially implying a sort of condensation of fermions. A brief consideration, though, reveals the trick: each of the $N_c$ quarks is tagged by a different value of the color index. Hence, the baryon density is amenable to standard bosonic Hartree-Fock methods with exact or near exact accuracy.~$^{\footnotemark[3]}$\footnotetext[3]{Reference~\cite{Witten1979} provides a detailed and enlightening exposition of this topic.}

 Studying the landscape of the effective quark potential at the mean field level will certainly provide a broad, general understanding of the physics near the QCP. But, in order to probe the subtleties of the QCP, we require a full quantum mechanical treatment of the problem that incorporates fluctuations of the quark field. As previously mentioned in passing, the microdetails of the QCP are resolved by examining small fermionic fluctuations in the baryonic effective potential near the potential minima. These ground-state fluctuations are obtained by expanding the many-body baryon wavefunction in terms of separated density, spin, and orbital degrees of freedom. This approach allows us to probe potential regions where spin-charge separation~\cite{Tomonaga1950} may occur, a foresight which proves particularly effective near the meson-diquark phase transition. There, the effect is rather striking at zero temperature where interactions are relatively strong and one should not be surprised to observe exotic collective behavior.

 The massless modes in these expansions, orbital and spin phases $\vartheta$ and $\varphi$, describe rotations along the symmetry directions of the ground state. In addition to the massless phases, two massive modes appear, density and chiral amplitudes $\eta$ and $\gamma$. The former is a density wave around a nonzero minimum average that reflects the system's resistance towards compression or expansion, i.e., competition between quark attraction and repulsion. The latter, a chiral fluctuation that distorts the local chirality of the ground state in the direction of the underlying discrete chiral symmetry that mixes left and right chirality, hence its appearance as a massive mode. The full zero-temperature quantum field theory (QFT) is then constructed by writing down the effective action for fluctuations in $\vartheta$, $\varphi$, $\eta$, and $\gamma$. The method of steepest descent can then be applied to the resulting partition function 
 \begin{eqnarray}
 Z = \int \mathcal{D}\!\left[ \delta \vartheta , \delta \varphi , \delta \eta , \delta  \gamma \right]  e^{i S \left[  \delta \vartheta , \delta \varphi , \delta \eta , \delta  \gamma \right]   }   \, . 
 \end{eqnarray}
 The partition function provides a readily accessible probe of the physics for the fluctuations $\delta \vartheta$ and $\delta \varphi$, upon standard Gaussian integration over the massive modes, through the effective reduction
  \begin{eqnarray}
 S \left[  \delta \vartheta , \delta \varphi , \delta \eta , \delta  \gamma \right]   \to  S^\mathrm{eff}\left[  \delta \vartheta , \delta \varphi  \right]_{m_\eta, m_\gamma} \, , 
 \end{eqnarray}
 with the masses $m_\gamma$ and $m_\eta$ parametrizing the resulting effective action. Quantum effects due to the massive modes are then enfolded into the coefficients for the remaining fields from which one obtains the critical behavior in Table~\ref{table1}. The critical behavior of bilinears listed in Table~\ref{table1} are particularly affected by $\delta \gamma$ due to the vanishing mass, $\lim_{m \to m_c} m_\gamma = 0$, at the QCP. The density fluctuations $\delta \eta$ retain their mass through the phase transition $m_\eta \ne 0$, as one should expect. No reckless hyperbole is risked by stating that most of the interesting physics at the QCP derives from the behavior of the chiral fluctuations $\delta \gamma$ near and at criticality.

The approach we have described up to this point provides detailed information about fermionic and bosonic fluctuations of quark bound states near the QCP. Another method we use to probe the QCP focuses directly on the behavior of the Fermi surface near the critical point. This is done by studying changes in the topology of the fermion inverse Green's function from the BCS side as the system is tuned towards the QCP~\cite{Volovik2003}. Effects coming from $\delta \gamma$ fluctuations manifest at the level of quasiparticle dispersion by encoding them into the ground-state chiral wavefunctions through the parametrization
 \begin{eqnarray}
\Psi_{R, L} = \eta^{1/2} \exp(i \vartheta) \left[ \,  \cos \gamma_{R,L} , \,\pm  \exp(i \varphi ) \sin \gamma_{R,L}  \, \right]^T , \label{FirstAnsatz1}
\end{eqnarray}
from which one readily observes the distorting action of the $\gamma$ field which twists between left and right chiral states. Here, $\gamma_{R} = (n + 1/4 )\pi$ and $\gamma_{L} = (n + 3/4) \pi$, with the scalar meson an equal admixture of left and right chirality $\gamma_S =  (\gamma_R + \gamma_L)/2 = (n+1/2) \pi$, $n \ge 0  \in \mathbb{Z}$. Competition between quark mass and baryon chemical potential plays out in this scenario by modulating the positions of average values $\langle \gamma \rangle_\mathrm{min}$ in the effective potential. Left and right chiral minima are gradually brought in close enough proximity to one another that a respectable model can no longer omit the effects of $\delta \gamma$-tunneling. The resulting dispersion relation boldly displays a transition from a fully gapped ground state when $\delta \gamma \ll 1$, to one with two isolated Fermi points when $\delta \gamma \simeq  1$ at $m = m_c$, precisely where the perturbative ansatz begins to break down. This method dramatically confirms the dissolution of gap states into free quarks as the system is tuned towards criticality from the diquark side. Notably, the appearance of Fermi points (versus lines) reflects an enlarged symmetry group at the QCP due to the emergence of an additional massless direction for $\delta \gamma$ at the QCP. In fact, a dramatic consequence of this is the emergence of a critical $(3+1)$-dimensional conformal theory, which we demonstrate in detail.

\subsection{Topological order and Berezinskii-Kosterlitz-Thouless transitions}

We investigate the topological sector of our model by constructing the classical finite energy representation consistent with continuous symmetries of the ground state. The structure of the large $N_c$ baryon wavefunction provides a natural scaffolding support for topological solitons~\cite{Witten1979,Weigel2007} in the form of spatially extended vortices that wrap around $U(1)$ subgroups of the full ground-state symmetry group. These vortices, we show, are localized tunneling events between left and right chiral ground states, passing through an intermediate virtual meson cloud. Hence, from the diquark side, they represent nucleations of the scalar meson condensate. At exceptionally low temperatures in the diquark regime, such objects are pairwise bound due to their inaccessibly large excitation energy, yet proliferate freely at the meson-diquark QCP where thermo/quantum dynamics begins to favor quark-anti-quark bound states over diquarks.

The mechanism responsible for vortex dissociation at the meson-diquark QCP is multifold in nature, but fundamentally rooted in the $\delta \gamma$ fluctuations. First, note that these fluctuations are inherently just wiggling/loosening of the tight spin-orbit locking that characterizes left or right pure chiral states as the system approaches a chiral phase transition: the $\gamma$ field is the angle in spin-space that rotates between left and right chiral states, passing through a highly entangled intermediate left-right mixed state. This mixed state marks the point where spin and orbital currents are exactly orthogonal to one another. At the current level then, a picture of spin-charge separation naturally emerges near the QCP: fermionic spin (spinon) current is found to be expelled by each member of a pair (liberation of a slave fermion), leaving behind the bosonic orbital (chargon) ``guts'' of the baryonic vortex.

Consider, further, that our analysis of vortex-anti-vortex binding energy through application of renormalization group techniques shows that the bosonic parts of vortices participate in the same dance as that of paired defects in a Coulomb gas. With however one important distinction: dissociation of baryonic vortices is driven by quantum fluctuations, rather than thermal ones. From a thermodynamic viewpoint, the shift in relative size of the diquark pairing field to that of the quark mass that occurs through the QCP, $m < \bar{\Delta}_d \to \bar{\Delta}_d < m$, signals an inverted energetic advantage from diquarks favored over mesons to mesons over diquarks. But, this switch occurs at zero temperature so that the additional critical entropy associated with diquark-to-meson restructuring must reside in the singular quantum entanglement entropy of some underlying variable. This entropy is in fact carried by the $\delta \gamma$ fluctuations mentioned in our discussion.

\begin{table*}[]
\resizebox{18cm}{!}{
\begin{tabular}{lllll}
\multicolumn{5}{c}{} \\
 \hline \hline
Parameter &    Symbol  & \hspace{.5pc} Definition &  \hspace{.5pc}  Regime   &  \hspace{.5pc} Critical Behavior  \\
\hline
Vector meson pairing  & \;  $\Delta_v$ \; & $- g_V^2  \langle \bar{\psi}\gamma^\mu  \psi  \rangle$ &\;   $$  &\;   \; $\sim   \left| m^2  - m_c^2 \right|^{\nu}$  \\
 Diquark pairing  & \;  $\Delta_d$ \;   & $- g_D^2  \langle \psi^T i C \psi \rangle$  &\;   &\;   \; $\sim   \left| m^2  - m_c^2 \right|^{\nu}$   \\
 Effective diquark pairing  & \;  $\bar{\Delta}_d$ \;   & $\Delta_v + \Delta_d$  &\;    &\;   \; $\sim   \left| m^2  - m_c^2 \right|^{\nu}$   \\
Effective chem. pot.  & \;  $\tilde{\mu}_B$ \; & $\mu_B - \bar{\Delta}_d$ & \;  $$    & \; \;  $\sim m$   \\
Quark density & \;  $\rho$ \; & $\langle \psi^\dagger \psi  \rangle$ & \;$$ & \; \;  $\sim$ const.  \\
 Scalar condensate  & \;  $\Delta_s$ \; & $ - g_S^2\langle \bar{\psi} \psi  \rangle$ &\;   $\bar{\Delta}_d < \mu_B< m$  &\;  \;   $\sim \left| \rho^2 - C_s ( m^2  - m_c^2 )\right|^{\nu}$ \\
Pseudogap  & \;  $\Delta_\mathrm{pg}$   &  \; $\bar{\Delta}_d$ \;    &  \; $m  <  \bar{\Delta}_d < \mu_B$  &\;  \;  $\sim   \left| m^2  - m_c^2 \right|^{\nu}$  \\
BCS gap & \;  $\Delta_\mathrm{BCS} $ \; & \;  $| \tilde{\mu}_B|$ &\;   $m < \mu_B <  \bar{\Delta}_d$   &\;  \;  $\sim   \left| m^2  - m_c^2 \right|^{\nu}$ \\
Running quark mass  & \;  $m$ \; & $m_0 + \Delta_s$ & \;  $$ \;  & \; \; $\sim 1/\mu_B$  \\
Chiral mixing angle  & \;  $\gamma$ \; & \;  $\langle \gamma \rangle_\mathrm{min} $ &        &\;  \;  $\sim \pi/2 + C_\gamma \left| m - m_c \right|^{\nu}$    \\
Chiral mass & \;  $m_\gamma$ \; & \;  $ m_\gamma \,  \delta \gamma^2 $ &        &\; \;   $\sim \left| m - m_c \right|^{\nu}$    \\
B-field gen. scalar cond. & \;  $\lim_{m \to 0} \langle \bar{\psi} \psi  \rangle_\mathrm{mag}$ \; & \;  $- |e B|/2\pi$ &        &\; \;   $\sim \tilde{\mu}_B$    \\
Fermi surface fluct. & \;  $E^2_\pm({\bf p})$ \; & \;  $m_\mathrm{eff}^2 + |\mathrm{\bf p}|^2  + b^2 \pm 2 b \left[ m_\mathrm{eff}^2 + ( {\bf p} \cdot \hat{\bf b} )^2 \right]^{1/2}$ &        &\;  \; gap $\to$ Fermi points \\
\hline \hline
\end{tabular}    }
\caption{\emph{Order parameters, extensive quantities, and quantum critical behavior.} The regions near the superconducting phase display a range of order parameters with critical exponent $\nu = 1/2$. The bare quark mass $m_0$ and baryon chemical potential $\mu_B$ are the two independent parameters in the system.The critical point at zero-temperature is reached as the running quark mass increases, and the baryon chemical potential decreases towards the QCP at the lower edge of the superconducting region. The three main phases are characterized by dominant scalar condensate, pseudogap, and BCS gap, respectively, with associated regimes for each delineated by the relative strengths of mass, chemical potential, and diquark pairing. The first nine quantities listed are first encountered in Sec.~\ref{CondensateFormation}. The definitions for $\gamma$, associated mass $m_\gamma$, and critical behavior in the far right column are derived in Sec.~\ref{QPT}. The line second from the bottom refers to a dynamically generated quark mass through an external applied magnetic field, which connects directly to holographic systems in Sec.~\ref{Connection}. The last line refers to our analysis of Fermi surface fluctuations in the last part of Sec.~\ref{QPT}. Note that the critical behavior in the far right column does not account for exponential BKT scaling from topological degrees of freedom, which we address thoroughly in Sec.~\ref{QBKTPT}.  } \label{table1}  
\end{table*}

Another feature of vortex dissociation relates to structural changes in the symmetry group of the ground state near the QCP. Let us examine this. At large chemical potential, far from criticality, the ground state exhibits a near-perfect $\mathbb{Z}_4^{L, R}$ chiral symmetry, only weakly broken down to $\mathbb{Z}_2^{L,R} \times \mathbb{Z}_2^{L, R}$ by the presence of a small quark mass. Both groups are discrete subgroups of the continuous (unrealized) chiral symmetry parametrized by $\gamma$ in Eq.~(\ref{FirstAnsatz1}). There are, moreover, continuous $U(1)_R \times U(1)_L$ factors multiplying the discrete symmetries. These are associated with some combination of orbital and spin quark phases and support the $\pi_1\left( U(1)   \right)  \cong  \mathbb{Z}$ homotopy structure necessary for vortex formation. So, far from criticality the ground state lives in some effective potential landscape that supports both chiral left-left and right-right bound baryonic vortices.  As the system is tuned towards the QCP (increasing $m$, decreasing $\mu_B$), the height of the central potential peak along the diagonal directions, $\gamma_R$ and $\gamma_L$ in Eq.~(\ref{FirstAnsatz1}), gradually smooths out, finally falling away at criticality where the original four-fold structure of the ground state has coalesced into a two-fold structure described by a discrete $\mathbb{Z}_2^\sigma$ symmetry times some $U(1)$ factor. The main takeaway from all this is that at criticality the potential central peak flattens out only along the diquark direction in spin space, allowing once bound baryonic vortices to proliferate freely in position space.

It is not surprising, then, that our model exhibits a breakdown in the superconducting phase through an infinite-order Berezinskii-Kosterlitz-Thouless (BKT) type phase transition~\cite{Berezinskii1971,Berezinskii1972}. This occurs as a conventional finite temperature BKT transition along the meson-diquark transition curve, yet more importantly surviving as a \emph{quantum BKT transition} (QBKT) at zero-temperature. As our discussion has brought to light, the QBKT mechanism is fundamentally rooted in competition between the running quark mass and baryon chemical potential. The QBKT vortex degrees of freedom are simply the bosonic parts of baryonic solitons, pairwise bound on the diquark side of the QCP, proliferating freely beyond the critical point where mesons are energetically favored over diquarks. The meeting of superfluid and dissociation temperature curves at the quantum critical point and the absence of thermal fluctuations there, means that both topological and smooth long-wavelength excitations dominate the BEC-BCS transition. Hence, at the meson-diquark QCP the BCS gap dissolves through short-distance molecular dissociation of fundamental quarks as well as baryonic vortices, rather than long-distance superfluid phase decoherence, without an intermediate pseudogap bridging the BEC and BCS limits. We will demonstrate explicitly exponential decay of long-range order at zero-temperature, fully characterizing the BKT type transition.

An intriguing connection begins to form out of all this which betrays our deeper interest in $(2+1)$d models of QCD. The $\gamma$ field appears in every sense to act as an additional energy scale, a quantum ``temperature'' that dominates the QCP, frozen out when the system is far from criticality. In fact, this bears out in the expansion of the action using the form Eq.~(\ref{FirstAnsatz1}), wherein one finds that single factors of $\gamma$ multiply the lowest-order phase fluctuations, thus $\gamma$ acts as an effective momentum scale of the system. We will also show that by including the effects of an applied external magnetic field we obtain the same QBKT, but now driven by the ratio of baryon density to magnetic field strength, even in the absence of a quark mass. Hence, our results point inexorably towards the study of holographic systems~\cite{Witten1998,Maldacena1999,Zaanen2015,Herzog2009} through what have been called \emph{holographic} Berezinskii-Kosterlitz-Thouless phase transitions~\cite{Jensen2010,Jensen2010.2,Iqbal2010,Evans2011.3}.

 The first holographic BKT transition, discovered for an effectively $(2+1)$d D3/D5 brane system~\cite{Jensen2010}, is of particular relevance to the results presented in the present article. The initial holographic results were subsequently extended to include various other driving parameters and systems~\cite{Jensen2010.2,Iqbal2010,Evans2011.1,Evans2011.3,Grignani2012,Bigazzi2012}. The phase transition in~\cite{Jensen2010} was found to be driven by competition between applied magnetic field strength and quark density, in contrast to standard thermally driven BKT.  As mentioned, this scenario bares a striking resemblance to our problem, a parallel which attests anew to the versatility and richness of four-fermion theories. Indeed, in our work we show that the four-fermion approach has the advantage of providing deeper insight by elucidating the microphysics underlying the BKT mechanism, a point on which holography has so far fallen short. In essence, the approach to holography that most have taken, studies, indeed defines, the boundary field theory purely in terms of its bulk gravity dual at the unfortunate expense of a clear physical description for the zero-temperature BKT mechanism, from the field theory side.

\section{The model}
\label{Model}

We begin our mathematical analysis working directly from the microscopic dynamics of quark degrees of freedom interacting locally through Lorentz scalar, vector, and diquark quartic terms in $(2 + 1)$-dimensions. Ultimately, properties of the CSC diquark condensate and related transition into the scalar condensate will be computed from transitions of bound states comprised of quarks of the same chirality into ones of opposite chirality. A few words should be said regarding spinors and space-time symmetries in (2+1)-dimensions. Poincar\'e invariance demands symmetry under three space-time translations, two boosts, and one spatial rotation generated by the algebra
\begin{eqnarray}
\left[ P_\mu , \, P_\nu \right]  &=& 0 \, ,  \label{P1}\\
\left[ J^\mu , \, J^\nu \right]  &=& i \, \epsilon^{\mu \nu \rho} J_\rho  \, ,  \label{P2}   \\
\left[ J^\mu , \, P^\nu \right]  &=& i \, \epsilon^{\mu \nu \rho} P_\rho  \, . \label{P3}
\end{eqnarray}
In the present work we construct the spinor representation with a time-like signature, $g^{\mu \nu}= \mathrm{diag}\left( 1 , - 1 , -1 \right)$, using the $2 \times 2$ gamma matrices 
\begin{eqnarray}
\gamma^0 =  \sigma_3 \; , \;\; \gamma^1 = i  \sigma_1 \;, \;\; \gamma^2 =  i \sigma_2 \, ,  \label{myalgebra}
\end{eqnarray}
which satisfy the Dirac algebra  
\begin{eqnarray}
\left\{ \gamma^\mu , \, \gamma^\nu \right\} = 2 g^{\mu \nu}\, , 
\end{eqnarray}
with the identification $J^\mu =   \gamma^\mu/2$ in Eqs.~(\ref{P1})-(\ref{P3}), and the $\sigma_i$ are the Pauli matrices. Thus, in order to satisfy the Dirac algebra in $2+1$ dimensions one is forced into using two antihermitian generators. Two-spinors transform under planar rotations through the time-like rotation matrix 
\begin{eqnarray}
S\left[ \varphi \right] =  \exp\left( i  \varphi  \sigma_3   /2 \right) = \cos\left(  \varphi /2 \right)  + i \sigma_3  \sin\left(  \varphi /2 \right) \, , 
\end{eqnarray}
where $\phi$ is the planar polar angle. 

The Lagrangian density for our model is given by
\begin{eqnarray}
\mathcal{L} = \mathcal{L}_0 +  \mathcal{L}_S +  \mathcal{L}_V  +   \mathcal{L}_D \, , \label{FullLagrangian}
\end{eqnarray}
with the kinetic, scalar, vector, and diquark contributions given explicitly by 
\begin{eqnarray}
\hspace{-1pc} \mathcal{L}_0 \! &=&  \!  \sum_{n=1}^{N_c} \bar{\psi}^{(n)} \left( i \gamma^\mu  \partial_\mu - m_0  + \mu_B \gamma^0  \right)   \psi^{(n)}  ,  \label{kinetic} \\
 \hspace{-1pc}  \mathcal{L}_S\! &=& \!   \frac{g_S^2 }{2}  \! \left(  \sum_{n=1}^{N_c} \bar{\psi}^{(n)}   \psi^{(n)} \right)^2  ,  \label{scalar} \\ 
 \hspace{-1pc} \mathcal{L}_V \!&=& \! - \frac{ g_V^2}{2} \! \left(  \sum_{n=1}^{N_c} \bar{\psi}^{(n)}   \gamma^\mu \psi^{(n)}  \right)^2   , \label{vector} \\
  \hspace{-1pc} \mathcal{L}_D \! &=& \!  \frac{ g_D^2}{2} \! \left(  \sum_{n =1}^{N_c} \bar{\psi}^{(n)}  i C   \bar{\psi}^{(n) T} \!  \right) \! \! \!  \left(  \sum_{n'=1}^{N_c}    \psi^{(n') T}  \! i C \psi^{(n')} \delta_\mathrm{A}\! \!  \right)\! \!.   \label{diquark1}
\end{eqnarray}
Here and throughout, we will work in a single-flavor formulation. Note in $\mathcal{L}_0$ the inclusion of a bare quark mass $m_0$ and baryon chemical potential $\mu_B$. Note also that the vector interaction is repulsive in contrast to the scalar and diquark terms. The superscript index ($n$) indicates our preliminary formulation in terms of the fundamental representation for $N_c$ species of quarks with couplings scaling like $g_{S, V, D}^2\sim 1/N_c$, which we have yet to decompose into two-index quark states.

In terms of this fundamental representation, we take the quark field to be in the two-dimensional Weyl representation 
\begin{eqnarray}
\psi^{(n)}= \left(  \psi_\uparrow^{(n)} , \, \psi_\downarrow^{(n)} \right) \, , 
\end{eqnarray}
with dimensions of $[\mathrm{length}]^{-1}$, which forces the couplings $g_{S, V, D}^2$ to have dimensions of $[\mathrm{energy}] \cdot [\mathrm{length}]^2$. Momentarily omitting the color index, the diquark term contains the 2D charge conjugation matrix $C = \gamma^2 \gamma^0$, where the Dirac adjoint is $\bar{\psi} \equiv \psi^\dagger \gamma^0$. This interaction couples the quark field $\psi^T$ with one of the same chirality rotated by 180 degrees and charge conjugated: $C \left[ \sigma_3  \exp\left( i  \pi  \sigma_3   /2 \right) \psi        \right] = C \left( \sigma_3   i    \sigma_3   \psi        \right) =  i  C   \psi $, where we include the second factor of $\sigma_3$ to recover the original chirality after the parity rotation. With regards to the color index, first, an anitsymmetrization factor 
\begin{eqnarray}
\delta_A = \delta^{ac} \delta^{bd} -  \delta^{ad} \delta^{bc} \, , 
\end{eqnarray}
appears in order to isolate the attractive channel in the diquark interaction, where $a, b, c, d$ are stand-ins for general quark color indices. A second important point worth mentioning is that in the present form of Eqs.~(\ref{kinetic})-(\ref{diquark1}) our model implies an internal two-index antisymmetric structure defined by the mapping from the fundamental to the two-index representation 
\begin{eqnarray}
 \psi^{(n)}   \to  \sum_{j , k}  c^{(n) j k }  \,   \psi_{[j k ]}   \, .  \label{2indexA}
\end{eqnarray}
Here, the two-index representation in Dirac notation is 
  \begin{eqnarray}
 | \psi_{[ j k]} \rangle = \frac{1}{\sqrt{2}} \left(  | \psi^{(j )} \rangle \,  |  \psi^{(k )} \rangle  - |  \psi^{(k)} \rangle  \,  |  \psi^{(j)} \rangle    \right) \, . 
  \end{eqnarray}
The $N_c$-dependent scaling along with the two-index structure in Eq.~(\ref{2indexA}) will be crucial in our particular extrapolation towards the semiclassical regime.

There are two regions of the three-dimensional coupling space $\left\{ g_i^2 \right\}_{i = S, V, D}$ that exhibit distinctly different symmetries of particular interest to us. They are the planes demarcated by the conditions: 1) $g_V^2  - 4 g_S^2 + 4 g_D^2 = 0$; and 2) $g_V^2  + 2  g_S^2 - 4 g_D^2 = 0$. In the special case $\mu_B = m_0 = 0$, Eq.~(\ref{FullLagrangian}) is symmetric under an $SU(2N)$ chiral transformation in region 1), which gets broken explicitly to $SU(N)_R$ $\times$ $SU(N)_L$ in region 2) wherein the discrete $\mathbb{Z}_4$ subgroup plays an important role in the quantum phase transition driven by the effective quark mass or baryon chemical potential, as we will see. Moreover, there are two regimes determined by the relative size of the vector coupling. These are the vector-dominant and vector-suppressed regimes: $(4/3) g_D^2  < g_S^2  <  g_V^2$ and $g_V^2  <  g_S^2  < (4/3) g_D^2$ for region 1), respectively, and $g_S^2   < (4/3) g_D^2  <  g_V^2$ and $g_V^2   < (4/3) g_D^2  <  g_S^2$ for region 2). Thus, we see that the presence of the repulsive vector interaction enhances chiral symmetry and the diquark interaction restores it, at least at the classical level. For the rest of our work we will focus on region 2) with reduced discrete chiral symmetry and remain within the vector-dominant regime. Note also that setting $g_D^2$, $g_V^2 = 0$ retrieves the Gross-Neveu model; $g_D^2$, $g_S^2 = 0$, the Thirring model; and $g_D^2 = 0$, $g_V^2 = g_S^2$, the Nambu-Jona-Lasinio model.

\section{Condensate Formation}
\label{CondensateFormation}

Several possible condensed phases emerge associated with formation of a Fermi surface in our model, driven by competing short-range attractive forces. A classical assessment of the Lagrangian Eq.~(\ref{FullLagrangian}) shows that quark-antiquark and quark-quark attractions favor chiral and diquark condensates at high quark densities, respectively, with a finite but small coexistence mixed phase region. In contrast, vector meson repulsion favors diquarks over scalar mesons but only at moderate quark densities due to density-density repulsion coming from the time-like part of this interaction. In our work we are particularly interested in the effect of strong coupling in diquark and vector meson channels on the BEC-BCS crossover, at both finite and zero temperature. 

\subsection{General Mean-Field Fermion Pairing}

Semiclassical methods become exact in the large $N_c$ limit and provide a readily tractable approach to solving for the ground state. Mean-field analysis of our model, Eqs.~(\ref{kinetic})-(\ref{diquark}), yields several condensates and fermionic pairing in the superconducting portion of the QCD phase diagram, whose definitions and properties are summarized in Table~\ref{table1}. In what follows, we will omit the color index for clarity. First, diquark pairing $\Delta_d$ is given by the expectation value of the diquark field 
 \begin{eqnarray}
 \langle \phi \rangle = - g_D^2  \langle \psi^T i C \psi \rangle =  \Delta_d   e^{- i \theta} \, ,  \label{diquark}
 \end{eqnarray}
 where $\Delta_d \equiv   - 2 g_D^2 \sqrt{ \langle  \rho_\uparrow \rho_\downarrow   \rangle}$ and $\theta \equiv   \langle \theta_\uparrow  + \theta_\downarrow  \rangle + \pi/2$, expressed in terms of the density and phase of the individual spinor components. Here we note that $\Delta_d$ is a stand-in for uncondensed Cooper pairs in a BCS regime, when $\Delta_d < \mu_B$, and condensed pairs in the BEC phase, when $\mu_B < \Delta_d$. In the BEC phase, breaking the $U_B(1)$ symmetry implicit in Eq.~(\ref{diquark}) results in a nonzero value for $\Delta_d$ with Goldstone modes appearing as massless fluctuations around the direction of symmetry breaking indicated by the phase factor $\theta$. Contrasting this with a similar definition of the chiral or scalar condensate as the expectation of the chiral field, one obtains
\begin{eqnarray}
  \langle \sigma \rangle  = - g_S^2\langle \bar{\psi} \psi  \rangle =  - g_S^2  \langle  \rho_\uparrow  -  \rho_\downarrow   \rangle \, =  \Delta_s. \label{scalarcon}
  \end{eqnarray}
  Similarly, one may examine pairing in the space-like part of the vector meson field in spin space
  \begin{eqnarray}
  \langle V \rangle  &=& - g_V^2 \sum_{i = 1, 2}  \langle \bar{\psi}\gamma^i  \psi  \rangle  = \Delta_i \, , 
  \end{eqnarray}
  where the vector meson condensate is
   \begin{eqnarray}
 \hspace{-1pc}  \Delta_i &=&  \nonumber \\
   && \hspace{-3pc} - \frac{1}{2} g_V^2  \sqrt{ \langle  \rho_\uparrow \rho_\downarrow   \rangle} \left[ \cos\left(    \langle \theta_\uparrow  - \theta_\downarrow  \rangle \right) +  \sin\left(    \langle \theta_\uparrow  - \theta_\downarrow  \rangle \right)     \right] . 
  \end{eqnarray}
  Without loss of generality, we are free to choose the direction of symmetry breaking (i.e., the phase factors) in both the diquark and vector condensates such that $\Delta_d/g_D^2 = 4  \Delta_i/g_V^2= - 2 \sqrt{ \langle  \rho_\uparrow \rho_\downarrow   \rangle}$. Thus, the vector and diquark condensates occupy a region of spin space essentially orthogonal to that of the chiral condensate. Moreover, the time-like contribution from the vector interaction yields a term $\Delta_0 =  g_V^2 \left( \langle \rho_\uparrow \rangle +   \langle \rho_\downarrow   \rangle       \right)$, which provides a positive chemical potential shift such that the net effect of the vector interaction is encapsulated in $\Delta_v = \Delta_0 - \Delta_i$. An important relation can be obtained from Eqs.~(\ref{diquark})-(\ref{scalarcon}) and the quark density $\rho = \langle \psi^\dagger \psi \rangle =  \langle  \rho_\uparrow    \rangle + \langle  \rho_\downarrow \rangle$, which when combined yield 
  \begin{eqnarray}
 \left(\frac{\Delta_v - \Delta_d}{g_D^2}\right)^2   +  \left(\frac{\Delta_s}{g_S^2}\right)^2  = \rho^2  \, . \label{symmetryrelation}
  \end{eqnarray}
  Equation~(\ref{symmetryrelation}) relates the baryon chemical potential to the diquark, vector, and scalar condensates through $\mu_B \sim \rho$. Equation~(\ref{symmetryrelation}) expresses a $U(1)$ rotational symmetry that mixes the chiral symmetry breaking and restoring condensates at fixed chemical potential.

From this analysis we can introduce an effective quark mass $m$ as the dynamical scalar mass generated by a finite scalar (chiral) condensate 
\begin{eqnarray}
m = m_0+ \Delta_s \, , \label{massbreaking}
\end{eqnarray}
where $m_0$ is the bare mass. In this picture the chiral condensate in Eq.~(\ref{scalarcon}), $\Delta_s = - g_S^2  \langle \bar{\psi} \psi \rangle$, appears as a Hartree-Fock term for the meson field $\langle \sigma \rangle$ generated by the attractive scalar part of Eq.~(\ref{FullLagrangian}). Similarly, one may view the chemical potential in Eq.~(\ref{kinetic}) as being shifted by the space-like vector meson and diquark pairing fields. Including these shifts, we define an effective baryon chemical potential as 
\begin{eqnarray}
\tilde{\mu}_B = \mu_B  - \bar{\Delta}_d  \, ,    \label{diquarkbreaking}
\end{eqnarray}
where the net diquark pairing is $\bar{\Delta}_d =  \Delta_v + \Delta_d$, which reflects the fact that the vector meson field enhances diquark pairing. We will soon see how the sign change in $\tilde{\mu}_B$ signals the BEC-BCS crossover where $0 <\tilde{\mu}_B$ marks the pseudogap domain which accounts for pair fluctuations through the pseudogap $\Delta_\mathrm{pg}   = \Delta_v + \Delta_d$ with vanishing BCS spectral gap $\Delta_\mathrm{BCS} = 0$, whereas $\tilde{\mu}_B < 0$ determines the superfluid regime marked by the appearance of a finite spectral gap $\Delta_\mathrm{BCS} \sim |\tilde{\mu}_B|$.

\subsection{Low-Temperature Action}

 In this section we track the development of the Fermi surface and associated condensates as we tune the temperature to zero. We are interested in the moderate to high quark density regimes of the QCD phase diagram. Here, the effective quark mass decreases with increasing baryon chemical potential and density from its vacuum value at low densities to zero at high densities~\cite{Sun2007,Lianyi2010}. Note that this trend characterizes the expected onset of chiral symmetry restoration associated with asymptotic freedom.

We approach the problem by first treating the quark mass and chemical potential as independent parameters of the system, we then recover the more natural interdependence for the final step of constructing the QCD phase diagram in the $T-\mu_B$ plane. This approach is justified if we consider that fixing $m$ while tuning $\mu_B$ (or the reverse) requires changing one or more parameters that do not affect both simultaneously. This might be accomplished for example by tuning $g_V$, which favors the diquark condensate and hence a weaker chiral condensate affecting the size of the constituent quark mass but not the baryon chemical potential. 

Formation of the Fermi surface is addressed by first expanding the spinor fields in $\mathcal{L}$ in terms of the single-particle Dirac states 
\begin{eqnarray}
{\psi^{(n)}}({\bf r}, t) =  \sum_{\mathrm{\bf{k}}}  \,  e^{  - i \epsilon_{\mathrm{\bf{k}}} t /\hbar} \, {\psi^{(n)}_{\mathrm{\bf{k}}}}({\bf r})\, ,\,       \label{decomp1}
\end{eqnarray}
and
\begin{eqnarray}
{\psi^{(n)}}^\dagger({\bf r}, t) =  \sum_{\mathrm{\bf{k}}}  \,  e^{  i \epsilon_{\mathrm{\bf{k}}} t /\hbar} \, {\psi^{(n)}_{\mathrm{\bf{k}}}}^\dagger({\bf r})\, ,\,     \label{decomp2}
\end{eqnarray}
where the summation index labels the single-particle momentum and the superscript is the quark-color index. The spatial spin states for a homogeneous system are defined as 
\begin{eqnarray}
\! \! \psi^{(n)}_{\mathrm{\bf{k}}}({\bf r}) =  \frac{1}{\sqrt{2}} \,  \eta_{\mathrm{\bf{k}}} ^{(n)} e^{i {\mathrm{\bf{k}}}\cdot  {\bf r}} \left(  \,  \xi^{(n)}_{\mathrm{\bf{k}}, \uparrow} \,  e^{ -i \phi_{\mathrm{\bf{k}}}/2} , \;   \xi^{(n)}_{\mathrm{\bf{k}}, \downarrow   }  \,  e^{ i \phi_{\mathrm{\bf{k}}}/2 }   \right)^T \!  \! , 
\end{eqnarray}
where the $\eta_{\mathrm{\bf{k}}} ^{(n)}$ are species and momentum-dependent complex fields subject to bosonic commutation relations
\begin{eqnarray}
 \left[  \eta_{\mathrm{\bf{k}}}^{(n)} , \, {\eta_{\mathrm{\bf{k}'}} ^{(n')}}^* \right]  &=& \delta_{n , n'}    \delta_{\mathrm{\bf{k}} , \mathrm{\bf{k}'}} \, ,   \\
\left[ \eta_{\mathrm{\bf{k}}}^{(n)} , \, \eta_{\mathrm{\bf{k}'}}^{(n')}  \right]  &=&  \left[  {\eta_{\mathrm{\bf{k}}}^{(n)}}^* , \, {\eta_{\mathrm{\bf{k}'}}^{(n')}}^*  \right] = 0 \,  . 
\end{eqnarray}
This is in keeping with our previous discussion. The internal phase depends on the momentum through $\phi_{\mathrm{\bf{k}}} = \mathrm{tan}^{-1}\left(k_y/k_x\right)$, coupling the direction of momentum to the internal spin structure. The spin polarization transverse to the plane is encoded in the parameters $\xi^{(n)}_{\mathrm{\bf{k}}, \uparrow \downarrow}\,$. These are Grassmann variables that encode the fermionic degrees of freedom and adjust the chirality of the Dirac states. They therefore depend completely on the mass $m$ through the interactions, as left and right states comprise distinct subspaces only in the massless noninteracting theory. In particular, we note that for $\xi^{(n)}_{\mathrm{\bf{k}}, \uparrow } = 1$ and $ \xi^{(n)}_{\mathrm{\bf{k}}, \downarrow} =  \pm 1$, the states $\psi^{(n)}_{\mathrm{\bf{k}}}$ solve the right ($+$) and left ($-$) hand chiral spinor equations
\begin{eqnarray}
\big(\pm  i \hbar c \,  \boldsymbol{\sigma} \cdot    \boldsymbol{\nabla}  - \epsilon_{\mathrm{\bf{k}}} \big)   \psi^{(n)}_{\mathrm{\bf{k}}}({\bf r}) = 0 \, .  \label{Diracequation}
\end{eqnarray}

Substituting the field expansions Eqs.~(\ref{decomp1})-(\ref{decomp2}), we obtain 
\begin{widetext}
\begin{eqnarray}
         \hspace{-1pc}&&L_0 =  \sum_{ {\mathrm{\bf{k}}}}   \sum_{\sigma = \uparrow \downarrow }   \sum_{n} \,     \left(-  \omega_{\mathrm{\bf{k}}} +  \epsilon_{\mathrm{\bf{k}}} + \mathcal{M}_{\sigma}   \right)                        {\eta^{(n)}_{\mathrm{\bf{k}}, \sigma}}^*  {\eta^{(n)}_{\mathrm{\bf{k}}, \sigma}}                       \, ,  \label{S0integrated}   \\
         \hspace{-1pc}&&L_I =   \frac{g^2}{2}     \sum_{\mathrm{{\bf j}}, \mathrm{{\bf l }}, \mathrm{{\bf q}} }   \sum_{\sigma = \uparrow \downarrow }    \left[  \sum_{n}              \mathcal{S}_\sigma^{(n, n )}({\mathrm{{\bf j}}, \mathrm{{\bf l }}, \mathrm{{\bf q}} })       \;                                 {\eta_{\mathrm{\bf{ j }} - \mathrm{\bf{q}} ,  \sigma   }^{(n)}}^{\hspace{-.75pc}*} \; \;  \eta_{\mathrm{\bf{j}} ,  \sigma   }^{(n)} \; {\eta_{\mathrm{\bf{l + q}} ,  \sigma  }^{(n)}}^{\hspace{-.75pc}*}  \;\; \eta_{\mathrm{\bf{l}} ,  \sigma }^{(n)} \;  + \! \!  \sum_{n , n' ;  \, n \ne n' }    \mathcal{S}_\sigma^{(n,n')}({\mathrm{{\bf j}}, \mathrm{{\bf l }}, \mathrm{{\bf q}} })             \;                    {\eta_{\mathrm{\bf{ j }} - \mathrm{\bf{q}} ,  \sigma   }^{(n)}}^{\hspace{-.75pc}*} \; \;  \eta_{\mathrm{\bf{j}} ,  \sigma   }^{(n)} \; {\eta_{\mathrm{\bf{l + q}} ,  \sigma  }^{(n')}}^{\hspace{-.5pc}*}  \;\eta_{\mathrm{\bf{l}} ,  \sigma }^{(n')}      \right]  \, ,    \label{brokensymmetry1}
\end{eqnarray}
\end{widetext}
where $L_0$ and $L_I$ are the noninteracting and interacting parts, respectively, expressed in terms of the bare spin components. $L_0$ contains the kinetic, chemical potential, and mass contributions each quadratic in the quark fields for which the spin polarization is now implicit ${\eta^{(n)}_{\mathrm{\bf{k}} , \uparrow \downarrow }}   \equiv {\eta^{(n)}_{\mathrm{\bf{k}}}}      {\xi^{(n)}_{\mathrm{\bf{k}}, \uparrow \downarrow }}$. Here we have defined $\mathcal{M}_{\sigma} = \tilde{\mu}_B + p_\sigma m$, replacing the bare mass $m_0$ and baryon chemical potential $\mu_B$ with the constituent quark mass $m$ and the effective chemical potential $\tilde{\mu}_B$, respectively, from Eqs.~(\ref{massbreaking})-(\ref{diquarkbreaking}). Note in addition the condensed notation for the spin-up and spin-down phase $p_\uparrow = -1, \, p_\downarrow = +1$, induced by a nonzero mass.

The quartic interactions are contained in $L_I$ with $g^2 = g_V^2 - g_S^2$ the coupling strength, working in the vector-dominant regime of the $SU(N)_R \times SU(N)_L$ theory described in Sec.~\ref{Model}. Inside the square brackets, we have split the interactions into the two types: intra-species and inter-species contact terms. In each term the spin polarization is denoted by the subscript $\sigma$. Incident momenta are denoted by vectors $\mathrm{\bf{j}}$ and $\mathrm{\bf{l}}$, and transferred momentum by $\mathrm{\bf{q}}$ such that integration over the space and time variables is consistent with conservation of energy and momentum during collisions. 

Finally, the spin and momentum dependent structure factor $\mathcal{S}({\mathrm{{\bf j}}, \mathrm{{\bf l }}, \mathrm{{\bf q}} })$ arrises out of the coupling between the internal spinor phases of the incoming and outgoing particles. This factor is explicitly given by 
\begin{eqnarray}
&&\hspace{-2pc}\mathcal{S}_\sigma({\mathrm{{\bf j}}, \mathrm{{\bf l }}, \mathrm{{\bf q}} }) =  e^{ \pm i \left( \phi_{\mathrm{\bf{l}}} -  \phi_{\mathrm{\bf{l + q}}} + \phi_{\mathrm{\bf{j}}}  - \phi_{\mathrm{\bf{j - q}} }  \right) /2} =  \label{phasefactor} \\
&&\hspace{-2pc}\exp \frac{ \pm i}{2} \! \!  \left\{   \mathrm{cos}^{-1} \! \left[    \frac{\mathrm{{\bf l }} \cdot \left( \mathrm{{\bf l }} + \mathrm{{\bf q }} \right)}{ \left| \mathrm{{\bf l }}\right| \left| \left( \mathrm{{\bf l }} + \mathrm{{\bf q }} \right) \right| }  \right]  +       \mathrm{cos}^{-1} \! \left[    \frac{\mathrm{{\bf j}} \cdot \left( \mathrm{{\bf j }} - \mathrm{{\bf q }} \right)}{ \left| \mathrm{{\bf j }}\right| \left| \left( \mathrm{{\bf j }} - \mathrm{{\bf q }} \right) \right| }  \right]   \right\}  \! , \, \nonumber 
\end{eqnarray} 
with the plus and minus signs associated with spin up and down, respectively.

Working from the momentum-space Lagrangian in Eqs.~(\ref{S0integrated})-(\ref{brokensymmetry1}), we now want to focus on the low-temperature regime around the formation of a condensate. The first step is to recall that the mass and chemical potential depend on temperature. We would like to resolve the thermodynamic dependence of the system on $m$ and $\mu_B$. The way forward is to note that $m$ runs (scales) with $\mu_B$, which provides an additional energy scale at low temperatures. Fixing $m$ and lowering the temperature by considering a temperature-dependent continuous mass renormalization which holds $m$ constant throughout the process, we simultaneously invoke the large $N_c$ limit. This is accomplished by inserting the explicit dependence of the coupling on $N_c$, i.e., we want $g^2 N_c$ to remain constant (yet possibly with $g^2 N_c \gg 1$) as $N_c \to \infty$. This limit buys us considerable advantage by curtailing fluctuations in the quark amplitude fields ${\eta^{(n)}_{\mathrm{\bf{k}}, \uparrow \downarrow}}$.

\begin{figure}[h]
\centering
\subfigure{
\hspace{-.25pc}\includegraphics[width=.48\textwidth]{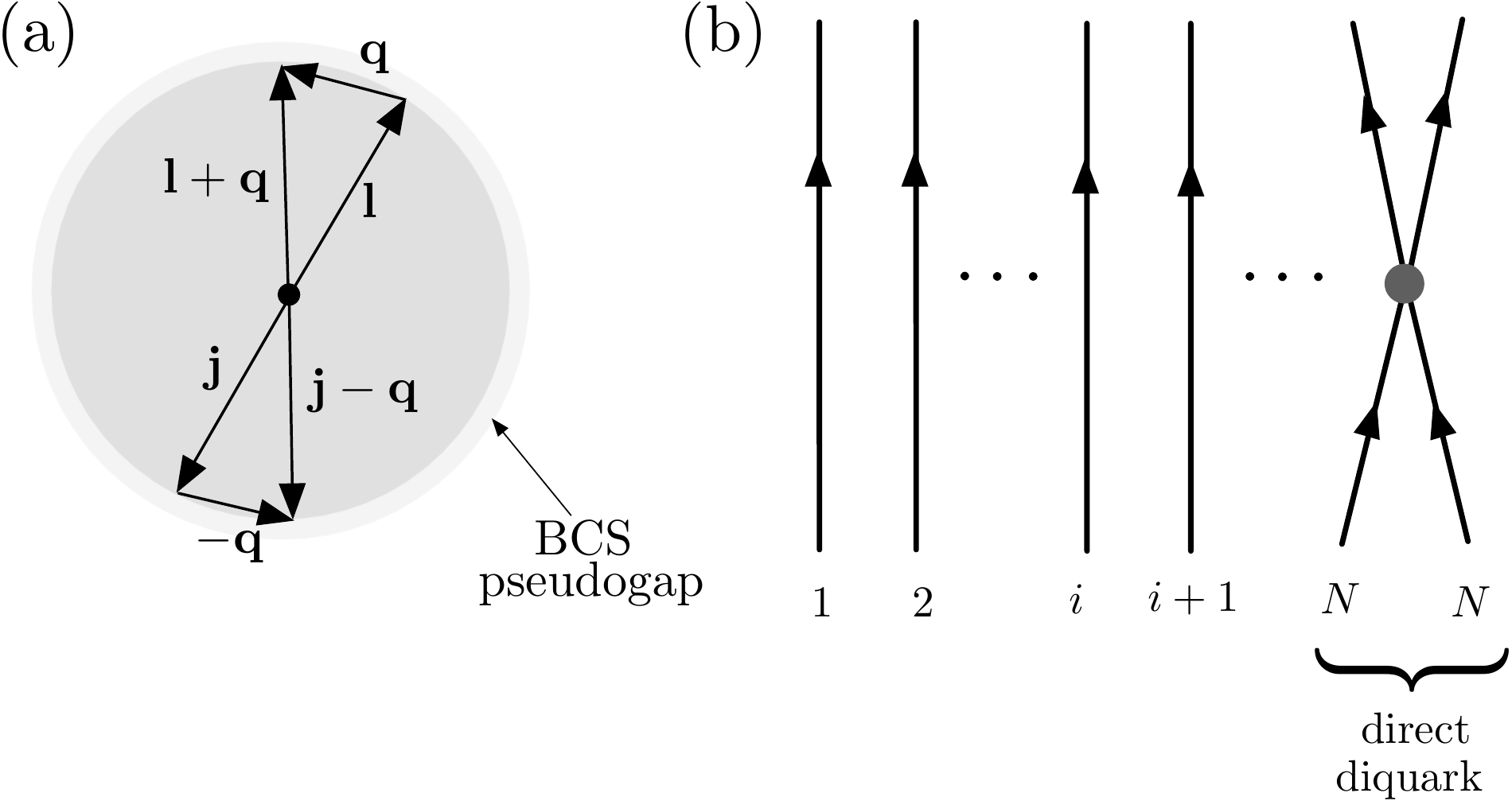}} \\
\caption[]{\emph{Direct color pairing at large N}. (a) Weakly bound BCS states near the Fermi surface resemble conventional superconductivity. (b) Diquarks with the same internal color charge interact with the baryon background. 
}
\label{FermiSurface}
\end{figure}

The behavior of the two sums inside the brackets of Eq.~(\ref{brokensymmetry1}) is now evident. For one, in the limit $N_c \to \infty$ the interaction between any two (fundamental) quarks becomes weak since the interaction goes like $g^2/N_c$, even though the overall system may be strongly interacting since the number of degrees of freedom grows with $N_c$. This has the effect of greatly reducing the momentum transfer $\mathrm{{\bf q }}$ between any two quarks with the same color charge, hence constraining their individual momentum states to lie near $\mathrm{{\bf k}}_F$. The effect of the remaining $N_c-1$ quarks on the pair in question can then be accounted for through an overall approximate Hartree contribution which becomes exact in the large $N_c$ limit. Note however that for $|\mathrm{{\bf q }}|  <<  | \mathrm{\bf{k}}_F|$, momenta for the direct terms (intra-species terms in Eq.~(\ref{brokensymmetry1})), cannot be identical due to Pauli blocking, yet these may be reduced sufficiently to approximate the number of terms from this contribution to be close to $N_c$. These reduced momentum transfer conditions are illustrated in Fig.~\ref{FermiSurface}, showing pairing of opposite momentum states near the Fermi surface. This is a universal property of low-temperature Fermi systems in the presence of arbitrarily weak attractive interactions.

In contrast, momentum transfer in the second sum containing the exchange terms (inter-species) cannot be reduced to lie near the Fermi surface. To see this, consider that while the impulse $\mathrm{{\bf q }}$ received by a single quark interacting with another quark may be small, the exchange interaction in Eq.~(\ref{brokensymmetry1}) includes many such impulses which must be summed over. The total momentum transfer to a single quark then scales as $N_c |\mathrm{{\bf j }}  - \mathrm{{\bf q }}|  , \,N_c | \mathrm{{\bf l}} + \mathrm{{\bf q }}| \simeq N_c |\mathrm{{\bf k }}_F|$. Moreover, this sum contains $N_c( N_c-1) \simeq N_c^2$ terms. This, combined with the overall factor $g^2/ N_c$, tells us that the first summation in Eq.~(\ref{brokensymmetry1}) can be neglected after taking $N_c \to \infty$: its dependence on $N_c$ is canceled by the factor of $1/N_c$ in the interaction, whereas all other terms grow at a rate proportional to $N_c$. It is important to keep in mind that our counting arguments translate to the two-index representation with the size of the color Hilbert space expanded by $N_c \to N_c (N_c -1)/2$.

\begin{figure}[h]
\centering
\subfigure{
\hspace{-.25pc}\includegraphics[width=.48\textwidth]{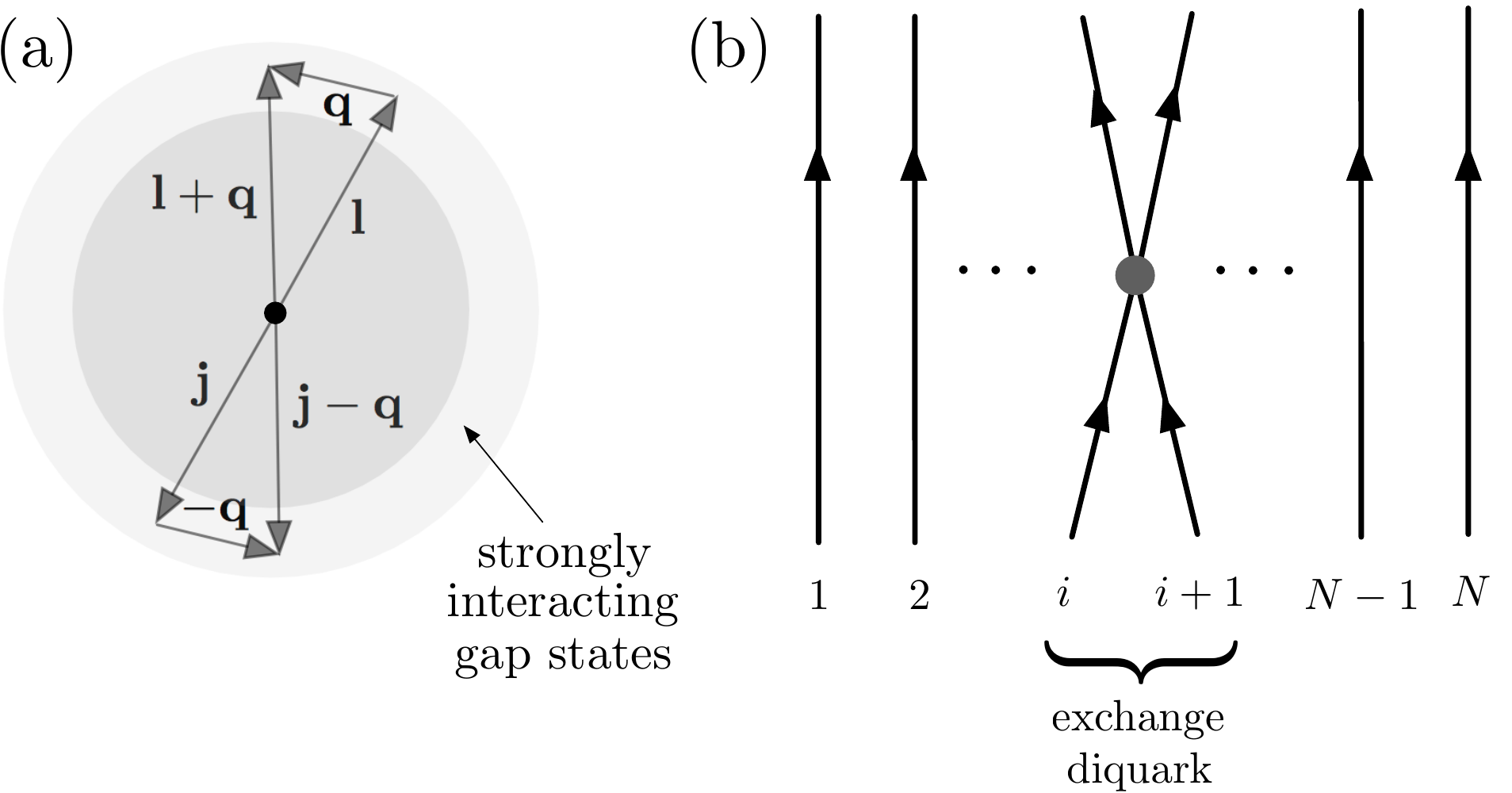}} \\
\caption[]{\emph{Exchange color pairing at large N}. (a) Summing over deep BEC gap states with different color charge produce a large gap. (b) Diquarks composed of quarks with different color charges internal to the baryon. 
}
\label{FermiSurface2}
\end{figure}

Thus, we find two emergent regimes that decouple at high and low energies (or equivalently, densities): 
\begin{enumerate}

\item The direct interaction in Eq.~(\ref{brokensymmetry1}) (first sum) contains a low-energy/density regime made up primarily of weakly interacting diquarks. At low baryon densities diquark pairing within the same color charge dominates the spectrum, similar to the standard result obtained for mesons. In the large $N_c$ limit these should be described by a weakly interacting classical system. 

\item The exchange interaction in Eq.~(\ref{brokensymmetry1}) (second sum) describes the high-energy/density regime which accounts for interactions within and between heavy baryons whose masses scale like $M_B \sim N_c$. Baryon self-interactions can be viewed as individual constituent quarks within the baryon interacting via diquark exchange. Diquark pairing in this case involves quarks of different color charge whose resulting condensate produces a strongly bound BCS states. 

\end{enumerate}

\noindent We will see that regimes 1. and 2. above are also associated with weak and strong coupling, respectively, since the baryon mass $M_B$ scales like $1/(1/N_c)$, and the interaction is $g^2/N_c$. In the large $N_c$ limit, type 2. diquarks produce a robust condensate, whereas type 1. diquarks will dissociate at a much lower temperature.

We now focus on quark fluctuations in the baryon sector with internal diquarks in the large $N_c$ limit of Eq.~(\ref{brokensymmetry1}). In this limit, fields can be approximated by their classical limits. The Euler-Lagrange equations give the equations of motion
\begin{widetext}
\begin{eqnarray}
   \hspace{-2pc}&&  \sum_{n, \,  \mathrm{\bf{k}} ,  \, \sigma,  \, p_{\uparrow, \downarrow} }  \left[   \left( - \omega_\mathrm{\bf{ k}} +  \epsilon_\mathrm{\bf{ k}}          +  \mathcal{M}_{\sigma}      \right)    \eta_{\mathrm{\bf{ k}},  \sigma   }^{(n)}          +    \,   2  \pi   g^2  \!\! \sum_{n', \, \mathrm{\bf{l}}, \, \mathrm{\bf{q}}  }   \Delta_{ \mathrm{\bf{ k}} -   \mathrm{\bf{q} },  \mathrm{\bf{l}} + \mathrm{\bf{q}}, \sigma  }^{(n , n')}  \,  \eta_{\mathrm{\bf{l}},  \sigma }^{(n')}   \right]  = 0  \,  .  \label{groundstateeq}
\end{eqnarray}
\end{widetext}
Momentum sums are over a volume of radius $\sim 2 \pi/\delta$, where $\delta$ is the characteristic thickness of the of the baryon boundary. For low-energy fluctuations around a Hartree mean-field background, momenta are much smaller than $2 \pi/\delta$. This allows us to approximate ground state fluctuations by taking $\mathrm{\bf{k}},  \mathrm{\bf{l}}, \omega_0,  \epsilon_0   \to 0$ in Eq.~(\ref{groundstateeq}), which gives
\begin{eqnarray}
   \hspace{-2pc}&&   \sum_{n ,  \, \sigma,  \, p_{\uparrow, \downarrow} }  \left[   \mathcal{M}_{\sigma}    \,       +    \,   2  \pi   g^2  \! \sum_{n', \, \mathrm{\bf{q}}  }   \Delta_{ \mathrm{\bf{q} }, - \mathrm{\bf{q}}, \sigma  }^{(n , n')}    \right]  \,  \eta_{ 0 ,  \sigma }^{(n)}    = 0   ,   \label{groundstateeq2}
\end{eqnarray}
where anti-symmetrization with respect to the color index allows all $N_c$ quarks to occupy the same ground state wavefunction, i.e., $\eta_{ 0 ,  \sigma }^{(n)} = \eta_{ 0 ,  \sigma }^{(n')}$ for all $n$ and $n'$. In diagonalized form Eq.~(\ref{groundstateeq2}) describes fluctuations over the background pairing field. Notably, the character of the fluctuations in Eq.~(\ref{groundstateeq}) changes abruptly when the sign of $\tilde{\mu}_B$ in the quadratic term changes from positive to negative as the diquark pairing exceeds the baryon chemical potential in Eq.~(\ref{diquarkbreaking}). Recall here that in the presence of attractive interactions pairing occurs in either a weakly bound BCS regime, $0< \tilde{\mu}_B$, or a strongly bound BEC regime, $\tilde{\mu}_B <  0$, consistent with Eq.~(\ref{diquarkbreaking}), as previously discussed. Note also that the interaction is positive when expressed in terms of the bare spinor components, hence a nonzero spectral gap appears in the ground state solution of Eq.~(\ref{groundstateeq}) at the superfluid transition temperature $T_c$. 

 Let us examine the pairing field in Eq.~(\ref{groundstateeq2}). It is given by
\begin{eqnarray}
\Delta_{ \mathrm{\bf{ q}} , - \mathrm{\bf{ q}}, \sigma  }^{(n , n')}  =  \langle {\eta_{- \mathrm{\bf{ q}},  \sigma  }^{(n')}}^{\hspace{-.2pc}*}   \eta_{\mathrm{\bf{ q}},  \sigma   }^{(n)}  \rangle \, , 
\end{eqnarray}
 such that
\begin{eqnarray}
\sum_{\mathrm{\bf{ q}}, \sigma}  \Delta_{ \mathrm{\bf{q}} , - \mathrm{\bf{q}}, \sigma  }  = \frac{|\mathcal{M}_\sigma|}{2 \pi N_c^2 g^2}   \, ,  \label{groundsolution}
\end{eqnarray}
where we have omitted a factor of $N_c^2$ that comes from summing over the color indices on the left hand side: $N_c^2 \Delta_{ \mathrm{\bf{q}} , - \mathrm{\bf{q}}, \sigma  }  \equiv  \sum_{n , n'}  \Delta_{ \mathrm{\bf{q}} , - \mathrm{\bf{q}}, \sigma  }^{(n , n')}$. Note that the phase difference between upper and lower spin components of right and left chiral states in the two-spinor representation are $+1$ and $-1$, respectively. Equation~(\ref{groundsolution}) for the pairing field gives us an explicit expression for the spectral or superconducting gap in terms of the baryon chemical potential and the quark mass. To see this, consider the pairing equations for up and down spin polarizations in Eq.~(\ref{groundsolution}) 
  \begin{eqnarray}
  &&\sum_{\mathrm{\bf{q}}} \Delta_{ \mathrm{\bf{q}} , - \mathrm{\bf{q}}, \uparrow  } =  \frac{| \tilde{\mu}_B + m |}{2 \pi g^2}  , \, \\
   &&\sum_{\mathrm{\bf{q}}}\Delta_{ \mathrm{\bf{q}} , - \mathrm{\bf{q}}, \downarrow  } = \frac{ | \tilde{\mu}_B - m |}{2 \pi g^2}  \, . \; \; 
  \end{eqnarray}
  Multiplying these gives 
   \begin{eqnarray}
 4  g^4  \sum_{\mathrm{\bf{q}} , \, \mathrm{\bf{q}}'   } \! \! \Delta_{ \mathrm{\bf{q}} , - \mathrm{\bf{q}}, \uparrow  } \,  \Delta_{ \mathrm{\bf{q}}' , - \mathrm{\bf{q}}', \downarrow  }   = \frac{\tilde{\mu}_B^2 - m^2}{ \pi^2 }   \, .\label{explicitdiquark1}
  \end{eqnarray}
   The left hand side of Eq.~(\ref{explicitdiquark1}) is the square of a spatially uniform BCS gap $\Delta_\mathrm{BCS}$, consistent with the analysis around Eq.~(\ref{diquark})-(\ref{symmetryrelation}):
  \begin{eqnarray}
 \Delta_\mathrm{BCS}({\bf r})^2    \equiv  \Delta_\mathrm{BCS}^2 =  4  g^4 \! \sum_{\mathrm{\bf{q}}, \, \mathrm{\bf{q}}'} \!  \Delta_{ \mathrm{\bf{q}} , - \mathrm{\bf{q}}, \uparrow  } \,  \Delta_{ \mathrm{\bf{q}}' , - \mathrm{\bf{q}}', \downarrow} \, , 
   \end{eqnarray}
  which allows us to express Eq.~(\ref{explicitdiquark1}) in terms of the mass-dependent gap
  \begin{eqnarray}
 \Delta_\mathrm{BCS}(m)  \,  =    \, \frac{ |\tilde{\mu}_B|}{\pi} \sqrt{ 1 - \left(\frac{m }{\tilde{\mu}_B} \right)^2       }  \, . \label{explicitdiquark2}
  \end{eqnarray}
Equation~(\ref{explicitdiquark2}) identifies the BEC-BCS transition at the location where the gap dissolves due to the quark mass $m$ approaching the value of the effective chemical potential $|\tilde{\mu}_B|$ at zero temperature. Thus, $|\tilde{\mu}_B|$ gives the asymptotic value of the superconducting gap $\Delta_\mathrm{BCS}$ when the quark mass vanishes far from the critical point. We will address this quantum phase transition associated with CSB in detail in Sec.~\ref{QPT}.

 \section{Finite-Temperature Phase Transitions}
\label{FiniteTPT}

 Starting at some temperature $T$ between the critical superfluid phase transition temperature $T_c$ and that of scalar meson and diquark dissociation $T^*$, i.e., $T_c <  T < T^* $, the spectral and pseudogaps are given by $\Delta_\mathrm{BCS}(T) = 0$ and $\Delta_\mathrm{pg}(T) = \bar{\Delta}_d(T)$. Then tuning the temperature below the superfluid critical point, $T > T_c$ $\to$ $T < T_c$, where the pseudogap vanishes and a finite spectral gap appears. To study this transition, we assume the following critical behavior for diquark pairing taken with respect to the dissociation temperature 
\begin{eqnarray}
  \bar{\Delta}_{d}(T)  =   \bar{\Delta}_{d}(0)            \sqrt{ 1 - \left(T/T^* \right)^2}\,.  \label{pseudogapcriticality1}
\end{eqnarray}
Note that by introducing a quadratic form for the temperature dependence we are inherently assuming finite $1/N$ corrections, since otherwise one should expect a more squared off, flat, and singular temperature dependence for condensates. Since we deal only in moderate to large chemical potential $\mu_B$, the superfluid critical point $T_c$, where $\tilde{\mu}_B =0$, occurs when Eq.~(\ref{pseudogapcriticality1}) nears its maximum value, i.e., where $T/T^* <<1$. We thus use a binomial expansion of Eq.~(\ref{pseudogapcriticality1}) to obtain the lowest-order finite temperature behavior from Eq.~(\ref{diquarkbreaking})
\begin{eqnarray}
\tilde{\mu}_B(T) \simeq   - |\tilde{\mu}_{B}^{(0)}|   + \frac{1}{2}  \left(      |\tilde{\mu}_{B}^{(0)}|     + \mu_B \right) \left( \frac{T}{T^*} \right)^2 + \dots  , \label{binomial1}
\end{eqnarray}
where $|\tilde{\mu}_{B}^{(0)} | = |  \mu_B  - \bar{\Delta}_{d}(0) |$, which yields a relation between the superfluid and dissociation critical temperatures obtained through $\tilde{\mu}_B(T_c) = 0$:
\begin{eqnarray}
  T_c =  \sqrt{\frac{ 2 \Delta_\mathrm{BCS}^{(0)}}{ \Delta_\mathrm{BCS}^{(0)}   + \mu_B   }  } \;  T^*       \,.  \label{pseudogapcriticality2}
\end{eqnarray}
where we have used the fact that the spectral gap and effective chemical potential are equal at zero temperature and zero quark mass, $\Delta_\mathrm{BCS}^{(0)} =  |\tilde{\mu}_{B}^{(0)}|$, with the BCS and BEC regimes corresponding to the regions $T_c < T < T^*$ and $T< T_c  < T^*$, respectively. Note that the superscript on $\Delta_\mathrm{BCS}^{(0)}$ indicates the spectral gap at zero temperature but not necessarily at zero mass, so that Eq.~(\ref{pseudogapcriticality2}) generalizes to any value of the quark mass, consistent with Eq.~(\ref{explicitdiquark2}). 

The result in Eq.~(\ref{pseudogapcriticality2}) applies only in the limit $T_c << T^*$, which implies 
\begin{eqnarray}
  \sqrt{\frac{ 2 \Delta_\mathrm{BCS}^{(0)}}{  \Delta_\mathrm{BCS}^{(0)}   + \mu_B   }  }  << 1    \,.  \label{pseudogapcriticality3}
\end{eqnarray}
There are two regimes for $\mu_B$ that satisfy the inequality Eq.~(\ref{pseudogapcriticality3}). One occurs for moderate $\mu_B$ when $\Delta_{sg}(0)$ vanishes at $|\tilde{\mu}_{B}| = m$, due to increasing quark mass according to  Eq.~(\ref{explicitdiquark2}). We will soon show how a full quantum mechanical treatment of bound states in our system reveals a dissociation temperature curve vanishing at the same critical point as that of the superfluid critical temperature, $T^* = T_c = 0$. Hence, the critical point in Eq.~(\ref{explicitdiquark2}) describes a sharp BEC-BCS quantum phase transition versus a smooth crossover. The second occurs for asymptotically large values of $\mu_B$ for which $\mu_B >>  \Delta_\mathrm{BCS}(0)$. A similar calculation of bound states in this regime reveals a non-vanishing pseudogap, thus describing a smooth BEC-BCS crossover at large quark density.

\subsection{Ginzburg-Landau Theory for Large $N_c$ Quarks}
\label{GLT}

In the large $N_c$ limit quantum mechanics simplifies considerably, since sums over large numbers of fermion fields self-average, resulting in asymptotically small fluctuations with $1/N_c$ suppression of quark loops. This argument applies to fundamental quarks and we would now like to extend as much of it as possible to the two-index antisymmetric representation. In the latter case, though, a key difference is that quark loop are not suppressed. Introducing the additional antisymmetry in the two-index formalism forces antisymmetry upon $\psi_\mathrm{space} \times \psi_\mathrm{spin}$. We would like to model light quarks at high densities, thus retaining Dirac structure seems reasonable. We will also see that at high densities it is energetically favorable to allow $U(N_c)$ symmetry breaking in the external quark amplitude and phase, associated with its energy, and taking the anticommuting fermionic structure to reside purely in the internal spin degree of freedom, which couples to the direction of quark momentum. This paradigm coincides precisely with that of spin-charge separation in which the fundamental spin and charge of the quark field decouple in the strongly interacting collective state: the fermionic nature of the quark field resides in the spin degree of freedom; often referred to as a spinon; with the bosonic charge degrees of freedom, called chargons, available for condensation into a macroscopic quark amplitude. Thus for our large $N_c$ limit, the baryonic wavefunction finds all $N_c$ quarks residing in the same spatial mean field single-particle state, with regards to amplitude and overall phase, which must be solved for self consistently, with the direction of the quark current coherently spread over all values the internal spinor phase. It is important to point out that, although each quark provides only a small contribution to the baryon energy, the total interaction energy is large. Hence, the strongly interacting nature of the theory means that the semi-classical Hartree wavefunction will be some averaging over highly entangled underlying Dirac scattering states, with exact coherence in some overall amplitude and phase and maximal decoherence in the direction of quark current.

\subsubsection{Mean Field Theory with Spin-Charge Separation}

We now shift the focus of our discussion to the two-index framework for quarks, retaining the Dirac spin structure combined with spin-charge separation. Choosing the Hartree single-particle states to be some strongly interacting combination of the original scattering states 
\begin{eqnarray}
 \Phi_{\mathrm{\bf k}, \sigma}          \Psi_{  \mathrm{\bf k}, \sigma} = \sum_{  \mathrm{\bf j}, ( \alpha, \beta)  }   {c_\mathrm{\bf j}}^{ (\alpha, \beta) }   \,   e^{i p_\sigma \phi_\mathrm{\bf j}^{ ( \alpha, \beta) }/2}     \,   \eta^{(\alpha, \beta)}_{  \mathrm{\bf j}, \sigma} \, ,  \label{HartreeSoln}
\end{eqnarray}
where ${\mathrm{\bf j}}$ is the momentum of the two-index quark labeled by $\alpha$ and $\beta$, with $\mathrm{\bf k} = \sum_n  \mathrm{\bf j}_n$, the total baryon momentum. On the left hand side of Eq.~(\ref{HartreeSoln}), we have introduced the fermionic spinon and bosonic chargon fields, $\Phi_{\mathrm{\bf k}, \sigma}$ and $\Psi_{  \mathrm{\bf k}, \sigma}$, respectively, such that 
\begin{eqnarray}
\left\{  \Phi_{\mathrm{\bf k}, \sigma}, \, \Phi_{\mathrm{\bf k}', \sigma'}^*    \right\}    =    \delta_{\mathrm{\bf k}, \mathrm{\bf k}'}  \delta_{\sigma , \sigma'} \; , \;\;\; \left[  \Psi_{\mathrm{\bf k}, \sigma}, \, \Psi^*_{\mathrm{\bf k}', \sigma'}    \right]  = \delta_{\mathrm{\bf k}, \mathrm{\bf k}'}  \delta_{\sigma , \sigma'} \, , \;\;\; \left[  \Phi_{\mathrm{\bf k}, \sigma}, \, \Psi_{\mathrm{\bf k}', \sigma'}   \right]  = \left[  \Phi^*_{\mathrm{\bf k}, \sigma}, \, \Psi^*_{\mathrm{\bf k}', \sigma'}   \right]    =   0 \, . 
\end{eqnarray}
We can recast the original Hamiltonian of our system in terms of the collective bosonic states $\Psi_{  \mathrm{\bf k}, \sigma}$ by splitting off the fermionic spin portion and renormalizing the energy, chemical potential, mass, and coupling, which leads to
\begin{eqnarray}
   \!  \!   \! H_0 &=&  \sum_{\mathrm{\bf k} , \sigma }    \left( \bar{\epsilon}_\mathrm{\bf k} + \bar{\tilde{\mu}}_B + p_\sigma  \bar{m} \right)  \,  \left| \Psi_{\mathrm{\bf k},\sigma }\right|^2 \, ,   \label{bs1red}   \\
   \!  \!  \! H_I &=&  \frac{\bar{g}^2}{2}  \sum_{\mathrm{\bf j}, \mathrm{\bf q}, \mathrm{\bf l} , \sigma }       {\Psi_{\mathrm{\bf{ j }} - \mathrm{\bf{q}} ,  \sigma   }}^{\hspace{-1.75pc}*} \hspace{1.5pc}  \Psi_{\mathrm{\bf{j}} ,  \sigma   } \; {\Psi_{\mathrm{\bf{l + q}} ,  \sigma  }}^{\hspace{-1.75pc}*}  \hspace{1.5pc}  \Psi_{\mathrm{\bf{l}} ,  \sigma }  \,  .   \label{brokensymmetry2Red} 
\end{eqnarray}
The bar notation indicates the renormalized parameters. In particular, the interaction strength gets renormalized in a non-trivial way, $g \to \bar{g}$, where now $\bar{g}$ reflects the weakly interacting nature of the Hartree approximate single-particle wavefunction $\Psi_{  \mathrm{\bf k}, \sigma}$. The renormalization of Eqs.~(\ref{bs1red})-(\ref{brokensymmetry2Red}) has an interesting physical interpretation. The $U(N_c)$ charged scalar field $\Psi$ is a collective state that encodes dynamics of the color charged quark current. The full dynamics for the spin degrees of freedom does not appear in Eqs.~(\ref{bs1red})-(\ref{brokensymmetry2Red}). Rather, a complete picture for $\Phi$ requires solving the problem of a strongly correlated quantum spin liquid formed out of quark spins separated from the $U(N_c)$ current~\cite{Balents2010}. Thus, from the perspective of the field $\Psi$ the spin liquid manifests as a background renormalization of the parameters in Eqs.~(\ref{bs1red})-(\ref{brokensymmetry2Red}). Lowering the temperature, the wavefunction tends toward its ground state $\lim_{T \to 0} \Psi_{  \mathrm{\bf k}, \sigma} \to  \Psi_{ 0, \sigma}$, the shifted ground state energy taken to $\epsilon_\mathrm{\bf k} \to \epsilon_0 = 0$, for which the Hamiltonian reduces to 
 \begin{eqnarray}
H =     { \bf \mathcal{M}}  |\Psi_0|^2   +   \frac{\bar{g}^2}{2}      |\Psi_0|^4     \, ,   \label{brokensymmetry3Red}
 \end{eqnarray}
with the definitions ${\bf  \mathcal{ M}}$ $\equiv$ $\left[  \left(  \tilde{\mu}_B - m \right)  , \, \left(  \tilde{\mu}_B + m   \right) \right]$ and $|\Psi_0|^2$ $\equiv$ $( \,   \left| \Psi_{0, \uparrow}\right|^2 \, , \,  \left| \Psi_{0,\downarrow}\right|^2 \,)^T$. Here, for clarity, we have omitted the bar notation for all parameters except the renormalized coupling, a convention we adhere to for the remainder of our work. If we examine first the zero-mass limit of Eqs.~(\ref{bs1red})-(\ref{brokensymmetry2Red}) in the presence of quark-quark attraction combined with repulsion from vector mesons dissolved inside the baryon, lowering the temperature drives the effective chemical potential $\tilde{\mu}_B$, in Eq.~(\ref{brokensymmetry3Red}), through a sign change ${\bf  \mathcal{ M}} >0$ $\to$ ${\bf  \mathcal{ M}} <0$. Thus, a second-order phase transition occurs wherein all $N_c\left( N_c -1 \right)/2$ quarks condense into a single spatial and chiral state $\Psi_{0, R (L)}$ with finite macroscopic amplitude and phase determined by the minimum of the effective potential
\begin{eqnarray}
U_\mathrm{eff}  \equiv   -  |{ \bf \mathcal{M}}| \,   |\Psi_0|^2   +   \frac{\bar{g}^2}{2}      |\Psi_0|^4 \, .  \label{EffPot}
\end{eqnarray}

\begin{figure}[]
\centering
\subfigure{
\includegraphics[width=.54\textwidth]{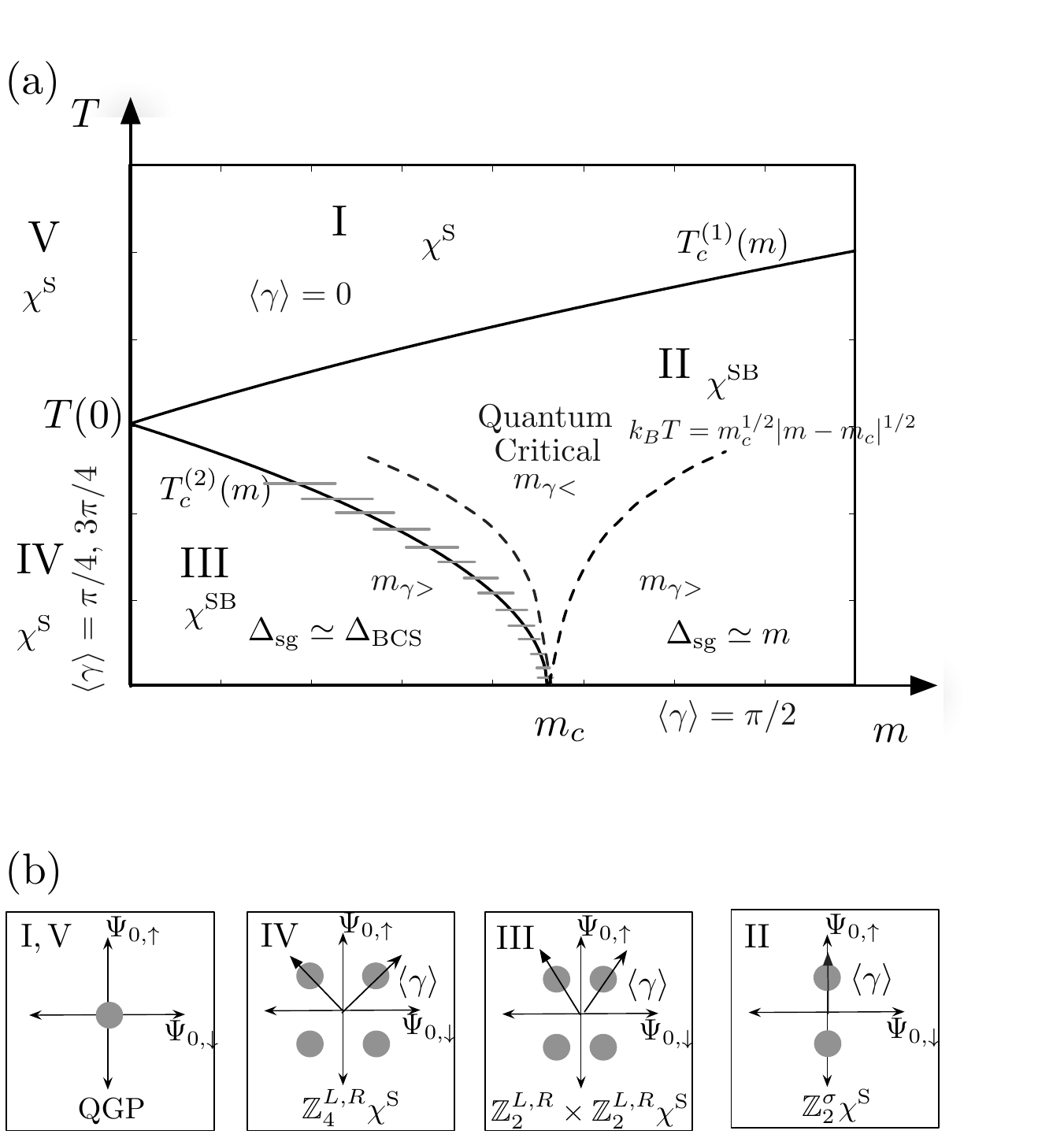}} \\
\caption[]{\emph{Temperature-mass phases of (2+1)d QCD}. (a) Phases that appear in the formation of a Fermi surface are separated by Landau-Ginzburg-Wilson type transitions. Chiral symmetry of the full theory in each region is indicated as either broken ($\chi^\mathrm{SB}$) or retained ($\chi^\mathrm{S}$). In regions I and V ($T > T_c^{(1)}$), no well-defined Fermi surface exists and the system is described by a quark-gluon plasma (QGP). In region II ($T_c^{(2)}< T< T_c^{(1)}$), the Fermi surface begins to form in the presence of a large quark mass or a small mass at moderately low temperatures. Region III ($T< T_c^{(2)}$, $m \ne 0$) corresponds to a Fermi surface with a finite spectral gap in the presence of a small, but non-zero, quark mass. In region IV ($T< T_c^{(2)}$, $m = 0$), the Fermi sea is comprised of nearly asymptotically free massless quarks. The quantum critical region is indicated between the dashed curves emanating from the quantum critical point $m = m_c$. (b) Minima of the ground state (GS) effective potential in Eq.~(\ref{brokensymmetry1}) are depicted as disks in the spin-space coordinate plane for each region in (a) with spin up and down along the vertical and horizontal axes, respectively. The inherited chiral symmetry degeneracy group of the ground sate is also indicated. }
\label{PseudospinDomain}
\end{figure}

To determine the finite-temperature phase transitions for general values of the quark mass $m$, we must now compute the Hessian matrix for the effective potential in Eq.~(\ref{EffPot}). This yields two regimes defined by the following inequalities: 

\begin{enumerate}

\item  for $m    -  | \tilde{\mu}_B|$ $>$ $0$ and $- m  -  | \tilde{\mu}_B| <0$, two minima occur at $|\Psi_{0, \uparrow}|$ $=$ $\pm \sqrt{ (| \tilde{\mu}_B| + m)/\bar{g}^2}$, $|\Psi_{0, \downarrow}|$ $=$ $0$;

\item for $m    -  | \tilde{\mu}_B|$ $< $ $0$ and $- m    -  | \tilde{\mu}_B| < 0$, four minima occur at $|\Psi_{0, \uparrow}|$ $=$ $\pm \sqrt{(| \tilde{\mu}_B| + m)/\bar{g}^2}$, $|\Psi_{0, \downarrow}|$ $=$ $\pm \sqrt{(|\tilde{\mu}_B| - m)/\bar{g}^2}$.

\end{enumerate}
To delineate the temperature dependence for the regimes given by the above inequalities we use the expansion Eq.~(\ref{binomial1}) with $\Delta_{sg}(0)$ the spectral gap at zero mass and zero temperature, for which the two regimes lead to two critical temperature (mass-dependent) curves:
\begin{eqnarray}
T_c^{(1)}(m) &=&  C T^* \,   \sqrt{ 1 + \frac{m}{m_c}  }  \, , \label{CTemp1} \\
T_c^{(2)}(m)  &=&   C T^* \,   \sqrt{ 1 -  \frac{m}{m_c}  } \, ,  \label{CTemp2}
\end{eqnarray}
where the phase corresponding to four minima is delineated by the condition $T < T_c^{(2)}(m)$, and that with two minima by the condition $T_c^{(2)}(m) < T < T_c^{(1)}(m)$. The overall constant in Eqs.~(\ref{CTemp1})-(\ref{CTemp2}) is 
\begin{eqnarray}
C = \sqrt{\frac{2 |\tilde{\mu}_B^{(0)}|}{|\tilde{\mu}_B^{(0)}| + \mu_B}} \, , 
\end{eqnarray}
but more importantly that the critical mass in Eq.~(\ref{CTemp2}) is given by $m_c = |\tilde{\mu}_B(0)|$, in line with the previous result Eq.~(\ref{explicitdiquark2}). Also, the dissociation temperature $T^*$ is evaluated at $m=0$. Figure~\ref{PseudospinDomain} depicts the various finite-temperature regimes discussed so far. 

\subsubsection{Quantum Criticality}

If we examine that part of the phase diagram that extends just above $m_c$ in Fig.~\ref{PseudospinDomain}(a), for $T>0$, we encounter the quantum critical region. The dynamics in this region is characterized by the local thermal equilibration time $\tau_\mathrm{eq}(|m - m_c|, T)$ as a function of temperature and the deviation from the critical point $m_c$. Moving from $T=0$ to $T > 0$ in Fig.~\ref{PseudospinDomain}, a second energy scale $k_B T$ characterizes the system in addition to $m_\gamma$. One finds two distinct regimes, $m_\gamma > k_B T$ and $m_\gamma < k_B T$, associated with different equilibration times: $\tau_\mathrm{eq} \ll \hbar /k_B T$, for $m_\gamma > k_B T$, where the dynamics is essentially classical, and $\tau_\mathrm{eq} \sim \hbar /k_B T$, for $m_\gamma > k_B T$, which defines the quantum critical region. Quantum criticality displays rich complexity through the interplay between thermal and quantum fluctuations. The classical and quantum critical regions are shown in Fig.~\ref{PseudospinDomain}(a) separated by smooth crossovers (dashed curves) defined by $T = (16 \sqrt{2} m_c^{3/2}/k_B g^2) | m - m_c |^{1/2}$. Note that the horizontal hash marks covering the line $T_c^{(2)}(m)$ indicate the region for which the theory of classical (thermally driven) phase transitions may be applied.

\subsection{Symmetry Analysis}

The superfluid transition curves $T_c^{(1)}$ and $T_c^{(2)}$ divide the $T-m$ phase diagram in Fig.~\ref{PseudospinDomain}(a) into several phases associated with two temperature-driven stages of $U(1)$ symmetry breaking in the bosonic fields $\Psi_{R(L)}$. These two stages are a direct consequence of the formation of a coherent ground state associated with two chiral degrees of freedom which acquire finite expectation values $\langle \Psi_{R(L)} \rangle =   \sqrt{\rho_{R(L)}} \exp \left[ - i \theta_{R(L)} \right]$, expressed in terms of a macroscopic density and phase for the two chiral states. The various chiral combinations for elementary quarks translate into diquark and meson bound states which identifies $T_c^{(1)}$ and $T_c^{(2)}$ as the scalar meson and diquark superfluid transitions, $T_c^{(s)} = T_c^{(1)}$ and $T_c^{(d)} = T_c^{(2)}$, respectively, with the diquark molecule dissociation curve obtained through the relation Eq.~(\ref{pseudogapcriticality2}).

To organize our symmetry analysis, we should first delineate the three low-temperature regimes determined by the value of the running quark mass. The first of these lies along the temperature axis in Fig.~\ref{PseudospinDomain}, where the quark mass vanishes, $m=0$. Here, the mean-field approach developed so far is inadequate in formulating an accurate picture of the physics: we will show that the full quantum mechanical treatment reveals a BEC-BCS crossover in the region near the temperature axis, culminating in completely dissociated massless quarks along the axis. This contrasts with a robust BEC regime predicted by the mean-field calculation. Furthermore, the zero-mass limit corresponds to the onset of asymptotic freedom at large baryon chemical potential and restoration of the original chiral symmetry that distinguishes the theory at high energy. The second regime occurs for small to intermediate quark mass: $0< m < m_c$, where $m_c  = |\tilde{\mu}_B^{(0)}|$ is the size of the spectral gap. This regime is characterized by the coexistence of diquark and scalar condensates. In this mass range, the mean-field calculation reveals a diquark-meson mixed phase characterized by a finite BCS spectral gap in the diquark-dominant regime ($m << m_c$) and a vanishingly small BCS gap in the meson-dominant regime ($m \lesssim m_c$) with fully dissociated diquarks at the critical point ($m =  m_c$). Note that the diquark superfluid and dissociation temperature curves coincide at the zero-temperature critical point $m = m_c$. Finally, the third regime is defined for large quark mass, $m_c < m$, associated with a purely mesonic phase. In the following, we will detail the full symmetries and temperature-dependent symmetry breaking for each of these regimes. 

\begin{figure}[h]
\centering
\subfigure{
\hspace{-.5pc}\includegraphics[width=.52\textwidth]{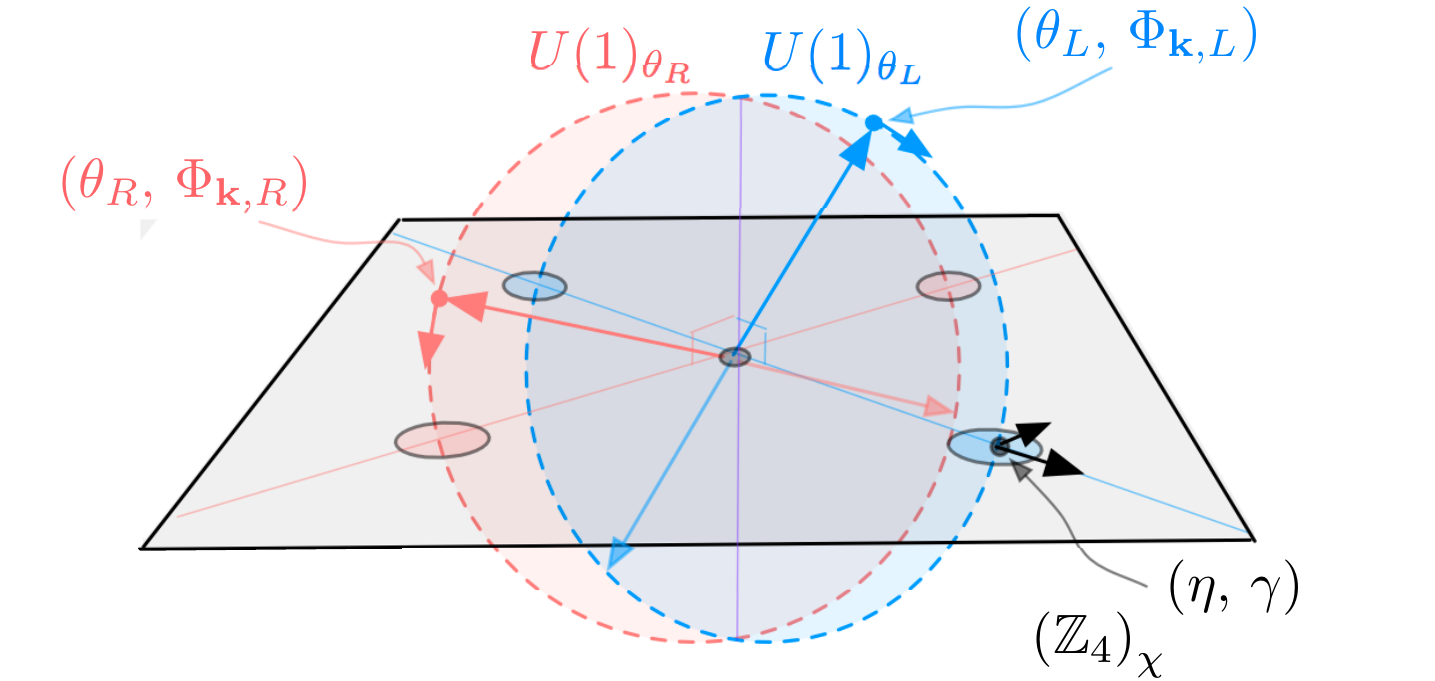}} \\
\caption[]{\emph{Symmetries of the single-quark ground-state manifold}. The right and left chiral ground states are $U(1)_{\theta_{R, L}}$ vertical circles (large red and blue, isomorphic to $S^1 \times S^1$), corresponding to broken symmetries in the original theory, times an internal $U(1)_{\phi, S}$ unbroken symmetry of the background spin liquid. This results from the structure of the quark states, left-hand side of Eq.~(\ref{HartreeSoln}), inside the Hamiltonian Eq.~(\ref{brokensymmetry3Red}). A typical state is labeled by its coordinate in the space of mean-field overall phase angle and the wavefunction of the spin liquid $\left( \theta_{R, L} , \, \Phi_{\mathrm{\bf k}, R,L} \right)$, indicated by small red or blue points along each circle. The horizontal planar slice (grey) corresponds to Fig.~\ref{PseudospinDomain}(b) (second panel from left), with the small red and blue disks indicating locations of minima. The grey plane contains the massive directions, quantum and thermal fluctuations orthogonal to the ground state: the radial direction $\eta$ is related to the baryon density; the polar angle $\gamma$ mixes left and right chiral states, related to the discrete generator of $\left(\mathbb{Z}_4 \right)_\chi$ chiral symmetry. Note that bound quarks in this picture correspond to diametrically opposite points on either the right or left chiral circles. }
\label{ContSym}
\end{figure}

In addition to a somewhat conventional analysis of symmetries, the symmetry structure of the Hamiltonian Eq.~(\ref{brokensymmetry3Red}) lends itself naturally to a framing in terms of the individual single-quark mean-field wavefunctions. This is a top-down approach that constructs the low-energy symmetries (e.g., condensation) from the detailed high energy structure. This approach is possible (and natural) since the theory at large $N_c$ results in a Hamiltonian that retains the individual quark structure of the diquarks, displayed in Eq.~(\ref{brokensymmetry3Red}), where bound states appear as non-local correlations. The general symmetry structure of Eq.~(\ref{brokensymmetry3Red}) is shown in Fig.~\ref{ContSym}. Let us now take a more detailed look at symmetries in the various limits of our model.

\subsubsection{BEC-BCS Crossover, Asymptotic Freedom in the Limit of Large Baryon Chemical Potential and Massless Quarks}
\label{Symmetries1}

 This limit corresponds to the region along the temperature axis in Fig.~\ref{PseudospinDomain}(a), charctareized by asymptotic freedom and massless quarks. As mentioned, treating this regime at the mean-field level reveals a spectral gap $\Delta_\mathrm{BCS} = |\tilde{\mu}_B|$ (omitting factors of $1/\pi$ throughout), from Eq.~(\ref{explicitdiquark2}). However, the substantial size of the baryon chemical potential in this region requires a full quantum mechanical treatment of the problem. At $m=0$ the full quantum calculation reveals diquark dissociation at $T= 0$, thus implying a superfluid critical point for $0 \lesssim T$. There, we find that taking the present viewpoint of coherent quarks as the fundamental degrees of freedom, i.e., as a starting point for perturbation expansions or a full quantum treatment, predicts a BEC-BCS crossover as the baryon chemical potential exceeds both the diquark pairing and the quark mass, where the vanishing spectral gap gives way to a finite pseudogap followed by complete molecular dissociation. Nevertheless, in keeping with the spirit of our work so far we will discuss here symmetry breaking from a purely mean-field perspective, and address the full quantum mechanical calculation in later in this paper.

 The mean-field calculation reveals finite and equal values for the meson and diquark superfluid transitions in the limit of zero quark mass. For temperatures well above this point, i.e., $T >> T_c^{(1)}\!(0) = T_c^{(2)}\!(0)$, both mesons and diquarks dissociate into a quark-gluon plasma (QGP) defined by the symmetry group $SU(N_c)_R \times SU(N_c)_L$. Yet, it is important to point out that this is the quark-dominant regime wherein, by relation Eq.~(\ref{symmetryrelation}) and full quantum calculations, we find that mesons give way completely to diquarks at much larger (intermediate) values of the running quark mass. The relevant Hamiltonian to our system is given by Eq.~(\ref{brokensymmetry3Red}), for which the dependence on the number of species is implicit, hence the relevant symmetry group is given by the quotient 
 \begin{eqnarray}
 \hspace{-1pc} G_\mathrm{QGP}  &\cong&    \frac{SU(N_c)_R \times SU(N_c)_L}{SU(N_c-1)_R \times SU(N_c-1)_L}   \, , 
 \end{eqnarray}
 where
 \begin{eqnarray}
  \hspace{-1pc} \frac{SU(N_c)_{R(L)}}{SU(N_c-1)_{R (L)}}  \hspace{0pc}  \xrightarrow{\mathrm{4D \to 3D}}    \hspace{.5pc}       U(1)_{\theta_{R(L)}} \times Spin(2)_{R(L)}        \\
                               \cong    \hspace{.5pc}         U(1)_{\theta_{R(L)}} \times   U(1)_{\phi, S,R(L)} \times  \left( \mathbb{Z}_{2 } \right)_{S, R(L)}   , 
 \end{eqnarray}
 where the special unitary group gets reduced when projected down to $(2+1)$-dimensions and we have used the fact that the Spin group in $(2+1)$-dimensions forms a double cover of the rotation group and $\theta$, $\phi$, and $S$ refer, respectively, to the overall gauge, polar angle, and spin degrees of freedom. In what follows, we examine symmetries mainly from the point of view of the quarks residing in the classical large $N_c$ single-particle states, and symmetry breaking in terms of condensation of bosonic bilinear operators of these.

The range of temperatures between the diquark dissociation and superfluid transitions, $T_c^{(d)}\!(0) = T_c^{(2)}\!(0)  < T < T^*$, is marked by the appearance of a pseudgap, $\Delta_\mathrm{pg}$. This defines the BCS limit in which quarks belonging to the same chiral state are loosely bound into diquark Cooper pairs $\langle q_R q_R \rangle$ and $\langle q_L q_L \rangle$. The wavefunction of each member of a bound pair contributes two phase angles associated with the $U(1)_\theta \times U(1)_{\phi, S}$ subgroup of the full symmetry group for a total of four $U(1)$ degrees of freedom. One of these is fixed due to the oppositely correlated momentum directions between bound quarks. Another is fixed by the internal p-wave spatial form of the bound pair, which depends on the binding energy. Thus, two phase symmetries remain: the relative phase $\phi_\mathrm{rel}$ and center of mass phase $\Phi_\mathrm{cm}$, identified with the  unbroken $U(1)_\theta \times U(1)_{\phi, S}$ symmetry. In addition, both quarks in a particular chiral bound state belong to the same classical ground state, but are different excitations with a specific phase relationship. This comes from noting that the condition of oppositely correlated momenta within a pair contains two constraints: that the magnitude of momentum for both quarks be equal, which removes one phase degree of freedom (as previously noted), and that the momentum directions are oppositely correlated, which removes an additional discrete degree of freedom denoted by $\left( \mathbb{Z}_2\right)_\pm$, which simultaneously flips the directions for both momentum and spin. Finally, we take the diquark to be spin-singlet with overall zero spin. This, combined with the momentum correlation condition, implies the equivalence for the discrete part $\left( \mathbb{Z}_2\right)_S \cong \left( \mathbb{Z}_2\right)_{L/R}$, where the cyclic group on the right interchanges left and right chirality.

The BEC limit is associated with temperatures below the superfluid critical temperature, $T < T_c^{(1)}\!(0) = T_c^{(2)}\!(0)$, at which point a finite spectral gap $\Delta_\mathrm{BCS}$ appears in place of the pseudogap. Here, the average separation between bound pairs and the characteristic length scale $\lambda = \hbar c \left[ (k_B T)^2 - m^2 c^4 \right]^{-1/2}$ are much larger than the internal separation $r_\mathrm{rel}$ between quarks within a pair. The system remains symmetric with respect to the relative phase, $\phi_\mathrm{rel}$, which nevertheless becomes an internal symmetry similar to the species number $N_c$. However, the presence of a spectral gap in the BEC regime is synonymous with breaking of $U(1)$ symmetry. Indeed, the remaining symmetry, $\Phi_\mathrm{cm}$, gets broken in the BEC limit, since the gradient satisfies $\nabla \Phi_\mathrm{cm} \sim \lambda^{-1} << r_\mathrm{rel}^{-1}$, marking the regime wherein resolution of individual fermions becomes untenable. Condensation in the center of mass phase $\Phi_\mathrm{cm}$, associated with overall spatial translations of the diquark, amounts to spontaneous breaking of one $U(1)$ symmetry.

\subsubsection{Meson-Diquark Mixed Phase for Intermediate Baryon Chemical Potential and Massive Quarks}

Here we consider the region in Fig.~\ref{PseudospinDomain}(a) for which $0 < m < m_c$. Lowering the temperature, we encounter first a phase transition from the quark-gluon plasma to the meson phase with the meson superfluid critical temperature given by $T_c^{(s)}(m) = T_c^{(1)}(m)$. Scalar mesons condense below this temperature with $T_c^{(s)}(m)$ monotonically increasing with the quark mass, consistent with the fact that the running quark mass is dynamically generated.

The analysis in Sec.~\ref{GLT} shows that condensation in the regime $T_c^{(d)}(m) < T < T_c^{(s)}(m)$ is purely mesonic: the two minima of $U_\mathrm{eff}$ reside along the circle $\bar{\phi}_{\mathrm{ \bf k}, \uparrow}$ on the left hand side of Eq.~(\ref{HartreeSoln}), oriented along the $\Psi_{0,\uparrow}$ direction. By Eqs.~(\ref{diquark})-(\ref{scalarcon}), this corresponds to finite scalar but vanishing diquark condensates for a ground state with $\left(\mathbb{Z}_2\right)_\chi$ chiral symmetry that transforms quarks into antiquarks. Here, scalar condensation is characterized by $U(1)$ symmetry breaking for the translational motion of the meson center of mass. Continued reduction in temperature leads to a second phase transition, this time into a diquark-meson mixed phase. Here, chiral symmetry is partially restored to $\left(\mathbb{Z}_2\right)_{\chi, R} \times \left(\mathbb{Z}_2\right)_{\chi, L}$ by the presence of diquarks, as shown in Fig.~\ref{PseudospinDomain}. The spectral gap is comprised of contributions from both BCS and mass gaps such that the total gap is $\Delta_\mathrm{BCS}^2 + \Delta_s^2 = \tilde{\mu}_B^2$. In the meson-diquark mixed phase the broken continuous symmetry is $U(1)_s \times U(1)_d \cong  \mathrm{\bf T}^2$, topologically isomorphic to the toroid. Note that from the left hand side of Eq.~(\ref{symmetryrelation}), one deduces an underlying broken full $SU(2)$ symmetry that mixes both amplitudes and phases of scalar and diquark condensates, for fixed quark density, acting on the complex two-dimensional space $\left( \langle \phi \rangle , \,  \langle \sigma \rangle \right)^T$. However, we note again that chiral symmetry, i.e., that which mixes right and left quark states (effectively mixing mesons and diquarks), is dynamically broken at the discrete, not the continuous level. Thus, no associated Goldstone mode is observed: meson and diquark condensates remain distinct. Analysis of the internal symmetries follows closely that of Sec.~\ref{Symmetries1}. 

\subsubsection{Quark Confinement in the Meson Limit of Small Baryon Chemical Potential and Large Quark Mass}

This is the region defined by $m > m_c$, characterized by strong confinement, CSB, and complete absence of diquarks. A single phase transition occurs along the curve $T^{(s)}_c(m) = T^{(2)}_c(m)$ from the QGP into the meson condensate with ${\left(\mathbb{Z}_2\right)}_\chi$ chiral symmetry with the scalar condensate given by $\Delta_s = |\tilde{\mu}_B|$. Following similar reasoning to our previous discussion, each member of a bound state is now in a superposition of left and right chirality, consistent with breaking the left/right $U(1)$ symmetries under $q_{L(R)} \to e^{i \theta_{L(R)}} q_{L (R)}$. Thus, bound states acquire an additional random internal phase due to the vanishing of spin-momentum locking in this limit, in contrast to strong spin-momentum locking in the $m=0$ limit. For diquarks, these internal $U(1)$ spin fluctuations provide one source for decoherence of the diquark condensate, the other coming from thermal fluctuations in the overall phase of the condensate. However, a loss in coherence from the standpoint of diquarks is replaced by additional coherence in the meson field.

\subsubsection{Discrete Chiral Symmetry}

It is instructive to elaborate on discrete internal symmetries retained for single quarks within bound states in the BEC limit in order to gain a deeper understanding of chiral symmetry breaking, the BKT nature of the BEC-BCS transition, and the general microphysics of quark bound states. The Hamiltonian expressed in terms of the single-particle wavefunction containing $N_c$ quarks has a discrete $\left(\mathbb{Z}_4\right)_\chi$ chiral symmetry embedded in the full symmetry group~\cite{hladikJ1999}. To see this, consider that through the large $N_c$ quark wavefunction, the potential Eq.~(\ref{EffPot}) maps the full $SU(2)_R \times SU(2)_L$ symmetry onto three-space with our two-dimensional right and left chiral theories residing on separate orthogonal planes that connect diagonal minima in Fig.~(\ref{PseudospinDomain})(b). In the presence of a spectral gap, the effective potential Eq.~(\ref{EffPot}) develops four minima at the corners of a square along a central planar slice (normal to those of our right and left theories) through our embedding of spin space into coordinate space. Rotations within this plane are associated with the time-like direction in spin space generated by transformations that mix right and left chiral states, hence are chiral rotations. The symmetry group in the chiral direction, associated with the minima of Eq.~(\ref{EffPot}), is the cyclic group of order eight, a subgroup of the full symmetry group, $\left(\mathbb{Z}_8\right)_\chi \leq SU(2)_R \times SU(2)_L$. This is because rotations out of the chiral plane, connecting minima along diagonals of the square, are associated with the double cover of the rotation group in 2D, $U(1)_{\phi, S} \to SO(2)$, for separate right and left chiral theories. Thus, eight discrete rotations by $\pi/4$ in the chiral plane return the wavefunction to its original state. However, a finite quark mass breaks chiral symmetry (which we address next) in a way that respects the double-cover redundancy. Thus, it is more appropriate to represent chiral symmetry by the four-fold quotient group that removes this redundancy.

Detailing this point, consider the two-dimensional unitary representation of the cyclic group $U(\mathbb{Z}_8) = \left\{ g,b ,a \in \mathbb{Z}_8 , \, n \in \mathbb{Z} \vert g =e^{i  \pi /4} \sigma_3 , \,b = g^n , \,  a =  g^4 \right\}$, where $g$ is the group generator and $\sigma_3$ is the third Pauli matrix, which is time-like in our theory. Hence, $g$ is the eighth root of unity for two-by-two matrices. The automorphism defined by $f: b \to a b$ identifies elements of $\mathbb{Z}_8$ related through multiplication by four factors of the group generator $g$. Thus, the double-cover redundancy is removed by taking the chiral group to be the quotient $\left( \mathbb{Z}_4\right)_\chi = \left( \mathbb{Z}_8 \right)_\chi/ f$. It is important to keep in mind that this discrete chiral symmetry does not result from spontaneous breaking of some larger symmetry group, but characterizes the Lagrangian of the QGP at higher temperatures as well, explicitly broken by defining our theory at the outset. In the present massless limit, the largest possible chiral symmetry is manifest, fixed by the particular form of the interaction that defines our model.  
 
\subsubsection{Continuous Symmetries and Homotopy Structure}

Computing the fundamental group for a single baryon quark from the group structure of Eq.~(\ref{EffPot}) yields the direct sum 
\begin{eqnarray}
\pi_1\!  \left(G_\mathrm{BEC}^\mathrm{(int)}\right)   \cong  \mathbb{Z}_{\theta, R(L)}  \oplus  \mathbb{Z}_{2 , \phi, R(L)}  \, ,  \label{homotopy}
\end{eqnarray}
where $G_\mathrm{BEC}^\mathrm{(int)}$ refers to the internal symmetry group of a single bound quark in the presence of a spectral gap, depicted in Fig.~\ref{ContSym}. The two homotopy groups in Eq.~(\ref{homotopy}) describe available phases for mapping nonlinear quark states to the right or left circle at spatial infinity. We will see that at zero temperature the superfluid and molecule dissociation temperature curves coincide at the BEC-BCS critical point $m = m_c$, with the Kosterlitz-Thouless temperature generally satisfying $T^{(d)}_c < T_\mathrm{KT} < T^*$, allowing for the possibility of a BKT driven BEC-BCS transition through the unbinding of topological baryonic vortices: the energy scale of the BKT vortices coincides with that of the bound quarks.

 The nature of relativistic BKT-type transitions at zero temperature departs fundamentally from conventional BKT in several important ways. First, we note the fact that the meson and diquark condensates break chiral and Lorentz symmetry, respectively. In two dimensions, $\Delta_s$ and $\bar{\Delta}_d$ compete for bound states through the running quark mass and baryon chemical potential. At the quantum critical point where both values meet, the system of bound states undergoes qualitative restructuring from mesons into diquarks, or vice-versa depending on the direction of approach into the critical point. This occurs in such a way that both chiral and Lorentz symmetry are restored at the point of phase transition through enlargement of the $SU(2)$ group from the two-dimensional reduced form into the full three-dimensional group structure, at which point the theory becomes conformal. Thus, in relativistic BKT transitions, restoration of large-scale bosonic $U(1)$ symmetry is a derivative effect, secondary to restoration of the more fundamental microscopic $SU(2)$ symmetry.

\section{Quantum Phase Transitions}
\label{QPT}

It is a general result from analysis of three-dimensional relativistic Fermi systems that quarks will condense into mesons or diquarks when the baryon chemical potential is less than the quark mass or half the mass of the bound state in accord with $\mu_B  <  m_{d, \sigma, \pi}/2, \, m$. Thus it is useful to define a nonrelativistic chemical potential $\mu_N = \mu_B - m$, which is negative in the BEC region and positive in the BCS region, a result which comes directly from the associated dispersion relation. In this section we examine the the BEC-BCS crossover at zero temperature in (2+1)-dimensions based on the large $N_c$ semi-classical picture developed up to this point.

So far, we have seen that at zero temperature our model undergoes qualitative changes in its properties as a function of the quark mass. In particular, abrupt changes in chiral symmetry at $m=  \tilde{\mu}_B \equiv m_c$ and at $m = 0$ occur for moderate and large densities and baryon chemical potential, respectively. A full quantum mechanical treatment shows diquark molecular dissociation occurring in both limits.

The BEC-BCS transition at $m= m_c$ is driven by competition between the quark mass and the size of the BCS gap in accord with Eq.~(\ref{explicitdiquark2}). In the meson-diquark mixed region, the spectral gap remains constant with the size of the mass and BCS gap interchanging as $m \to m_c$, shown in Fig.~\ref{BECBCS}. Thus, here the spectral gap is comprised of both meson and diquark contributions. The regimes $m << \Delta_\mathrm{BCS}$ and $m >> \Delta_\mathrm{BCS}$ correspond, respectively, to the BEC and BCS limits: for $m << \Delta_\mathrm{BCS}$, quarks minimize their energy by combining into tightly bound pairs to form a diquark condensate; for $m >> \Delta_\mathrm{BCS}$, quarks prefer to bind with antiquarks to form a scalar condensate, i.e., scalar mesons are favored over diquarks which identifies a BCS state from the diquark standpoint. At the quantum critical point where $m = m_c$ and $\Delta_\mathrm{BCS} \to 0$, the lower endpoints of the superfluid phase transition and diquark dissociation curves, $T_c^{(d)}$ and ${T^{(d)}}^*$, coincide: the pseudogap vanishes resulting in a BEC-BCS transition versus a smooth crossover.

The second zero-temperature phase transition occurs at higher densities where the quark mass vanishes, associated with the onset of asymptotic freedom. There we encounter an actual smooth crossover in the region between the curves $T_c^{(d)}$ and ${T^{(d)}}^*$, with a finite pseudogap in the zero-temperature limit. This occurs when the baryon chemical potential exceeds the size of diquark pairing $\Delta_d < \mu_B$, which drives superfluid decoherence followed by dissociation of bound quarks, connected by a finite region wherein quarks form loosely bound Cooper pairs. Quantum criticality is marked by a vanishing spectral gap, $|\tilde{\mu}_B| = |\mu_B - \Delta_d| = 0$, with the quantity $\mu_B - \Delta_d$ changing sign through the critical point: $\mu_B - \Delta_d <0$ corresponds to the BEC regime; $\mu_B - \Delta_d > 0$, the BCS regime.

The mechanism driving the BEC-BCS transition at $m = m_c$ is a unique feature of two-dimensional relativistic Fermi systems. It is characterized by a restructuring of the ground state through a change in its discrete symmetry involving augmentation to a higher dimension in parameter space at $m_c$. The reason is that spinors in (2+1)-dimensions do not map to the full $SU(2)$ group, as this projects down to $U(1) \times Spin(2)$ under reduced dimensionality, losing a degree of freedom associated with one of the three generators of the full $SU(2)$ group. However, the full symmetry is recovered at the critical point $m_c$ through a mechanism revealed by parametrizing left and right chiral spinor fields according to 
\begin{eqnarray}
\Psi_{R, L} = \eta^{1/2} \exp(i \theta) \left[ \,  \cos \gamma , \,\pm  \exp(i \phi ) \sin \gamma  \, \right]^T \, . \label{FirstAnsatz}
\end{eqnarray}
The fermionic fluctuations in the amplitude, $\delta \eta$, and spin polarization or chiral angle, $\delta \gamma$, are generally massive and frozen out except near the quantum critical point at which point $\delta \gamma$ becomes massless. Hence fluctuations in the polarization field $\delta \gamma $ provides a signature for tracking the micro-dynamics of the BEC-BCS transition at the quantum critical point $m_c$. Specifically, the dominant contribution to the fermionic dispersion near the critical point, where $\delta \gamma  \lesssim 1$, takes the form
\begin{eqnarray}
\! E^2_\pm({\bf p}) = \Delta_\mathrm{BCS}^2 + |\mathrm{\bf p}|^2  + b^2 \pm 2 b \sqrt{\Delta_\mathrm{BCS}^2 + \left( {\bf p} \cdot \hat{\bf b} \right)^2 }  , \,  \label{Spectrum}
\end{eqnarray}
characterized by the appearance of a CPT symmetry breaking term $b = | {\bf b}| = m \, \delta \gamma$, where $\hat{\bf b} =  \boldsymbol{ \sigma }/3$, near criticality where $\delta \gamma$ fluctuations become significant, i.e., $|\delta  \gamma| \sim1$. Note that the minus sign pertains to particle excitations. The spectrum in Eq.~(\ref{Spectrum}) reveals a quantum phase transition at $m = \Delta_\mathrm{BCS}$ that separates the fully gapped ground state, for $m < \Delta_\mathrm{BCS}$, from a ground state with two isolated Fermi points, for $m > \Delta_\mathrm{BCS}$, located at ${\bf p}_1 = + \hat{\bf b} \sqrt{m^2 - \Delta_\mathrm{BCS}^2}$, ${\bf p}_2 = - \hat{\bf b} \sqrt{m^2 - \Delta_\mathrm{BCS}^2}$, associated with right and left-handed fermions, respectively. The process of moving through the critical point, $m < \Delta_\mathrm{BCS} \to m >\Delta_\mathrm{BCS}$, depicts CSB with the formation of a Fermi surface indicated by the change in minimum $\mathrm{\bf p}_\mathrm{min} = 0 \to \pm  \hat{\bf b} \sqrt{m^2 - \Delta_\mathrm{BCS}^2}$ when condensed bound quarks (BEC) become loosely bound Cooper pairs (BCS). Note that Eq.~(\ref{Spectrum}), along with Eq.~(\ref{explicitdiquark2}), predicts the critical point for the quantum phase transition at $m = |\tilde{\mu}|/\sqrt{2}$. The meaning of the result Eq.~(\ref{Spectrum}) is clear: the chiral fluctuations $\delta \gamma$ are identified with fluctuations in the scalar field which induce formation of a small Fermi surface as diquarks dissociate, a precursor to the formation additional mesons.

\subsection{Small and Large Quark Mass Descriptions}

 Here we want to examine the Hamiltonian description on either side of the critical point $m = m_c =  |\tilde{\mu}_B|$. We first reformulate our theory for the small quark mass regime, $m \ll |\tilde{\mu}_B|$. At zero temperature the Hamiltonian for fluctuations in the coherent ground state for the two spin polarizations can be expressed in terms of the right and left quark coherent states with the substitution $\Psi_{\uparrow \downarrow } = \left(\Psi_{R} \pm \Psi_{L} \right)/\sqrt{2}$. Shifted to each chiral minimum of the effective potential, we obtain the modified Hamiltonian 
\begin{widetext}
\begin{eqnarray}
      H  =    \sum_{ \mathrm{\bf k}}  \left[  \left(  \epsilon_\mathrm{\bf k} +  | \tilde{\mu}_B|  \right)   \sum_{\alpha } \Psi_{ \mathrm{\bf k}, \alpha}^* \Psi_{ \mathrm{\bf k},  \alpha}   -   m     \sum_{\alpha \ne \alpha'}    \Psi_{ \mathrm{\bf k}, \alpha}^*  \Psi_{ \mathrm{\bf k}, \alpha'}      -   \frac{\bar{g}^2}{2}  \sum_{\alpha }  \left(   \Psi_{ - \mathrm{\bf k}, \alpha}^*  \Psi_{ \mathrm{\bf k}, \alpha}  \right)^2  \right]  \, ,  \label{brokensymmetryFree3}   
\end{eqnarray}
\end{widetext}
where $\alpha = R, L$. The first term on the right hand side of Eq.~(\ref{brokensymmetryFree3}) respects chiral symmetry and reflects the appearance of a mass gap due to the diquark condensate. Note that the sign of the chemical potential term is now positive since Eq.~(\ref{brokensymmetryFree3}) describes excitations relative to the potential minima in Fig.~\ref{PseudospinDomain}(b). The second term, proportional to the quark mass, breaks chiral symmetry by mixing left and right states. This is made explicit by noting that this term is not invariant under $\alpha$-independent phase transformations $\Psi_{\alpha} \to e^{i \theta_\alpha}\Psi_{\alpha}$. The third term in Eq.~(\ref{brokensymmetryFree3}) respects chiral symmetry and provides the attractive diquark channel within each of the right and left chiral channels. Note that quarks with opposite in-plane spin and momentum (same chirality) are coupled through the quartic interaction. Rotation of the quark field through a half turn results in the complex phase factor that flips the sign of the interaction term. The Hamiltonian expressed in the form Eq.~(\ref{brokensymmetryFree3}) provides a suitable starting point for studying perturbations of the system due to a small quark mass. This is essentially what we have described in Fig.~\ref{ContSym}, when $m =0$. 

Interestingly, the chiral symmetry breaking part of Eq.~(\ref{brokensymmetryFree3}) is reminiscent of the kinetic term in the attractive Fermi double-well or two-state system where the ``hopping'' term in our problem corresponds to tunneling between left and right chiral states. From the standpoint of in-plane spin, the second term encourages tunneling between spin-aligned (right chiral) and spin-anti-aligned (left chiral) states: fluctuations in the particle number of each ``well'' corresponds to dissociation of bound quarks. Pursuing the analogy, the third term acts as an attractive ``on-site'' contact interaction, with the average total number of particles fixed by the mean-field constraint $\sum_{\alpha }   \langle  \Psi_{\alpha} \Psi_\alpha \rangle =  4 \tilde{\mu}_B/\bar{g}^2$. Being attractive, it tries to retain both particles in the same well, i.e., to maintain a well-defined particle number in each well, or in our case, in each bound quark pair. Completing the analogy, the first term places a gap in the spectrum. 

\begin{figure}[]
\centering
\subfigure{
\hspace{-3pc}\includegraphics[width=.46\textwidth]{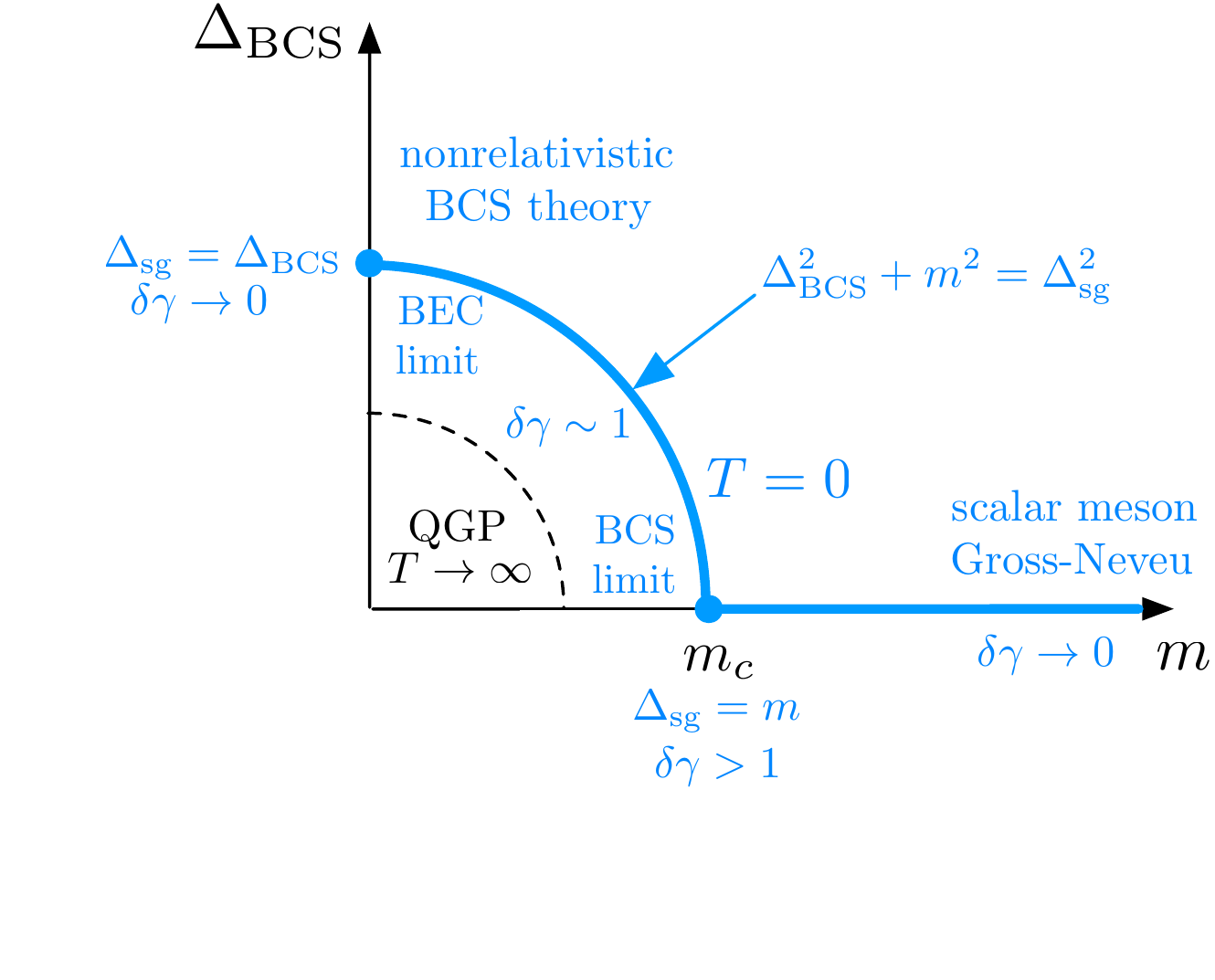}} \\  \vspace{-2.8pc}
\caption[]{\emph{Preliminary look at the diquark-scalar meson quantum phase transition}. The zero-temperature curve is shown (blue) for the behavior of the BCS gap with increasing quark mass near the phase transition. The arc tracks the BEC-BCS transition with increasing mass $m = 0 \to m_c^+$. The phase transition technically occurs at $m_c$, but the BCS gap is gradually transformed throughout the arc into a pseudogap, towards the lower edge near the BCS limit. This phase transition is driven by quantum fluctuations $\delta \gamma$ in the chiral angle $\gamma$, defined in Fig.~\ref{PseudospinDomain}(b) and Eq.~(\ref{FirstAnsatz}), corresponding to rotations in the horizontal plane of Fig.~\ref{ContSym}. Fluctuations are small near the limiting theories described by Eqs.~(\ref{brokensymmetryFree3})-(\ref{brokensymmetryFree4}) at the far left and far right of the $T=0$ curve, respectively, and reach maximum at the critical point $m_c$. Note that the spectral gap remains constant along the arc. }
\label{BECBCS}
\end{figure}

Now let us examine the large mass regime for which $m \gg |\tilde{\mu}_B|$. Here we find that it is more suitable to cast the Hamiltonian in terms of quarks and antiquarks, since a large quark mass favors bound states of these with the potential minima in this limit residing along the spin polarization axis $\Psi_\uparrow$. In this limit, the low-temperature Hamiltonian takes the form
\begin{widetext}
\begin{eqnarray}
         \hspace{-1pc} H &=&     \sum_{ \mathrm{\bf k} }  \left[  \left( \epsilon_\mathrm{\bf k}  +  m  \right)   \bar{\Psi}_\mathrm{\bf k}  \Psi_\mathrm{\bf k}   -  | \tilde{\mu}_B|  \,     \bar{\Psi}_\mathrm{\bf k} \gamma^0 \Psi_\mathrm{\bf k}      -   \frac{\bar{g}^2}{2}  \left( \bar{\Psi}_\mathrm{\bf k} \Psi_\mathrm{\bf k} \right)^2   \right] \, ,   \label{brokensymmetryFree4}
\end{eqnarray}
\end{widetext}
where we have converted to the quark-antiquark form using $\Psi^*_{\uparrow \downarrow } \Psi_{\uparrow \downarrow } = \left(\bar{\Psi} \gamma^0 \Psi  \pm \bar{\Psi} \Psi \right)/\sqrt{2}$. The first and third terms in Eq.~(\ref{brokensymmetryFree4}) together give the attractive Gross-Neveu model, with the second term now acting as a perturbation. The system in this limit is mainly described by a scalar Lorentz invariant theory with a small Lorentz symmetry breaking perturbation coming from the finite, but small (relative to $m$), quark density. Symmetry breaking comes from the fact that the second term is the time component of a four-vector, not itself invariant under generic Lorentz transformations. From the perturbative vantage point of Eq.~(\ref{brokensymmetryFree4}), where $|\tilde{\mu}_B| \ll m$, excitations of the coherent ground state form a scalar condensate, $\langle \bar{\Psi}  \Psi \rangle  \ne 0$, consistent with the presence of a mass gap $m$. Here, the perturbation encourages decoherence of the scalar condensate by introducing small baryonic fluctuations that restore chiral symmetry, becoming significant at the criticality $m = |\tilde{\mu}_B|$, at which point the quark-antiquark picture no longer gives a simple, convenient description of the physics. As this discussion suggests one may examine the quantum phases in our theory using a standard mean-field pairing approach (e.g., Hubbard-Stratonovich), which we will address this in future work. The results of our analysis are organized in Fig.~\ref{BECBCS}.

Summarizing, the relative strengths of the parameters $|\tilde{\mu}_B|$ and $m$ delineate the different phases in our system, as we have seen throughout our work up to this point. Namely, from the viewpoint of right and left chiral states displayed in Eq.~(\ref{brokensymmetryFree3}), for $|\tilde{\mu}_B| \gg m$, the system is in a gapped, insulating, phase. Left and right chiral modes remain bound in separate, distinct, chiral condensates. When $m$ $=$ $|\tilde{\mu}_B|$, the system undergoes a transition from insulating to conducting phase. Here, fluctuations between left and right chiral states are large enough to dissolve the diquark condensate: it becomes energetically favorable to form a limited ``Fermi liquid'' in that quarks are able to flow freely between left and right chiral states. One must keep in mind here that actual quarks are still bound in pairs, but now in the form of mesons, as one can see by the fact that the mean-field form of the symmetry breaking term in Eq.~(\ref{brokensymmetryFree3}) is just the meson condensate, $\langle \bar{\Psi} \Psi \rangle = \langle  \Psi_R  \Psi_L \rangle +  \langle  \Psi_L  \Psi_R \rangle$.

\subsection{Zero-Temperature Chiral Symmetry Breaking and Restoration}

The usual notion of chiral symmetry breaking occurs due to a dynamical quark mass through the formation of the scalar condensate. However, consider that at fixed temperature $T =0$ the quark mass runs inversely with the chemical potential. This means that if we tune the chemical potential towards large values, we should observe a restoration of chiral symmetry. Conversely, a small chemical potential leads to a large quark mass and thus broken chiral symmetry. We would like to better understand the dynamics of this process within the present context. 

\begin{figure*}[]
\begin{center}
 \subfigure{
\label{fig:ex3-a}
\hspace{0in} \includegraphics[width=.9\textwidth]{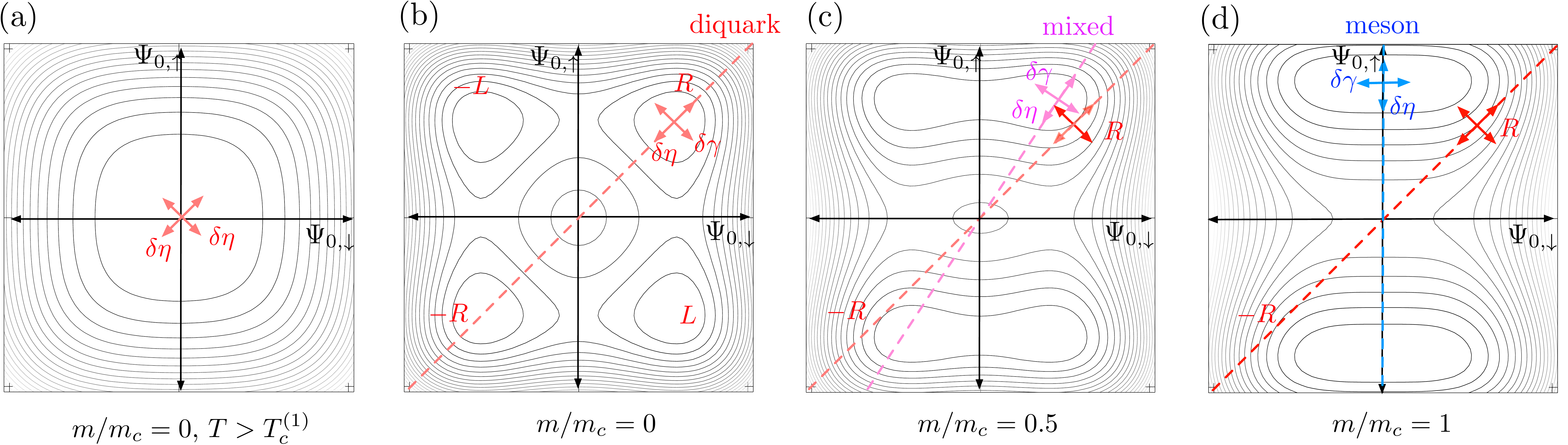}  } 
\vspace{0in}
\end{center}
\caption[]{(color online) \emph{Effective potential in spin-space. } Local minima in the mean-field effective potential. Note that $T= 0$ for (b)-(d). Panels coincide, from left to right, with those in Fig.~\ref{PseudospinDomain}(b). As the quark mass is tuned $m \to m_c$ the two potential minima approach each other, significantly increasing the tunneling rate between right and left chiral ground states. Consequently, near criticality, Cooper pairs undergo extreme fermion fluctuations until finally dissociating at $m = m_c$. Dashed lines depict directions of quark-quark (red), quark-antiquark (blue), and intermediate mixed state (purple) binding. Massive amplitude and chiral fluctuations $\delta \eta$ and $\delta \gamma$, respectively, are shown for each type of bound state.  }
\label{Minima}
\end{figure*}

Consider the contour plots of the quark effective potential in Fig.~\ref{Minima} corresponding to the schematic diagrams in Fig.~\ref{PseudospinDomain}(b). Each stage is characterized by the ratio of the quark mass to the quantum critical value, $m_c = \tilde{\mu}_B$, and the value of the mean chiral angle $\langle \gamma \rangle$, where $\gamma$ is the polar angle in spin space such that $\Psi \sim \eta ( \mathrm{cos} \gamma , \, \mathrm{sin} \gamma )^T$, as previously discussed. At higher temperatures and zero quark mass, shown in Fig.~\ref{Minima}(a), right and left handed chirality is preserved and expressed by the four-fold symmetry of the potential, where $\langle \gamma \rangle_R = \pi/4, 5\pi/4$ and $\langle \gamma \rangle_L = 3\pi/4, 7\pi/4$. Only amplitude/density fluctuations, $\delta \eta$, are massive since the minimum of the effective potential lies at $\Psi = ( \Psi_{0, \downarrow} , \, \Psi_{0, \uparrow} )  = 0$. In contrast, lowering the temperature to $T=0$ in the $m \to 0$ limit, which corresponds to Fig.~\ref{Minima}(b), the ground state inherits the four-fold symmetry of the full theory, with minima now located at left and right chiral angles $\langle \gamma \rangle_R =  \pi/4, \, 5\pi/4$ and $\langle \gamma \rangle_L = 3 \pi/4, \, 7\pi/4$. It should be understood that this is the mean-field prediction for the onset of asymptotic freedom in the extremely high density limit. Next, we see in Fig.~\ref{Minima}(c) that when the quark mass approaches half the critical mass, potential minima shift to $\pi/4 < \langle \gamma \rangle <  3\pi/4, \,7\pi/4$ and $5\pi/4 < \langle \gamma \rangle <  7\pi/4$, breaking the symmetry $\left(\mathbb{Z}_4\right)_\chi \to \left( \mathbb{Z}_2\right)_\chi \times   \left( \mathbb{Z}_2\right)_\chi$. Finally, $m=m_c$ in Fig.~\ref{Minima}(d), minima merge so that $\langle \gamma \rangle = \pi/2 , \, 3\pi /2$, where chiral symmetry of the ground state is further broken $\left( \mathbb{Z}_2\right)_\chi \times   \left( \mathbb{Z}_2\right)_\chi \to   \left( \mathbb{Z}_2\right)_\chi$. Thus, chiral symmetry is broken (restored) in two stages as the quark mass runs from zero ($m_c$) to $m_c$ (zero).

 \subsection{Order Parameters, Bound States, Quantum Chiral Fluctuations}

 From this discussion, it should be clear that $\langle \gamma \rangle$ characterizes each of the phases in Fig.~\ref{PseudospinDomain}. However, a nonzero value of $\langle \gamma \rangle$ does not in itself constitute an order parameter in the strict sense, as $\langle \gamma \rangle \ne 0$ is not identical to condensation. The natural order parameter is, as one would expect, the diquark condensate $\Delta_\mathrm{BCS}$. But the mass for the $\delta \gamma$ fluctuations $m_\gamma$ dual to the bound states of these fluctuations must play a fundamental role. In addition, it is informative to study the exact point of diquark dissociation at zero temperature. This would provide insight into the character of the diquark pseudogap $\Delta_\mathrm{pg}$ which is essentially the diquark pairing function in the absence of actual condensation. Consider that in the intermediate region $0 < m \lesssim m_c$, corresponding to the meson-diquark mixed phase, both $\Delta_\mathrm{BCS}$ and $m_\gamma$ are nonzero. At criticality, $\Delta_\mathrm{BCS}$ and $m_\gamma$ both vanish, with $\Delta_\mathrm{BCS}$ remaining zero and $m_\gamma$ returning to its maximum value beyond the critical point. It is significant to note that molecular dissociation occurs at the same critical point $m = m_c$, coinciding with superfluid criticality where $\Delta_\mathrm{BCS}$ vanishes. Let us gain deeper insight into the nature of the order parameter $\Delta_\mathrm{BCS}$ near the critical point, its relationship to the quark mass and chemical potential, and the meaning of the mass $m_\gamma$ by studying bound states and fluctuations within the potential energy landscape of the large $N_c$ semi-classical mean field quark ground state depicted in Fig.~\ref{Minima}.

 \subsubsection{Order Parameters and Bound States}

 Consider again the different configurations of the effective potential in Figs.~\ref{Minima}(b)-(d) lying along the $T=0$ axis in Fig.~\ref{PseudospinDomain}(a). Representing the planar slice in spin space by the polar coordinates $\eta$ and $\gamma$, the locations of the potential minima as functions of $m$ and $|\tilde{\mu}_B|$ are 
 \begin{eqnarray}
\hspace{-2pc} &&\left(\eta, \, \gamma \right)_\mathrm{min}=    \nonumber \\
\hspace{-2pc}  &&\left( 2\sqrt{|\tilde{\mu}_B|/\bar{g}^2}, \,  \mathrm{tan}^{-1}\sqrt{(|\tilde{\mu}_B|+m)/\left||\tilde{\mu}_B|- m \right|} \right) \, , 
 \end{eqnarray}
 and the height of the central peak relative to each minimum is 
 \begin{eqnarray}
 \Delta P = \left(m^2 + \tilde{\mu}_B^2 \right)/\bar{g}^2 \, .  \label{height}
 \end{eqnarray}
 Looking now at the upper right minimum labeled $R$ in Fig.~\ref{Minima}(b), for which 
 \begin{eqnarray}
 \left(\eta, \, \gamma \right)_\mathrm{min} = \left( 2\sqrt{|\tilde{\mu}_B|/\bar{g}^2} , \, \pi/4 \right) \, .  \label{finitefield}
 \end{eqnarray}
 Let us obtain a better understanding of pairwise bound states within the baryon ground state described by the Hamiltonian in Eq.~(\ref{brokensymmetryFree3}) with the quark mass $m$ set to zero. These are tightly bound quarks whose description in terms of free scattering states involves a highly entangled superposition of amplitudes and phases spread around the Fermi circle and smearing it out to form the gap. A single quark occupies the mean-field wavefunction $e^{i p_\sigma  \bar{\phi}_\mathrm{\bf k}/2}      \,    \Psi_{  \mathrm{\bf k}, \sigma}$ (left side of Eq.~(\ref{HartreeSoln})) at a single point along the manifold defined by either right or left chiral phase $\bar{\phi}_\mathrm{\bf k}$ in Fig.~\ref{ContSym}. Yet the underlying single-particle scattering phases $\phi_\mathrm{\bf j}^{(n)}$ (right side of Eq.~(\ref{HartreeSoln})) are incoherently entangled around the central peak and effectively map the outer annulus of the Fermi disk (the gap region) to the $SU(2)$ structure of the quark field, their phase average represented by the single mean field phase $\bar{\phi}_\mathrm{\bf k}$. To visualize the effect of the attractive diquark interaction, consider that the quartic term in Eq.~(\ref{brokensymmetryFree3}) binds a (collective) quark with momentum $\mathrm{\bf k}$ to one with momentum $- \mathrm{\bf k}$. In the geometric picture of the potential landscape in Fig.~\ref{Minima}(b), such bound pairs lie at diametrically opposite points along the $\bar{\phi}_\mathrm{\bf k}$ circle minimum, pairing a quark at $R$ with one at $-R$, say. Thus, a diquark within the ground state is comprised of two single-quark mean field states bound non-locally (offset by a half cycle in the figure).

 The occurrence of bound quarks in our geometric picture is reflected in the presence of a large BCS gap in this regime, which constitutes the order parameter for the system. The size of the gap is given by the difference between the baryon chemical potential and the diquark pairing field, $\Delta_\mathrm{BCS} = | \mu_B - \bar{\Delta}_d | \equiv  |\tilde{\mu}_B|$. It is this gap that forces the average quark field to reside at some finite nonzero density given by Eq.~(\ref{finitefield}), specifically $|\eta_\mathrm{min}|^2 = (4/g^2)\,  \Delta_\mathrm{BCS}$. Long distances or low energies correspond to the BEC limit with excitations lying along the circle $\bar{\phi}_\mathrm{\bf k}$. Shorter distances, higher energies, or a lower potential central peak correspond to the BCS limit wherein fermionic excitations of the diquark are loosely bound. 
 
Let us now examine the meson-diquark mixed phase defined by $0 < m \lesssim  m_c$, depicted in Fig.~\ref{Minima}(c) and Eq.~(\ref{brokensymmetryFree3}) with the mass perturbation turned on. To understand this regime, we must study the microphysics by expanding the Hamiltonian around a minimum of the effective potential in terms of fluctuations in the mean-field amplitude and spin-polarization angle $\left( \delta \eta , \, \delta \gamma \right)$. Such an expansion accounts for quantum corrections (beyond large N mean field theory) to fluctuations of the polar coordinates in spin space and is produced in detail later in this work. The coefficients of the second order terms, $\delta \eta^2$ and $\delta \gamma^2$, yield the masses $m_\eta$ and $m_\gamma$ 
 \begin{widetext}
 \begin{eqnarray}
m_\gamma =  8 m_c   \left[ 1 - \left( \frac{m}{m_c} \right) \right]^{1/2} \left\{   2  \left[  1 + \left( \frac{m}{m_c}\right) \right]^{1/2}  - \left( \frac{m}{m_c} \right)  \left[ 1 - \left(\frac{m}{m_c}\right) \right]^{1/2} \left[3 - \left(\frac{m}{m_c}\right)  \right]    \right\}  \, ,  \;\;\;\;\; m_\eta = 8   m_c   \, . \label{mgamma}
\end{eqnarray}
\end{widetext}
The significance of the result in Eq.~(\ref{mgamma}) is that the mass for fluctuations in the spin polarization, $m_\gamma$, vanishes at the critical point $m = m_c$. The detailed behavior of the transition near criticality is obtained by expanding Eq.~(\ref{mgamma}) for $m/m_c \lesssim 1$, which gives 
 \begin{eqnarray}
 m_\gamma \simeq 16 \sqrt{2} \,  m_c   \left[ 1 - \left( \frac{m}{m_c} \right) \right]^{1/2} \, , 
 \end{eqnarray}
from which one can read off the critical exponent $\nu = 1/2$. The behavior of the diquark condensate at zero temperature has been calculated in Eq.~(\ref{explicitdiquark2}), which as we saw also vanishes at $m = m_c$. This result tells use that superfluidity breaks down at the quantum critical point. To probe more deeply we must calculate the diquark pairing function beyond mean field theory by integrating out the elementary quark amplitude and chiral fluctuations, $\delta \eta$ and $\delta \gamma$, near a minimum of the effective potential. This effectively casts the theory in terms of bound quark currents of equal chirality. Casting the theory into this form allows us to read off the diquark pairing function from the coefficient of the resulting diquark mass term. At zero temperature, this yields 
\begin{eqnarray}
\Delta_d =  4 m_c   \left[  1   - \left(\frac{m}{m_c} \right)^2 \right]^{1/2}  \, . \label{Gap}
\end{eqnarray}
In effect, integrating out the chiral fluctuation $\delta \gamma$ enfolds its mass structure into the pairing function. We point out that vanishing of $\Delta_\mathrm{BCS}$, $m_\gamma$, and $\Delta_d$ at $m_c$ is not coincidental. The important result in Eq.~(\ref{Gap}) tells us that dissolving of the diquark condensate and diquark dissociation occur at the same quantum critical point, i.e., even in the complete absence of thermal energy. We will show that large fluctuations in the chiral field $\delta \gamma$ at $m = m_c$ provide the needed dissociation energy at zero temperature, due in fact to a diverging Green's function for quantum fluctuations of $\delta \gamma$ when $m_\gamma \to 0$. 

To summarize our discussion, the various quantities that vanish at criticality, and thus characterize the meson-diquark quantum phase transition are:

\begin{enumerate}

\item BCS gap 
\begin{eqnarray}
\Delta_\mathrm{BCS} = m_c \left[ 1- \left(m/m_c\right)^2 \right]^{1/2}\, ;
\end{eqnarray}

\item  Diquark pairing function 
\begin{eqnarray}
\Delta_d = 4  m_c \left[ 1- \left(m/m_c\right)^2 \right]^{1/2}\, ; 
 \end{eqnarray}
 
 \item Mass of chiral fluctuations 
 \begin{eqnarray}
 m_\gamma = 16 \sqrt{2}  m_c \left[ 1- \left(m/m_c\right) \right]^{1/2}\, ; 
 \end{eqnarray}

 \item Expectation of the chiral mixing angle 
 \begin{eqnarray}
 \left( \gamma_\mathrm{min} \right)_c=  \pi/2 -  \mathrm{tan}^{-1}\sqrt{ (1 + m/m_c )/|1 - m/m_c|}  \, . 
\end{eqnarray}

\end{enumerate}

\subsubsection{Dissociation Through Quantum Chiral Fluctuations: Diquark-Meson Tunneling}
\label{Dissociation}

\begin{figure}[]
\begin{center}
 \subfigure{
\label{fig:ex3-a}
\hspace{0in} \includegraphics[width=.25\textwidth]{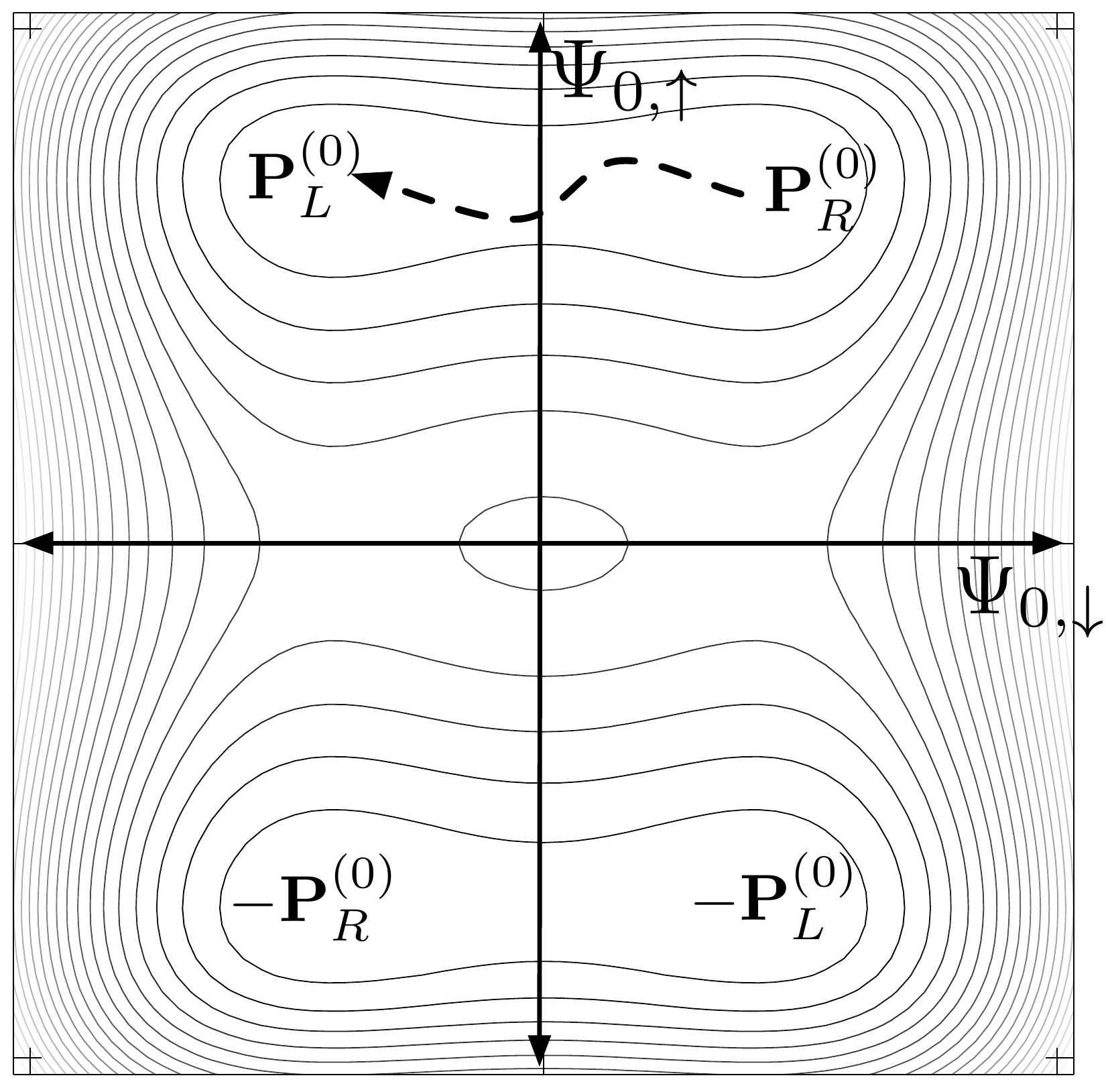}  } 
\vspace{0in}
\end{center}
\caption[]{(color online) \emph{Quantum tunneling between adjacent ground states of the effective potential. } Tunneling in spin-space between the right and left chiral ground states, ${\bf P}^{(0)}_R$ to the ${\bf P}^{(0)}_L$, passing through an intermediate virtual meson state along the vertical direction driven by the $\delta \gamma$ field. Note that the proximity to the critical point depicted here corresponds to panel (c) in Fig.~\ref{Minima}. }
\label{Tunneling}
\end{figure}

To better understand how the chiral fluctuations $\delta \gamma$ drive decoherence in the diquark condensate and molecular dissociation, we apply a Madelung decomposition of the Weyl spinor field which takes the form $\Psi_{R, L}$ $=$ $\eta^{1/2}$ $\exp(i \theta )$ $\left[ \mathrm{cos}\gamma , \, \pm \exp(i \phi ) \, \mathrm{sin}\gamma \right]^T$, where the various parameters are functions of the coordinates. Note that we present the main points in this section and put off the detailed mathematical exposition for Sec.~\ref{MadForm}. Here, as previously discussed, the additional field $\gamma$ is generally massive and therefore non-dynamical at zero temperature (except at the quantum phase transition). Its expectation value is related to the particular ground state of the system through the mean polarization vector ${\bf P}  \equiv (\mathrm{cos}\langle\gamma \rangle, \, \mathrm{sin}\langle\gamma \rangle)^T =  ( \langle \eta_1 \rangle , \, \langle  \eta_2 \rangle )^T$, described in Sec.~\ref{FermiSurface}. For illustration, consider that we are in the ${\bf P}^{(0)}_R = ( 1 , \, 1)^T$ ground state of the manifold with $\left( \mathbb{Z}_4 \right)_\chi$ symmetry, which corresponds to $\langle \gamma \rangle= \pi/4$ for the right hand chiral ground state in Fig.~\ref{Minima}(b). Consider then increasing the quark mass by a small amount so that the discrete chiral symmetry is broken $\left( \mathbb{Z}_4 \right)_\chi \to \left( \mathbb{Z}_2\right)_\chi  \times \left( \mathbb{Z}_2\right)_\chi $. To expand about the new minimum we take $\langle \gamma \rangle \to  \pi/4 + \delta \gamma$, so that 
\begin{eqnarray}
{\bf P} &\simeq& \left[ \mathrm{cos}(\pi/4 + \delta \gamma) , \,  \mathrm{sin}(\pi/4 + \delta \gamma)\right]^T \\
&\simeq&  (1 , \, 1)^T + \delta \gamma  \, ( 1 , \, -1 )^T  \\
&=& {\bf P}^{(0)}_R + \delta \gamma  \, {\bf P}^{(0)}_L \, . \label{SpinFluctuations}
\end{eqnarray}
Hence, a small fluctuation $\delta \gamma$ around a particular ground state, here ${\bf P}^{(0)}_R$, introduces a component of the opposite chiral ground state ${\bf P}^{(0)}_L$. Simply stated, the fluctuation $\delta \gamma$ describes quantum tunneling into an adjacent ground state as shown in Fig.~\ref{Tunneling}.

Functional integration over the massive fields, $\delta \eta$ and $\delta \gamma$, couples quarks of the same chirality, i.e., states of opposite momentum and spin polarization, forming singlet Cooper pairs. Chiral fluctuations in the second term of Eq.~(\ref{SpinFluctuations}) affect the overall spin of these Cooper pairs. Consider the diquark state $\langle \Psi_R \Psi_R \rangle$. The second term in Eq.~(\ref{SpinFluctuations}) induces the fluctuations 
\begin{eqnarray}
\hspace{-1.2pc}  \delta \langle \Psi_R \Psi_R \rangle \! \!  &=& \! \!    \langle \Psi_R \Psi_L \rangle   \delta \gamma   +    \langle \Psi_L \Psi_R \rangle   \delta \gamma   +    \langle \Psi_L \Psi_L \rangle   \delta \gamma^2   , \\
      &=&  \! \!    \langle \bar{\Psi}  \Psi \rangle   \delta \gamma   +    \langle \Psi_L \Psi_L \rangle   \delta \gamma^2  \, , 
\end{eqnarray}
 from which one sees that first-order corrections in the chiral field $\delta \gamma$ are mesonic, whereas the second-order (hence weaker) term describes tunneling into the opposite chiral state. Thus, the dominant effect of the $\delta \gamma$ field is to drive the individual chiral quarks that make up the diquark towards mixed chiral states by inducing fluctuations in the meson field. Indeed, by tuning $m \to m_c$, fluctuations in $\delta \gamma$ become more pronounced as adjacent minima in the effective potential are brought nearer to each other (see Fig.~\ref{Tunneling}), reducing the height of the barrier between them and introducing admixtures of the opposite chiral state, or equivalently by introducing meson fluctuations into the quark field. Finally, coherence of the diquark condensate and binding is completely destroyed when the meson field dominates at $m= m_c$.

It is interesting to visualize the unbinding of diquarks as we take $m$ $=$ $0$ $\to m_c$, by tracking the contours along the radial direction through the diagonal that connects the two right handed ground states in Fig.~\ref{Minima}. In Fig.~\ref{Minima}(c), the central peak along the diagonal (red dashed) has not yet vanished, so that bound states along that direction (diquarks) are still favored. However, the true eigenstates are the mixed states that lie along the line passing through the actual minima (purple dashed), with one minimum indicated by the new fluctuations in purple. When $m$ reaches its critical value $m_c$ in Fig.~\ref{Minima}(d), the central peak along the diagonal vanishes, at which point quark excitations are liberated from the condensate. Notice that now the stable bound states are those along the vertical axis (blue dashed), which are mesonic with minimal quantum fluctuations into the left and right hand chiral channels.

\subsubsection{Scaling of Quark Mass with Chemical Potential}

At this point we should discuss the relative scaling between the quark mass and chemical potential. We can deduce this relationship by examining the scaling behavior for $m_\gamma$ near criticality. First, in order to make sense of vertical lines of constant $m$ in the phase diagram Fig.~\ref{PseudospinDomain}, the temperature dependence in $\tilde{\mu}_B$ must cancel that in $m$ along such lines. To understand how the quark mass relates to the temperature-independent part of the effective chemical potential, (i.e., $\mu_B$), we can apply a scaling argument which determines precisely how the ground state characteristic energy scale, $m_\gamma$, scales with the chemical potential $\mu_B$. First, we note that increasing $\mu_B$ corresponds to increasing $m_c$. Thus, letting the chemical potential run corresponds to taking $m_c  \to \varepsilon m_c$, with $\varepsilon$ a dimensionless scaling parameter, so that the critical behavior for $m_\gamma$ becomes $m_\gamma \sim \varepsilon^2 m_c^2  \sqrt{ 1 - \left( m/\varepsilon m_c \right) }$. Taking $\varepsilon \gg 1$ drives $m_\gamma$ away from criticality, specifically towards lower values of $m$ in Fig.~\ref{PseudospinDomain}(a). Hence, the direction of increasing chemical potential correlates with decreasing mass. This argument applies similarly to our results for $\Delta_\mathrm{BCS}$ and $\Delta_d$.

\subsection{Topology of the Quantum Critical Point}

It is worthwhile to categorize the Fermi surface and the phase transitions near its formation in terms of topology. Transitions and formation of the Fermi surface at finite as well as at zero-temperature can be classified by the change in multiplicity and character of ground states consistent with discrete symmetry changes. Since we consider a finite chemical potential, generally, our system retains a Fermi surface (line) with co-dimension ${ 2} - { 1} = {1}$ (the dimension of ${\mathrm{\bf p}}$-space minus the dimension of the surface), which transforms to Fermi points in the special case $\mu_B =0$. Topological order related to the Fermi surface in our model can be studied through the Green's functions, from Eq.~(\ref{brokensymmetryFree3}), for the noninteracting right and left-handed chiral fluctuations taken at zero quark mass near the Fermi surface, as we now discuss. Good expositions of this topic may be found in\cite{Volovik1999,Volovik2003}.

 \subsubsection{Green's Function Formulation}

First, we recall that for $\tilde{\mu}_B  > 0$, the system is gapless and the Green's function is diagonal in the opposite polarizations labeled $\pm R (L)$ in Fig.~\ref{Minima}(b), which form the eight-component Nambu-Gorkov spinor $\Psi = \left( \Psi_{R +},  \Psi_{R -},  \Psi_{L +},   \Psi_{L -} \right)^T$. Note that a Lifshitz transition occurs here when $\tilde{\mu}_B$ changes sign from positive to negative, the sign change coming from a shift in $\tilde{\mu}_B$ due to attractive quark-quark interactions. The negative chemical potential term now couples positive and negative right and left polarizations in the Nambu-Gorkov spinor, consistent with a transition from the gapless to the fully gapped ground state, with gap size $= 2 |\tilde{\mu}_B|$. This is also associated with a transition to a fermion coherent state, as we have shown. In the gapped state (at zero temperature) the inverse Green's function develops off-diagonal terms and takes the $8 \times 8$ matrix form 
\begin{eqnarray}
&&G^{-1}\left( i \omega, \mathrm{\bf p} \right) =   \label{fullGreen}\\
&& \left( \begin{array}{cccc} i \omega - \boldsymbol{\sigma} \cdot \mathrm{\bf p}  \;   & \;   |\tilde{\mu}_B| \; & 0  & 0  \\
       |\tilde{\mu}_B| \;   & \;    i \omega + \boldsymbol{\sigma} \cdot \mathrm{\bf p}  \;  &  0 & 0  \\
       0  &  0 &  \;    i \omega +  \boldsymbol{\sigma} \cdot \mathrm{\bf p}  \;   &  \;    |\tilde{\mu}_B|  \; \\   
        0  &  0 &  \;     |\tilde{\mu}_B| \; &  \;  i \omega -  \boldsymbol{\sigma} \cdot \mathrm{\bf p}   \;       \end{array} \right)   \nonumber 
\end{eqnarray}
where the alternating signs of the kinetic terms account for fermionic particle and hole fluctuations near the gap. The corresponding Hamiltonian gives two identical energy spectra for the left and right chiral modes given by 
\begin{eqnarray}
E_{\alpha \pm}(\mathrm{ \bf p} ) = \pm \sqrt{ | {\bf p}|^2 + \tilde{\mu}_B^2 } \, , 
\end{eqnarray}
with $\alpha = L, R$. 

To see how the Fermi surface topology changes as the gap is formed, consider again that for $\tilde{\mu}_B > 0$ the right and left sub-matrices of the inverse Green's function are diagonal, and given by
\begin{eqnarray}
G^{-1}_{R, L}\left( i \omega, \mathrm{\bf p} \right) &&=  \label{FirstGreen} \\
&&\left( \begin{array}{cc} i \omega \pm \boldsymbol{\sigma} \cdot \mathrm{\bf p} +   |\tilde{\mu}_B|   & 0  \\
    0   &    i \omega \mp  \boldsymbol{\sigma} \cdot \mathrm{\bf p}    +   |\tilde{\mu}_B|  \end{array} \right)   \nonumber  \, . 
\end{eqnarray}
The components of the Green's function in each case can be expressed as $G_{n,n}\left( i \omega, \mathrm{\bf p} \right)  =  |G_{n,n}|  e^{i \Phi_{n,n }}$, where $n =1, 2$ relate to particles and holes, respectively. In its diagonal form, the Green's function corresponds to the gapless (BCS) state and defines four distinct Fermi surfaces for particles and holes with left and right chiral indices. In particular, for particles in the right hand chiral ground state, we have 
\begin{eqnarray}
G_{11} = \frac{  \sigma \cdot \mathrm{\bf p}    -   |\tilde{\mu}_B|   - i \omega    }{     \omega^2 + |\mathrm{\bf p}|^2  -   |\tilde{\mu}_B|^2  } \, , 
\end{eqnarray}
so that 
\begin{eqnarray}
\left| G_{11} \right| = \frac{ 1   }{  \sqrt{     \omega^2 + |\mathrm{\bf p}|^2  -   |\tilde{\mu}_B|^2  }} \, , 
\end{eqnarray}
and 
\begin{eqnarray}
\Phi_{11}  = \mathrm{tan}^{-1}\left(  \frac{ \omega    }{  \sqrt{ |\mathrm{\bf p}|^2  -   |\tilde{\mu}_B|^2  }} \right)  \, . 
\end{eqnarray}
The Fermi surface is a circle in the $(p_x, p_y)$ plane, but the energy $\hbar \omega$ adds a third dimension, enlarging the space to $(\omega , p_x, p_y)$. Hence, rotations along paths that encircle an element of the Fermi line ($2 \pi$ rotations of $\Phi_{11}$) are possible, allowing us to define the topological winding number associated with elements of the fundamental group $\pi_1$
\begin{eqnarray}
N_1 = \mathrm{tr} \oint_{C} \frac{dl}{2 \pi i } G_{11} \partial_l G_{11}^{-1} \, , 
\end{eqnarray}
where the trace is over spin indices and the contour $C$ arbitrarily encircles an element of the Fermi line. For the four Fermi surfaces, we obtain $N_{1, R\,  \pm } = + 1, -1$, for right-handed particles and holes, and $N_{1,L \, \pm } = - 1, +1$, for left-handed particles and holes, respectively. Note that in the gapless state, particles and holes are distinct species of fermions with opposite topological charges and vanishing right (left) total charges, $N_{1, R (L)\,  + }  + N_{1, R (L)\,  -  } = 0$. Changing the sign of the chemical potential term, $\tilde{\mu}_B > 0 \to \tilde{\mu}_B < 0$, drives the system into the fully gapped (BEC) state exhibited by the Green's function Eq.~(\ref{fullGreen}), where now annihilation of the particle and hole Fermi surfaces are allowed due to the trivial total topological charge.

Another key topological phase transition occurs when keeping $\tilde{\mu}_B < 0$ held fixed, while the quark mass is tuned from zero through $m = |\tilde{\mu}_B|$. The transition in this case is driven by competition between the meson and diquark channels, or equivalently, between the quark mass and superconducting gap. To resolve this transition in the Green's function one must include lowest-order contributions from the chiral fluctuations $\delta \gamma$, induced by the presence of the nonzero quark mass $m$. Incorporating these fluctuations, the matrix for the inverse Green's function, Eq.~(\ref{fullGreen}), expands to 
\begin{widetext}
\begin{eqnarray}
 G^{-1}   = \left( \begin{array}{cccc} i  \omega_-    -    \boldsymbol{\sigma} \cdot  \mathrm{\bf p}_-       &  \Delta_\mathrm{eff}  &  m_+   & -\,   \delta m   \\
      \Delta_\mathrm{eff}  &    i \omega_+   + \boldsymbol{\sigma} \cdot \mathrm{\bf p}_+     &  \delta m  &  m_-    \\
       m_+  & - \,  \delta m  &     i  \omega_+  +  \boldsymbol{\sigma} \cdot  \mathrm{\bf p}_+    &    \Delta_\mathrm{eff} \\   
        \delta m    &   m_-  &   \Delta_\mathrm{eff} &    i  \omega_-   -  \boldsymbol{\sigma} \cdot  \mathrm{\bf p}_-      \end{array} \right)       \,,  \label{FinalGreen}
\end{eqnarray}
\end{widetext}
where the shifted four-momentum and quark mass are, respectively,  
\begin{eqnarray}
\left( \omega_\pm ,  \, \mathrm{\bf p}_\pm \right)   &=&    \left(    \omega \pm  p_{\gamma , y}     , \,  \mathrm{\bf p}  \pm  { \bf  \sigma } \,  m \,  \delta \gamma/2  \right)  \, ,  \\
m_\pm &=& m \left( 1 - \delta \gamma^2 \right) \pm p_{\gamma , x}  \left( 1 + \delta \gamma^2 \right)  \, . 
\end{eqnarray}
In addition, the effective superconducting (diquark) gap and off-diagonal mass perturbation are
\begin{eqnarray}
\Delta_\mathrm{eff} &=&  |\tilde{\mu}_B| \left( 1 - \delta \gamma^2 \right)  \, , \\
\delta m &=&  |\tilde{\mu}_B| \delta \gamma \, .  
\end{eqnarray}
The effective quark mass is given by the determinant of the off-diagonal $4 \times 4$ sub-matrix
\begin{eqnarray}
m_\mathrm{eff} &=& \left[  \mathrm{det}    \left( \begin{array}{cc}   m_+   & -\,   \delta m   \\
       \delta m  &  m_-     \end{array} \right)  \right]^{1/2} \\
       &=&  \left( m_+ m_-  + \delta m^2 \right)^{1/2}  =  \\
    && \hspace{-3.5pc} \left[ m^2 \left( 1 - \delta \gamma^2 \right)^2 -  p_{\gamma , x}^2  \left( 1 + \delta \gamma^2 \right)^2  +    |\tilde{\mu}_B|^2  \delta \gamma^2    \right]^{1/2} \hspace{-1pc} . \hspace{1pc}
\end{eqnarray}
Note that by taking the system to be at zero temperature ($\omega = 0$) the chiral degree of freedom $\delta \gamma$ is suppressed for zero quark mass, activated only as the quark mass is tuned upward towards criticality.

 \subsubsection{Emergence of Critical $(3 +1)$-Dimensional Conformal Theory}

 Let us examine the limits of small and large $\delta \gamma$ quantum fluctuations. For $\delta \gamma , \,  |\mathrm{\bf p}_\gamma| \to 0$, we find that $\left( \omega_\pm ,  \, \mathrm{\bf p}_\pm \right) \to \left( 0,  \, \mathrm{\bf p} \right)$, $\delta m \to 0$, and $m_\pm, \, m_\mathrm{eff}  \to m \ll   |\tilde{\mu}_B|$, with the superconducting gap simplifying to $\Delta_\mathrm{eff} \to  |\tilde{\mu}_B|$. In this limit, we recover the superconducting state in Eq.~(\ref{FinalGreen}) with a small quark mass correction, consistent with our analysis up to this point. Things become more interesting when the (purely) quantum fluctuations $\delta \gamma$ become large, i.e., on the order of unity ($\delta \gamma \lesssim 1$) at the quantum critical point, due to an increasing quark mass. In this limit, we find the characteristic momentum strongly shifted by a spin correction to $\mathrm{\bf p}_\pm  = \mathrm{\bf p}  \pm  m \,  \boldsymbol{\sigma}/2$ (where, to be clear, $\boldsymbol{\sigma} = \sigma_x \,  { \bf \hat{i} }+ \sigma_y \,  { \bf \hat{j}}$), a vanishing superconducting gap, $\Delta_\mathrm{eff}  \to 0$, and the effective quark mass approaching $m_\mathrm{eff}  \to  |\tilde{\mu}_B|$. The shifted momentum describes modulation by quantum fluctuations and can be equivalently rewritten as a modulated effective chemical potential, $\mu_\mathrm{eff} = m \,  \delta \gamma$. Most significant in this limit is the fact that a new energy and momentum scale emerges at zero temperature, $\left(  \omega_{\gamma }  , \,   p_{\gamma}     \right) = \left(  p_{\gamma , y}  , \,   p_{\gamma , x}     \right)$. That is, at criticality, one of the planar components of $\mathrm{\bf p}_\gamma$ acts as an additional energy with the other in the role of an extra fermion momentum-space dimension with a factor of $\sigma_z$ attached.

 Thus, at the quantum critical point $m =   |\tilde{\mu}_B|$, particles and holes decouple, the diquark condensate dissolves, and a new higher dimensional theory emerges. This is displayed by the inverse Green's function for particle excitations near the Fermi surface 
 \begin{eqnarray}
 G^{-1}   = \left( \begin{array}{cc} i  \omega   - \boldsymbol{\sigma} \cdot   \mathrm{\bf p} +  \mu_\mathrm{eff}  & m_\mathrm{eff}   \\
        m_\mathrm{eff} &    i  \omega  + \boldsymbol{\sigma} \cdot  \mathrm{\bf p}    +  \mu_\mathrm{eff}    \end{array} \right)      \,,  \label{FinalGreen2}
\end{eqnarray}
 where we now have an effectively $(3 +1)$-dimensional theory with (leaving out factors of $\hbar$ and $c$) $\omega = p_{\gamma , y}$, $\boldsymbol{\sigma}= \left(\sigma_x ,  \, \sigma_y , \, \sigma_z \right)$,  $\mathrm{\bf p}  = \left( p_x , \, p_y , \,  p_{\gamma , x} \right)$, $\mu_\mathrm{eff} =  m_\mathrm{eff} =  |\tilde{\mu}_B|$. The emergent $(3 +1)$-dimensional Dirac equation associated with Eq.~(\ref{FinalGreen2}) is 
 \begin{eqnarray}
 \left( i \gamma^\mu \partial_\mu - m_\mathrm{eff} +  \mu_\mathrm{eff}   \gamma^0     \right) \psi = 0 \, , \label{EmergDirac} 
 \end{eqnarray}
 with $\psi = \left( \Psi_{L +} , \, \Psi_{R +} \right)^T$, such that $\psi : \mathbb{R}^4 \to \mathbb{C}^4$, and the $4 \times 4$ gamma matrices in the Weyl basis are 
 \begin{eqnarray}
 \gamma^\mu   = \left( \begin{array}{cc}0   &  \sigma^\mu   \\
         \bar{\sigma}^\mu  &    0   \end{array} \right)      \,,  
\end{eqnarray}
with $\sigma^\mu   \equiv \left( 1 ,   \boldsymbol{\sigma}\right)$, $\bar{\sigma}^\mu   \equiv \left( 1 , -  \boldsymbol{\sigma}\right)$. It is significant to point out that Eq.~(\ref{EmergDirac}) recovers the noninteracting version of our original finite-temperature theory in Eq.~(\ref{kinetic}), yet now time-dependent in one higher spatial dimension and at zero temperature. If we further take $|\tilde{\mu}_B|\to 0$ in Eq.~(\ref{EmergDirac}), which shifts $m_c \to 0$ in Fig.~\ref{PseudospinDomain}, the energy scales $m_\mathrm{eff}$ and $\mu_\mathrm{eff}$ disappear and the theory becomes scale invariant. Hence, at the quantum critical point, our original interacting theory is described by a free conformal theory in one higher dimension.

 \subsubsection{Fermi Surface Fluctuations at the Quantum Critical Point}

The topology of this quantum phase transition can be understood in terms of the Fermi surface. Expressing the inverse Green's function as 
 \begin{eqnarray}
 G^{-1}   = \left( \begin{array}{cc} i  \omega   - \boldsymbol{\sigma} \cdot  \left( \mathrm{\bf p} - {\bf b} \right) & m_\mathrm{eff}   \\
        m_\mathrm{eff} &    i  \omega  +   \boldsymbol{\sigma} \cdot  \left( \mathrm{\bf p}  + {\bf b} \right)  \end{array} \right)      \,,  \label{FinalGreen3}
\end{eqnarray}
obtained by rewriting the effective chemical potential in Eq.~(\ref{FinalGreen2}) as a CPT violating spin-current, ${\bf b } = m \,  \delta \gamma \,   \boldsymbol{ \sigma }/3$. The corresponding Hamiltonian then reads
 \begin{eqnarray}
 H  = \left( \begin{array}{cc}   \boldsymbol{\sigma} \cdot  \left( \mathrm{\bf p} - {\bf b} \right) & m_\mathrm{eff}   \\
        m_\mathrm{eff} &    -  \boldsymbol{\sigma} \cdot  \left( \mathrm{\bf p}  + {\bf b} \right)  \end{array} \right)      \,,  \label{CritHam}
\end{eqnarray}
which has the energy spectrum 
\begin{eqnarray}
E^2_\pm({\bf p}) = m_\mathrm{eff}^2 + |\mathrm{\bf p}|^2  + b^2 \pm 2 b \sqrt{m_\mathrm{eff}^2 + \left( {\bf p} \cdot \hat{\bf b} \right)^2 } \, , \label{Spectrum}
\end{eqnarray}
where $b = | {\bf b}| = m \, \delta \gamma$, $ \hat{\bf b} =   \boldsymbol{ \sigma }/3$, and $m_\mathrm{eff} =  |\tilde{\mu}_B|\, \delta \gamma$. From this viewpoint, the spectrum Eq.~(\ref{Spectrum}) reveals a quantum phase transition at $b = m_\mathrm{eff}$ that separates the fully gapped ground state, for $b < m_\mathrm{eff}$, and a ground state with two isolated Fermi points, for $b > m_\mathrm{eff}$, located at ${\bf p}_1 = + \hat{\bf b} \sqrt{b^2 - m_\mathrm{eff}^2}$, ${\bf p}_2 = - \hat{\bf b} \sqrt{b^2 - m_\mathrm{eff}^2}$, hence the liberation of free fermions from the diquark condensate. The location of the quantum critical point is at $m = m_c =   |\tilde{\mu}_B|$, precisely the same result that we have previously obtained. The Fermi points at ${\bf p}_1$ and ${\bf p}_2$ are just the original Fermi lines from the $(2+1)$-dimensional theory at $m =0$ in Eq.~(\ref{FirstGreen}), but expanded into the enlarged $(3+1)$-dimensional space at the critical point $m= m_c$. However, the codimension now becomes $3 - 0 = 3$, since each Fermi point is topologically equivalent to shrinking a two-dimensional Fermi surface down to a point. The fundamental group $\pi_1$ acting on this space is trivial, so that the loops encircling the original Fermi lines are no longer topologically stable. The relevant nontrivial homotopy group is now given by $\pi_2 \! \left(GL\left(n , \mathbb{C} \right)\right) = \mathbb{Z}$, with the topological invariant a surface integral in the four-dimensional frequency-momentum space $p^\mu = \left( \omega , \, {\bf p} \right)$ given by
\begin{eqnarray}
&&\hspace{-1pc}N_3 = \\
&&\hspace{-1pc}\frac{1}{24 \pi^2} \, \epsilon_{\mu \nu \rho \sigma } \mathrm{tr} \oint_{\Sigma_a}  \!    dS^\sigma  G \frac{\partial}{\partial p^\mu}  G^{-1}  G \frac{\partial}{\partial p^\nu}  G^{-1}   G \frac{\partial}{\partial p^\rho}  G^{-1} \, ,  \nonumber 
\end{eqnarray}
computed using the Green's function at criticality. A direct calculation yields topological charges $N_3 = +1$ and $N_3 = -1$, for Dirac points at ${\bf p}_1$ and ${\bf p}_2$, respectively.

This picture is crucial for understanding how bound quark states transform through the quantum critical point. Such states are fermion coherent states that wrap around the original Fermi line that only reappears either at short distances, on the order of the size of diquarks, or driven by competition between the quark density and mass. Thus, the quantum phase transition is fundamentally rooted in unwrapping a quark coherent state from the Fermi line by rolling it into the emergent $p_\gamma$ dimension and shrinking it to a vanishing point, equivalent to the trivial mapping $S^1 \to S^2$.

\vspace{2pc}
\section{Madelung Formalism for Spin-Charge Separated Excitations of Bilinear Quark Condensates}
\label{MadForm}

In this section we present the detailed method for expanding around the minima in Fig.~\ref{Minima} consistent with spin-charge separation which includes quantum corrections beyond the at large $N_c$ mean field limit. Here we present the general structure for quark excitations of bilinear condensates in greater mathematical depth for the regime defined by $m  > m_c$, or equivalently $\mu_B < \bar{\mu}_{B, c}$, or the reverse for the diquark condensate.

\subsection{General Formulation}

To proceed, we introduce the canonical Madelung decomposition of the mean field quark wavefunction that emerges at large N in terms of Weyl component amplitude and phase fields 
\begin{eqnarray}
\hspace{-1pc}  \Psi( {\bf r}, t   ) \! = \! e^{i \vartheta( {\bf r}, t  )}   \eta(  {\bf r}, t  )   \left[  \mathrm{cos}\gamma( {\bf r}, t  )      \, , \;    e^{i \varphi( {\bf r}, t  )} \mathrm{sin}\gamma( {\bf r}, t  )  \right]^T\! \! \! \! \!  , \label{Madelung}
\end{eqnarray}
where $\vartheta$ and $\varphi$ are the overall and relative phases, respectively, with the local spin-up(down) densities $\rho_{1(2)}$ expressed in terms of the fermion amplitude $\eta$ and chiral angle $\gamma$ by $\rho_1 = \eta^2  \mathrm{cos}^2\gamma$ and $\rho_2 =  \eta^2  \mathrm{sin}^2\gamma$. This decomposition into amplitude, orbital, and spin degrees of freedom fully displays the spin-charge separation of the system. Substituting the decomposition Eq.~(\ref{Madelung}) into Eq.~(\ref{FullLagrangian}), after some algebraic manipulation we extract the associated Hamiltonian
\begin{widetext}
\begin{eqnarray}
 &&\hspace{-1pc}H  =  \int \! d{\bf r} \left[ i \, \eta^2\,  \mathrm{sin}2 \gamma \, {\bf n}_\varphi \cdot \nabla \vartheta   + \frac{1}{2}  \,  \mathrm{sin}2 \gamma  \, {\bf n}_\varphi \cdot \nabla \eta^2    +    \eta^2  \,  \mathrm{cos}2 \gamma  \, {\bf n}_\varphi \cdot \nabla \gamma +  i\,   \eta^2  \,\vert  {\bf n}_\varphi  \times  \nabla \gamma  \vert +   i\, \frac{1}{2}     \eta^2 \,  \mathrm{sin}2 \gamma   \, {\bf n}_\varphi \cdot \nabla \varphi \right. \nonumber  \\
 && \left. -  \frac{1}{2}     \eta^2 \,  \mathrm{sin}2 \gamma  \, \vert {\bf n}_\varphi  \times  \nabla \varphi \vert  +      \eta^2 \,  {\bf n}_\gamma^T \, {\bf M}  \, {\bf n}_\gamma   +   \frac{g^2}{2}   \eta^4 \left( \mathrm{cos}^4\gamma  +    \mathrm{sin}^4\gamma \right) \right]  \! .  \label{landau}
\end{eqnarray}
\end{widetext}
From here forward we will work in this time-independent formalism which favors physical clarity at the expense of explicit Lorentz covariance. The derivative terms in Eq.~(\ref{landau}) appear as scalar and vector products of the unit vector ${\bf n}_\varphi({\bf r}) \equiv \left[ \mathrm{cos}\varphi({\bf r}   ) \,,  \; \mathrm{sin}\varphi( {\bf r} ) \right]$ with the various field fluctuations. The interaction and mass terms in the last line of Eq.~(\ref{landau}) depend on both the overall density $\eta^2({\bf r})$ as well as the spin polarization angle $\gamma({\bf r} )$, such that any finite interaction $g \ne 0$ or asymmetry in the coefficients of the quadratic terms may break the rotational symmetry associated with the polarization vector ${\bf n}_\gamma({\bf r})   \equiv \left[ \mathrm{cos}\gamma({\bf r}) \,,  \; \mathrm{sin}\gamma({\bf r}) \right]$. The temperature dependent mass matrix in Eq.~(\ref{landau}) is
\begin{eqnarray}
{\bf M}(T) \equiv \left(  \begin{array}{cc}
       m     - \tilde{\mu}_B(T)    &  0  \\
         0    &   - m     - \tilde{\mu}_B(T)   \\
             \end{array} \right)  \, , \label{massmatrix}
\end{eqnarray} 
where we indicate the chemical potential to be explicitly dependent on the temperature $T$.

It is important to note that we must also include terms in Eq.~(\ref{landau}) that account for parity reversal, i.e., both left and right chiral fields should be included. This is accomplished by following the same steps that lead to Eq.~(\ref{landau}) but starting from the parity reversed field  
\begin{eqnarray}
&&\psi( - {\bf r},  t) =   \\
&&e^{i \vartheta(-{\bf r}, t )}   \eta( - {\bf r}, t )   \left[  e^{ i \varphi( - {\bf r},  t)}  \, \mathrm{cos}\gamma(-{\bf r},  t )      \, , \;   \mathrm{sin}\gamma(- {\bf r}, t )  \right]^T  .\nonumber   \label{Madelung2}
\end{eqnarray}
For clarity, throughout most of our analysis we will focus on the terms in Eq.~(\ref{landau}). However, inclusion of parity reversed terms is crucial for defining chiral transformations and become important when we integrate out mediating fields to describe fermion pairing in the effective Hamiltonian.

We now focus on low-energy fluctuations of Eq.~(\ref{landau}). The quadratic and quartic terms in Eq.~(\ref{landau}) show that after symmetry breaking we generally have $\langle \, \gamma \, \rangle =  \gamma_\mathrm{min} \ne 0$ and $\langle \, \eta \, \rangle =  \eta_\mathrm{min} \ne 0$, in other words the average of the spin polarization and overall amplitude may be nonzero. Note, however, that $\langle \, \gamma \, \rangle \ne 0$ here is a result which does not depend on temperature, but is associated with the discrete chiral symmetry of quartic term and holds for all $T$. In contrast, the condition $\langle \, \eta \, \rangle \ne 0$ depends on temperature through the coefficient of the quadratic term. Expanding Eq.~(\ref{landau}) around the point $\left( \langle \, \gamma \, \rangle  ,  \langle \, \eta \, \rangle \right)$ to second order in the field fluctuations $\delta \gamma$ and $\delta \eta$ by shifted to the field minima $\eta = \langle \, \eta \, \rangle + \delta\eta $ and $\gamma = \langle \, \gamma \, \rangle + \delta\gamma$, we obtain the Hamiltonian for fermionic fluctuations 
\begin{widetext}
\begin{eqnarray}
  H   =  \int \! d{\bf r} \left[       s_m   \delta \eta  \nabla_\parallel \delta \eta +   \left( \rho^{1/2} + \delta \eta  \right)^2  \left(  i s_m  \vert \nabla \vartheta  \vert  +  c_m   \nabla_\parallel \delta \gamma   +   i   \nabla_\perp \delta \gamma  +    i \frac{1}{2}     s_m      \nabla_\parallel  \varphi  -  \frac{1}{2}    s_m      \nabla_\perp \varphi  \right)    \right]  \nonumber \\
    \hspace{-4pc}  +   \int \! d{\bf r}  \left[    4  g^2 \rho     \left(  \delta \eta^2  +  s_m  \rho \,  \delta \gamma^2  \right) - 2 \rho  s_m \left( g^2 \rho - 2  m \right)   \delta \gamma   \right]    \, ,  \label{landau3}
\end{eqnarray}
\end{widetext}
where we identify the average quark density as $\rho \equiv   \langle \, \eta \, \rangle^2$, and the mass-dependent coefficients are defined by $s_m \equiv  \mathrm{sin} \left(2 \langle \, \gamma \, \rangle\right)$, and $c_m \equiv  \mathrm{cos} \left(2 \langle \, \gamma \, \rangle\right)$. Note that we have used the fact that the variable $\varphi$ is identified with the polar angle $\phi_p \equiv \mathrm{tan}^{-1}\left( \partial_y \vartheta/\partial_x \vartheta\right)$, which describes the spatial orientation of $\nabla \vartheta$, i.e., the direction of the quark current, consistent with the internal angle of a fermionic field. The parallel and orthogonal notation in the other gradients refer to the direction vector $\nabla \vartheta/|\nabla \vartheta|$. Note that there are two massive modes, $\delta \eta$ and $\delta \gamma$, associated with discrete chiral symmetry breaking.

 Several features of Eq.~(\ref{landau3}) are worthy of discussion. There are several possible values for $\langle \, \gamma \, \rangle$, hence chiral ground states, corresponding to the minima for the effective potential in $H$, each defining a particular orientation of the average polarization $\langle {\bf n}_\gamma \rangle \equiv \left(  \mathrm{cos}\langle \, \gamma \, \rangle \, , \; \mathrm{sin}\langle \, \gamma \, \rangle\right)^T$ in spin space. The number of minima and the character of fluctuations around each of these is determined by the value of the scalar mass $m$ which appears in the last two terms of Eq.~(\ref{landau3}). In effect, tuning $m$ through a critical point $m_c$ at zero temperature drives the system between the meson and diquark phase.

Consider, next, the gradient terms in Eq.~(\ref{landau3}). From our understanding of the nature of $\mathrm{Spin}(2)$ states we deduce that the gradient terms in the phase $\varphi$ describe higher-order fluctuations in the (dressed) quark momentum $\nabla \vartheta$, thus parallel $\parallel$ and perpendicular $\perp$ notation refer to the direction of the quark current. Classifying here the fluctuations in our system, we find: 1) a longitudinal spin density wave (density compression wave) $\delta \eta$ parallel to the gradient of the quark phase $\vartheta$; 2) a massive spin-wave fluctuation in the local chirality encoded in the chiral fluctuation $\delta \gamma$; 3) a phase fluctuation (spin wave) associated with  the quark current $\nabla \vartheta$; and 4) fluctuations $\varphi$ in the direction of the quark current. Encoded in $\nabla \varphi$ are the longitudinal ``snake'' mode $\nabla_\parallel \varphi$, which potentially nucleates quantum turbulent flow in one of the bilinear condensates, and a transverse spin wave $\nabla_\perp \varphi$. This last mode introduces spatial divergence into the quark current, causing local fluctuations in the quark density.  

To deduce the dynamics of fluctuations in the bilinear condensates near the quantum critical point, we integrate over the field $\delta \gamma$ in Eq.~(\ref{landau3}), which implicitly incorporates all quantum fluctuations of the massive chiral field. Proceeding toward this end, we expand the coefficients $c_m$ and $s_m$ in Eq.~(\ref{landau3}) to first-order in $\delta \gamma$, which gives $c_m \simeq 1$, $s_m \simeq \delta \gamma$, and yields the Hamiltonian 
\begin{widetext}
\begin{eqnarray}
        && \hspace{-3pc}H  =      \int \! d{\bf r} \left\{   (1/2)    \delta \gamma       \nabla_\parallel \delta \eta^2 +   \left(  \rho^{1/2} + \delta \eta \right)^2    \left[ \left(  \nabla_\parallel  +  i   \nabla_\perp \right) \delta  \gamma +   i  \,  \delta \gamma  \vert \nabla \vartheta  \vert  +    (1/2)   \,    \delta \gamma     \left( i    \nabla_\parallel   -   \nabla_\perp   \right)  \varphi   \right]    \right\} \nonumber \\
       &&  +    \int \! d{\bf r}    \left[  4 (m + m_c ) \,   \delta \eta^2   + 2  | m- m_c|     \,  \delta \gamma^2 \right]       \label{landau4} \, ,   
\end{eqnarray}
\end{widetext}
where we have omitted higher-order interaction terms. We have expressed the coefficients of the mass terms in Eq.~(\ref{landau4}) in terms of the critical mass $m_c$ through the relations $4 g^2  \rho = 4 ( m + m_c )$ and $2 \left( 2 m  - g^2  \rho \right) = 2 \vert m - m_c \vert$. Contributions from phase fluctuations of the parity reversed field Eq.~(\ref{Madelung2}) do not appear in Eq.~(\ref{landau4}). To lowest order in the field $\delta \eta$, Eq.~(\ref{landau4}) becomes
\begin{eqnarray}
   &&\hspace{-1pc}H =  \nonumber  \\
   &&\hspace{-1pc}\int \! d{\bf r}    \,    \delta \gamma    \left[  (1/2)         \nabla_\parallel \delta \eta^2   +       i  \, \rho   \vert \nabla \vartheta  \vert  +    (1/2)   \rho    \left( i    \nabla_\parallel   -   \nabla_\perp   \right)  \varphi   \right]     \nonumber \\
      &&\hspace{-1pc}+    \int \! d{\bf r}    \left[   4 ( m + m_c )  \,   \delta \eta^2   + 2  | m- m_c|     \,  \delta \gamma^2  \right]     \label{landau5}  \, ,  
\end{eqnarray}
where now $\delta \gamma$ appears as a non-dynamical (auxiliary) field, upon integrating a total derivative in $\delta \gamma$.

To isolate the low-energy physics of Eq.~(\ref{landau5}) we integrate out the non-dynamical field $\delta \gamma$. This procedure retains the effect of varying $\delta \gamma$ implicitly through the mass $|m - m_c|$.~$^{\footnotemark[4]}$\footnotetext[4]{References~\cite{Stoof2009,Altland2010} provide thorough technical introductions to the field theory methods used in this article as well as condensed matter systems in general.}  We perform the functional integration using the Hubbard-Stratonovich transformation, applying essentially the continuous version of the identity
\begin{eqnarray}
  &&\hspace{-1pc}  \exp \left( \frac{1}{2} \int  \! d{\bf r} \,   x_{i } A_{i j } x_{j }   \right)   =   \label{Identity}  \\
    && \hspace{-1pc}  \int \frac{\mathcal{D}[ y_{1 }, y_{2 }, ... ]  }{  \sqrt{(2 \pi)^N \mathrm{det} A }} \;  \exp \left( \int  \! d{\bf r} \,\left\{ - \frac{1}{2}  y_{i }  [ A^{-1} ]_{i j } y_{j }   +  y_{i }  x_{i }  \right\}   \right)\, , \nonumber 
    \end{eqnarray}
where $[A]_{ij}$ is a real symmetric positive-definite matrix, $y_{i }({\bf r})$ and $x_{i }({\bf r})$ are fermionic fields, and the determinant factor in the denominator normalizes the integral to unity. Applying Eq.~(\ref{Identity}) at the level of the partition function transforms Eq.~(\ref{landau5}) into the effective Hamiltonian 
\begin{eqnarray}
     H_\mathrm{eff}  =       \frac{   1  }{16 | m- m_c|   }   \int \! d{\bf r}              \left(   \,    \vert    \nabla \delta \eta^2 \vert^2  +  \rho^2  \vert \nabla \zeta  \vert^2     \,    \right) \nonumber  \\
     + \,   4   \int \! d{\bf r}       \,  ( m + m_c )  \,   \delta \eta^2        \, ,     \label{landau5.2}
\end{eqnarray}
where we have combined phase fluctuations into a single overall phase defined as 
\begin{eqnarray}
 \vert \nabla \zeta  \vert^2 \equiv   \vert  \nabla_\perp    \varphi  \vert^2     +       \vert   \nabla_\parallel  \left( \vartheta    +   \varphi   \right)  \vert^2     \, . \label{totalphase}
\end{eqnarray}
The $\parallel$ notation on $\delta \eta$ has been dropped in Eq.~(\ref{landau5.2}), since the form of Eq.~(\ref{totalphase}) allows us to arbitrarily define the propagation direction of the $\delta \eta$ field.

\subsection{Scalar-Meson Quasi-Long-Range Order}

We would like to know more about the system as the mass is tuned towards the critical point: $\vert m - m_c \vert \to 0$. Equation~(\ref{landau5.2}) shows that the stiffness of the kinetic terms diverges towards critically, consistent with a second-order phase transition where density fluctuations become negligible. Since Eq.~(\ref{landau5.2}) is non-analytic at criticality, where the energy of phase fluctuations is unbounded, we should focus instead on the correlation function for the $\delta \gamma$ fluctuations; it is these fluctuations that are fundamentally responsible for breakdown of either bilinear condensate at criticality. To elucidate the physics of interest, we retain instead the $\delta \gamma$ fluctuations in Eq.~(\ref{landau4}) by expanding $c_m$ and $s_m$ to second order in $\delta \eta$. We then perform the functional integral over $\delta \eta$ in the partition function, as we have done for $\delta \gamma$ in our previous calculation, but now using a slightly different method with the aim of obtaining the desired long-range correlations.

Consider first the regime $m > m_c$ ($\mu_B < \bar{\mu}_B$). The single-quark mean field order parameter is given by $\langle {\bf n}_\gamma  \rangle  =  \pm  \left(  0 , \,1  \right)$ with the average fermion density $\langle \psi^\dagger   \psi \rangle$$=$$\left(  0 , \, \rho^{1/2}  \right)   \left(  0 , \, \rho^{1/2}  \right)^T$ $=$ $\rho$, where $\rho$ $=$ $\left(  \tilde{\mu}_B - m  \right)/\bar{g}^2$ $=$ $(m + m_c)/\bar{g}^2$. The large mass $m$ prevents any formation of a diquark condensate by canceling the quark-quark attraction, as we shall see in the next section. An overall macroscopic phase $\zeta$ appears when paired quarks and anti-quarks form a BEC. Note however that one cannot define an internal macroscopic phase here and that $\zeta$ is some combination of $\vartheta$ and $\varphi$. Moreover, spatial fluctuations in $\zeta$ occur only via coupling to mediating quantum fluctuations in the chiral quark field. This requires overcoming the large gap energy of the system, either thermally or through quantum mechanical tunneling (quantum fluctuations).

\begin{figure}[]
\centering
\subfigure{
\label{fig:ex3-a}
\hspace{-1pc} \includegraphics[width=.45\textwidth]{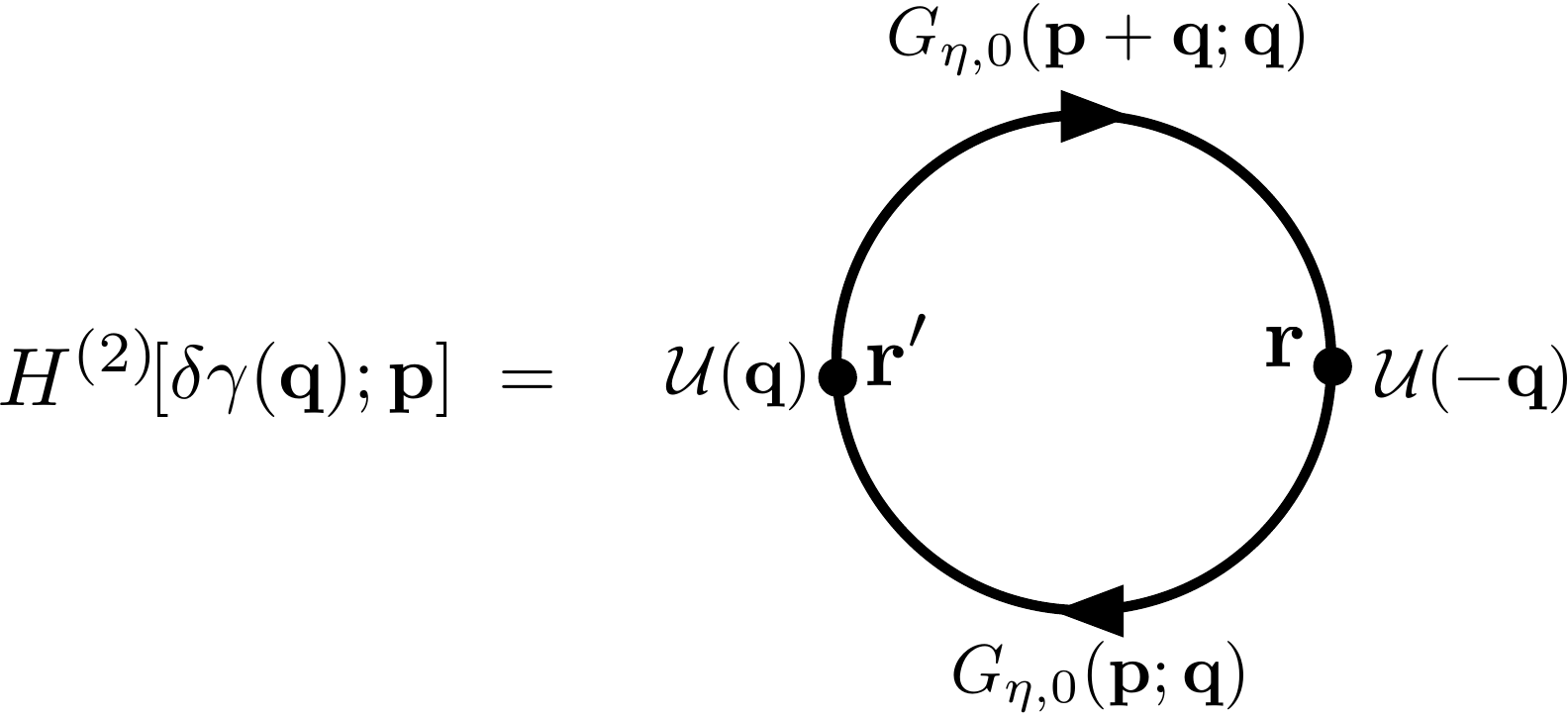}} \\
\caption[]{\emph{Low-energy coupling of the chiral current. } The second-order term in Eq.~(\ref{redlandau}) is depicted diagrammatically. Two factors of first-order gradient of the chiral field $\delta \gamma$ at positions ${\bf r}$ and ${\bf r}^\prime$ are coupled by massive density fluctuations $\delta \eta$. The coupling is attractive and becomes strong and local in the low-energy limit where characteristic momenta are much less than the mass scale associated with $\delta \eta$. In the low-energy limit, linear propagation of $\delta \gamma$ becomes quadratic due to the mass gap induced by density fluctuations, equivalent to a to quadratic Klein-Gordon dispersion. }
\label{BKTCurrents}
\end{figure}

We proceed first by expanding Eq.~(\ref{landau4}) to second order in $\delta \eta$. Focusing on the terms that involve only $\delta \eta$ and $\delta \gamma$, we obtain 
\begin{eqnarray}
      &&\hspace{-2pc} H_\mathrm{eff}  =   \nonumber   \\
         &&\hspace{-2pc} \int \! d{\bf r}\;      \delta \eta  \left[       \delta \gamma    \, \nabla_\parallel     +  4 (m + m_c ) +  \left(     \nabla_\parallel  +  i    \nabla_\perp \right)  \delta \gamma     \right]    \delta \eta .  
\end{eqnarray}
The non-interacting zero-temperature Green's function $G_{  \eta , 0 }$ is obtained by Fourier transforming the fields, e.g, $\delta \eta({\bf r}) = (2 \pi)^{-2} \int d{\bf p}\,  \delta \eta ({\bf p})  \exp \left(  i {\bf p} \cdot {\bf r}  \right)$, which leads to the inverse propagator 
\begin{eqnarray}
G_{  \eta , 0 }^{-1}({\bf p} ; {\bf q})  =  i (1/2)  \,     \, \delta \gamma( {\bf q})  \, \vert {\bf p}_\parallel \vert +  4 (m + m_c)   \, .  \label{EtaPropagator}
\end{eqnarray}
 Two momenta ${\bf p}$ and ${\bf q}$ are needed here because of the presence of the Fourier transformed polarization field $\delta \gamma( {\bf q})$. The effective Hamiltonian becomes
 \begin{widetext}
\begin{eqnarray}
    H_\mathrm{eff}  =    \int  \!   d{\bf p} \, d{\bf q }  \,  \delta \eta ( - {\bf p} - {\bf q} )           \left\{ G_{ \eta , 0 }^{-1}({\bf p} , {\bf q}) \left[ \,  \mathbb{1} \,  + \,  G_{  \eta , 0 }({\bf p} , {\bf q}) \, \mathcal{U} \! \left( {\bf q}  \right)   \right]  \right\}    \delta \eta ( {\bf p}  )    \, .   \label{landau6}  
  \end{eqnarray}
  \end{widetext}
From the viewpoint of $\delta \eta (  {\bf p} )$, the second term in Eq.~(\ref{landau6}) acts as a potential $\mathcal{U} \! \left( {\bf q}  \right)  =   \left( i \, {\bf q}_\parallel -  {\bf q}_\perp \right) \delta \gamma({\bf q}) \equiv {G_{  \gamma , 0 }^{(0)}}^{-1} \! \! \! ({\bf q}) \, \delta \gamma({\bf q})$. The functional integral over $\delta \eta $ can now be performed exactly by the method of Gaussian integration in order to arrive at an effective theory for $\delta \gamma$. Note that the quantity inside the braces in Eq.~(\ref{landau6}) is just the inverse of the full interacting Green's function $G_{\eta}^{-1}({\bf p} ; {\bf q})$. Applying the standard Gaussian prescription for functional integrals (saddle-point approximation/method of steepest descent), re-exponentiating the resulting determinant, then expanding about the extremum of the action using the relation $\mathrm{ln} \, \mathrm{det} \, M = \mathrm{tr}\,  \mathrm{ln} \, M$, we arrive at
\begin{widetext}
\begin{eqnarray}
     H_\mathrm{eff}  \simeq  \int  \!    d{\bf q} \left\{  \left[   G_{  \eta , 0}({\bf q} ; {\bf q}) \,    \mathcal{U} \! \left( {\bf q } \right)     \right]          +   \frac{1}{2}  \int  \!   d{\bf p}    \,   \left[      \mathcal{U}^* \! \left( {\bf q } \right)     G_{  \eta , 0}\!\left( {\bf p} ; -{\bf q}   \right)              G_{  \eta , 0}\!\left( {\bf p} + {\bf q} ; {\bf q} \right) \mathcal{U}    \! \left( {\bf q}  \right)  \right] \right\}  + \dots  \, .  \label{landau7}
  \end{eqnarray}
  \end{widetext}
In Eq.~(\ref{landau7}), we have taken the trace over the momentum variables. Also, we do not include the constant term $ \mathrm{tr} \left\{ \mathrm{ln}\! \left[  G^{-1}_{ \eta , 0} \right] \right\}$ which encodes the non-interacting information. The dots at the end of Eq.~(\ref{landau7}) indicate higher-order terms in the field expansion. The kernel of the second term in Eq.~(\ref{landau7}) is displayed diagrammatically in Fig.~\ref{BKTCurrents}. The low-energy expansion of the propagator in Eq.~(\ref{EtaPropagator}), for which $      {\bf p}_\parallel  \ll  | m + m_c| $, produces the momentum-space representation of the effective Hamiltonian for spin-polarization fluctuations $\delta \gamma$
\begin{widetext}
\begin{eqnarray}
     H_\mathrm{eff}[\delta \gamma]   &\simeq&       \frac{1}{2\left[4 (m + m_c ) \right]^2  }   \int  \!    d{\bf r}\,   d{\bf r}^\prime  d{\bf p} \,  d{\bf q} \,  \exp\left[- i {\bf q} \cdot ( {\bf r} - {\bf r}^\prime) /   \right]  \left[ { {G_{  \gamma , 0 }^{(0)}}^{-1}}\right]^*  \! \! \! (  {\bf q}) \,  \delta\gamma( - {\bf q}) \,    \label{lowenergyexp}   \\
       &&  \hspace{-3pc}  \times \left[   1     +    \frac{ 1 }{\left( 4 \vert m + m_c \vert \right)^2}  \delta \gamma(- {\bf q})   \, \vert   {\bf p}_\parallel \vert \,  \delta \gamma({\bf q})\,    \vert \left( {\bf p} + {\bf q} \right)_\parallel \vert  + \mathcal{O}\left(  ( m + m_c)^{-4} \right)  \right]       {G_{  \gamma , 0 }^{(0)}}^{-1} \! \! \! ({\bf q}) \, \delta \gamma({\bf q})  +  2  |m - m_c|  \int  \!    d{\bf r}  \,   \delta \gamma^2({\bf r})   \nonumber  \, . 
\end{eqnarray}
\end{widetext}
Note that we have retained explicit space and momentum dependence in order to distinguish between the two types of integration. The real-space representation of Eq.~(\ref{lowenergyexp}) amounts to a derivative expansion in the polarization field $\delta \gamma$ for which the lowest-order contribution yields
\begin{eqnarray}
     &&\hspace{-2pc}H_\mathrm{eff}[\delta \gamma]  \simeq   \nonumber   \\
     &&\hspace{-2pc}\frac{ 1 }{{ 2   \left[ 4 (m + m_c)\right]^2} }     \int  \!    d{\bf r}  \,  \vert \nabla  \delta \gamma \vert^2    \,    + \,  2  |m - m_c|  \int  \!    d{\bf r}    \,    \delta \gamma^2 \,   .    \label{redlandau}
  \end{eqnarray}

It is important to note at leading into Eq.~(\ref{redlandau}) we have intentionally omitted fluctuations in $\vartheta$ and $\varphi$ from our discussion, for clarity; following similar steps leads to second-order gradient terms for $\vartheta$ and $\varphi$ fluctuations as well. The inverse non-interacting Green's function for $\delta \gamma\left({\bf p}\right)$ can be read from Eq.~(\ref{redlandau}) as 
 \begin{eqnarray}
 G^{-1}_\gamma \left({\bf p}\right)  =   \frac{  1 }{\left[ 4 (m + m_c)   \right]^{2} }  \vert {\bf p} \vert^2       +     2   |m - m_c|    \, . \label{Ginverseint}
 \end{eqnarray}
 We are now in position to compute the contribution from the spin polarization field $\delta \gamma$ to the off-diagonal long-range order. This is done by starting from the single-particle density matrix $n({\bf r}, { \bf r}^\prime)$ defined in terms of fluctuations of the spinor field $\delta \psi({\bf r})  = \left[ \mathrm{cos}\delta  \gamma({\bf r}) , \, \mathrm{sin} \delta  \gamma({\bf r}) \right]^T$ as
 \begin{eqnarray}
n({\bf r}, { \bf r}^\prime)  &=&  \langle \delta\bar{\psi}({\bf r}) \delta \psi({\bf r}^\prime) \rangle \nonumber  \\
 &=&    \langle  \mathrm{cos}\left[  \delta \gamma({\bf r}) -  \delta \gamma({\bf r}^\prime )  \right] \rangle  \nonumber  \\
    &=&   \mathrm{Re}  \langle e^{ i \left[  \delta \gamma({\bf r}) -  \delta \gamma({\bf r}^\prime ) \right] }  \rangle  \nonumber   \\
     &=&    \mathrm{Re} \left\{   e^{ -  \langle \left[   \delta \gamma({\bf r}) -  \delta \gamma({\bf r}^\prime ) \right]^2 \rangle /2 }  \right\}   \, , 
     \end{eqnarray}
  where 
    \begin{eqnarray}  
      &&\hspace{-2pc} \langle \left[   \delta \gamma({\bf r}) -  \delta \gamma({\bf r}^\prime ) \right]^2 \rangle =  \nonumber \\
        &&\hspace{-1pc} \frac{1}{(2\pi)^2} \int \! d{\bf p}   \,  \frac{ 1 - e^{- i {\bf p} \cdot ( {\bf r} - {\bf r}^\prime )  }}{       \left[  4 ( m + m_c )  \right]^{-2}  \vert {\bf p} \vert^2 +  2  \lambda   | m -m_c|   } =  \nonumber \\
      &&   8  ( m + m_c )^2 \left( \frac{\lambda  }{  \pi       }\right)^2 \nonumber  \\
        && \times  \int_{ 1 / \Lambda}^{  1/\lambda } \! \frac{p \, dp}{ p^2    + \tilde{m}^2  }  \, \int_0^{2 \pi}  \! \! d\theta  \, \left(  1 -  e^{- i  p \, \vert {\bf  r}  - {\bf  r}^\prime \vert  \mathrm{cos}\theta   }   \right)   \, .   \label{correlation}
\end{eqnarray}
 In the last step we have simplified the notation by defining $\tilde{m}^2 \equiv 32  \lambda  ( m  + m_c)^2 | m - m_c|$. We have included upper and lower momentum cutoffs $1/\lambda$ and $1 / \Lambda$, where $\lambda$ and $ \Lambda$ are the respective length scales. The angle part of Eq.~(\ref{correlation}) can be performed by adapting Bessel's integral to our problem. Recalling Bessel's integral of order $n$, $J_n(x) = (1/2 \pi) \int_{- \pi}^\pi \! dz \exp i [ n z +  x\mathrm{cos}( z)]$, end expressing the angle integral in Eq.~(\ref{correlation}) in terms of the $n=0$ Bessel function leads to
 \begin{eqnarray}  
      &&\hspace{-2pc} \langle \left[   \delta \gamma({\bf r}) -  \delta \gamma({\bf r}^\prime ) \right]^2 \rangle =  \nonumber \\
       &&\hspace{-2pc}16  | m + m_c|^2 \lambda^2     \int_{1/ \Lambda}^{  1/ \lambda} \! \frac{p \, dp}{ p^2    + \tilde{m}^2  }   \left[    1 - J_0\! \left(p  \vert {\bf  r} - {\bf r}^\prime  \vert   \right) \right] \, .  \label{correlation2}
\end{eqnarray}
The asymptotic solution to Eq.~(\ref{correlation2}) is obtained by $J_0(x) \simeq   \sqrt{2/\pi x} \,  \mathrm{cos} \left( x - \pi/4 \right)$, for $\vert x \vert \gg 1$. Making the substitution $p^\prime \equiv p  \vert {\bf  r} - {\bf r}^\prime  \vert $ and taking the long-distance limit $p \vert {\bf  r} - {\bf r}^\prime \vert  \gg 1$, Eq.~(\ref{correlation2}) yields 
\begin{eqnarray}
 n({\bf r}, { \bf r}^\prime) \simeq  \left\{ 
  \begin{array}{ c  c   }
           C   \vert {\bf  r} - {\bf r}^\prime \vert^{ - \nu }                      &   \hspace{2pc}      \tilde{m} \ll   1/ \lambda  \, ,  \\
           \exp\left(  - \frac{ \vert {\bf  r} - {\bf r}^\prime  \vert^2}{ \xi^2 } \right)   &      \hspace{2pc}   \tilde{m} \gg   1/ \lambda     \,  .    
          \end{array}             \right.  \label{correlation3} 
\end{eqnarray} 
In the regime where $\tilde{m} \ll   1/ \lambda$ long-distance correlations decay algebraically with critical exponent $\nu = 1/2$. The coefficient is $C = 2 ( \lambda g     )^{1/2} (m + m_c)^{1/2} \vert m - m_c  \vert^{-1/2}$. A significant feature here is the factor of $\vert  m - m_c \vert^{-1/2}$ in $C$, which diverges towards criticality. Algebraic decay of correlations is associated with quasi-long-range order, i.e., the formation of a quasi-condensate, as one would expect for a two-dimensional system. At high temperatures we expect exponential decay of correlations. Exponential decay, however, relies on the presence of topological defects, independent of spontaneous symmetry breaking. We will show that the topological branch is crucial both at nonzero temperatures as well as for small values of the $\delta \gamma$ mass $\vert  m - m_c \vert$. For large values $\tilde{m} \gg   1/ \lambda$, exponential decay dominates for which a correlation length appears $\xi =  4 g^{-1} (m + m_c)  \vert m - m_c \vert^{-1}$, diverging towards criticality. Here, exponential decay of long-range order signals the breakdown of the mean-field theory for the chiral condensate when the mass energy of either the $\delta \eta$ or the $\delta \gamma$ fluctuations exceeds the energy scale set by upper momentum cutoff $1/ \lambda$.

\subsection{Fermionic Excitations of the Superconducting Phase}
\label{ExSC}

Up to this point we have focused on the meson phase with vanishing diquark condensate, for which $m >  m_c$. We now turn our attention to the BCS phase characterized by a non-zero diquark condensate associated with the regime $m <  m_c $. The BCS phase constitutes a disordered phase in that the average out-of-plane spin $\langle S_z  \rangle = \langle {\bf n}_\gamma^T  { \sigma_z} {\bf n}_\gamma \rangle  =0$. Note that this is in contrast to the meson phase for which $\langle S_z  \rangle = \pm 1$. In the BCS phase, the system fluctuates around one of four possible order parameters associated with the polarization vectors $\langle {\bf n}_\gamma \rangle  = \pm  \left(  1, \,  1\right)^T,  \left(  1, \, \pm 1\right)^T$. As one tunes $m \to  m_c$ (equivalently, $\mu_B \to \mu_{B, c}$), the $\delta \gamma$ fluctuations become massless, diverging in the infrared, driving fluctuations in the $z$-component of spin so that $\delta \langle S_z  \rangle \ne 0$. In contrast, $\delta\eta$ fluctuations remain massive throughout this process. 

\subsubsection{Low-Energy Fluctuations}

To better understand dynamics of the BCS phase at zero-temperature near criticality, we choose a particular broken symmetry background
\begin{eqnarray}
\langle { \bf n}_\gamma  \rangle =     \left(    \rho^{1/2}_1 , \,   \rho^{1/2}_2    \right)^T  \, , 
\end{eqnarray}
and consider tuning the mass from $m =0$ towards $m_c$. Thus, we take $m  <   m_c < 0$ throughout the present discussion. We thus have 
\begin{eqnarray}
    \rho_1  &=& \left( m -  | \tilde{\mu}_B|    \right)/\bar{g}^2 =  \vert m -  m_c \vert/\bar{g}^2  \,   , \label{SubDen1}\\
    \rho_2  &=&  \left(  m  +  |\tilde{\mu}_B|     \right)/\bar{g}^2  =  ( m + m_c )/\bar{g}^2 \, .   \label{SubDen}
\end{eqnarray}
From Eqs.~(\ref{SubDen1})-(\ref{SubDen}), we see that the spin order parameters are equal, $\rho_1 = \rho_2 = m_c/\bar{g}^2$, in the zero-mass limit $m \to 0$. It follows that the expectation value of the density polarization field is given by 
\begin{eqnarray}
\langle \gamma \rangle  \equiv  \mathrm{tan}^{-1}\sqrt{\frac{ m+ m_c  }{|m - m_c |}} \, , 
\end{eqnarray}
which leads to the following coefficients from Eq.~(\ref{landau3}) 
\begin{eqnarray}
&&\hspace{-2pc}c_m \simeq    \label{cm} \\
 &&\hspace{-2pc}\left( 1 - 2 \delta \gamma^2 \right)  \frac{m}{m_c}   - 2 \delta \gamma  \,   \frac{1}{ m_c }  (m + m_c )^{1/2}  |m - m_c |^{1/2}       \, , \nonumber \\
&&\hspace{-2pc}s_m  \simeq   \label{sm} \\
    &&\hspace{-2pc}\left( 1 - 2 \delta \gamma^2 \right)       \frac{1}{ m_c}  (m + m_c )^{1/2}  |m - m_c |^{1/2}     +  2 \delta \gamma  \,  \frac{m}{m_c}    \, .  \nonumber 
\end{eqnarray}
The Hamiltonian Eq.~(\ref{landau3}) becomes
\begin{widetext}
\begin{eqnarray}
      H &=& \int \! d{\bf r} \left(  \left[   \left(  1 - 2 \delta \gamma^2 \right)      \frac{1}{ m_c} (m + m_c )^{1/2}  |m - m_c |^{1/2}    +  2 \delta \gamma  \,  \frac{m}{m_c}    \right]   \delta \eta   \nabla_\parallel \delta \eta  \right.   \nonumber \\
   &&    \left. +   \left[  \left( \frac{2 m_c  }{\bar{g}^2} \right)^{1/2} \! \!  \! \! +  \, \delta \eta  \right]^2  \times  \left\{     i  \left[  \left(  1 - 2 \delta \gamma^2  \right)  \frac{1}{m_c}  (m + m_c)^{1/2}  |m - m_c |^{1/2}   +  2 \delta \gamma   \,  \frac{m}{m_c}    \right]  \vert \nabla \vartheta  \vert  \right. \right.  \nonumber \\
    &&   \left. \left. +  \left[  \left(  1 - 2 \delta \gamma^2  \right)  \frac{m}{m_c}   - 2 \delta \gamma   \,   \frac{1}{m_c} (m + m_c )^{1/2}  |m - m_c |^{1/2}    \right]  \nabla_\parallel \delta \gamma   +  i   \nabla_\perp \delta \gamma  \right. \right. \nonumber \\
    &&   \left. \left.  +   \left( \frac{1}{2} \right)   \left(     i           \nabla_\parallel  \varphi   -         \nabla_\perp \varphi  \right)   \times    \left[      \left(  1 - 2 \delta \gamma^2  \right)    \frac{1}{m_c}  (m + m_c )^{1/2}  |m - m_c |^{1/2}     +  2 \delta \gamma   \,  \frac{m}{m_c}  \right]   \right\}    \right) \nonumber \\
      && \;    +   \int \! d{\bf r}  \left(    8 m_c    \left\{   \delta \eta^2  +  \left[    \left(   1 - 2 \delta \gamma^2  \right)        \frac{1}{ m_c }  (m + m_c)^{1/2}  |m - m_c |^{1/2}     +  2 \delta \gamma   \,  \frac{m}{m_c}    \right]  \times   \left( \frac{2 m_c }{\bar{g}^2} \right) \delta \gamma^2  \right\}  \right. \nonumber  \\
       &&  \;  \left. -   \left[   \left(  1 - 2 \delta \gamma^2  \right)      \frac{1}{m_c}  (m + m_c )^{1/2}  |m - m_c |^{1/2}     +  2 \delta \gamma  \,  \frac{m}{m_c}   \right]   4   |m - m_c| \,   \delta \gamma   \right)    \, ,     \label{NematicLandauNext} 
\end{eqnarray}
\end{widetext}
where we have transformed the coefficients of the mass terms using the critical mass through $4 \bar{g}^2 \rho = 8 m_c$ and $\bar{g}^2 \rho - 2m = 2 |m - m_c|$, with $\rho = \rho_1 + \rho_2$ from Eq.~(\ref{SubDen}). Following the same procedure that we used to obtain Eq.~(\ref{landau5.2}) from Eq.~(\ref{landau4}), we retain factors up to first-order in $\delta \gamma$ and zeroth order in $\delta \eta$ for the gradient terms of Eq.~(\ref{NematicLandauNext}), which gives 
\begin{widetext}
\begin{eqnarray}
      H    &=&  \int \! d{\bf r}   \,   \delta \gamma   \left[  \frac{m}{m_c}       \nabla_\parallel \delta \eta^2  +  4 i \,   \frac{m}{\bar{g}^2}     \vert \nabla \vartheta   \vert  +   2  \frac{m}{\bar{g}^2}      \left(    i           \nabla_\parallel  \varphi   -         \nabla_\perp \varphi  \right)    \right]     \label{NematicLandauNext2} \\
         &+&  \int \! d{\bf r}  \left\{      8  m_c   \,    \delta \eta^2 +  16 \frac{m_c }{\bar{g}^2}    |m - m_c|^{1/2}  \left[     (m + m_c )^{1/2}   -     \frac{\bar{g}^2 m}{ 2 m_c^2 }       |m - m_c|^{1/2}  \, \right]   \delta \gamma^2   \right\}   \, . \nonumber     
\end{eqnarray}
\end{widetext}
Integrating over the $\delta \gamma$ fluctuations as we did to obtain Eq.~(\ref{landau5.2}), Eq.~(\ref{NematicLandauNext2}) becomes 
\begin{eqnarray}
    \! \! \! \! H_\mathrm{eff}    =    C   \int \! d{\bf r}             \left(    \vert    \nabla \delta \eta^2  \vert^2    +   \rho^2     \vert \nabla \zeta  \vert^2     \right)    +   8   m_c   \int \! d{\bf r}  \,        \delta \eta^2          ,   \,   \label{NematicLandauNext3}
\end{eqnarray}
where the coefficient $C$ is given by
\begin{widetext}
\begin{eqnarray}
C =    m^2   \,      \left\{ 16 m_c   |m - m_c|^{1/2}  \left[   2 m_c^2   (m + m_c )^{1/2}   -    \bar{g}^2  m  |m - m_c|^{1/2}  \, \right]    \right\}^{-1} \, .
\end{eqnarray}
\end{widetext}
 The condensate phase here is given by
\begin{eqnarray}
\vert \nabla \zeta  \vert^2  =  \vert \nabla_\perp  \varphi  \vert^2    +   \vert \nabla_\parallel \left[   \vartheta  +   \varphi  \right]  \vert^2 \, .  \label{effectivesf}
\end{eqnarray}
The second-order nature of fluctuations here reflects the presence of a finite mass gap which breaks the linear relativistic dispersion for low-momentum modes into positive and negative bosonic Klein-Gordon modes.

 \subsubsection{Quasi-Long-Range BCS Order}
 
 Next, we follow the same method as in Sec.~\ref{MadForm} to compute the break down of long-range order due to the $\delta \gamma$ fluctuations. Isolating terms linear in $\delta \gamma$ and those quadratic in $\delta \eta$ in Eq.~(\ref{NematicLandauNext}), gives 
\begin{eqnarray}
     &&\hspace{-.75pc}H   =   \hspace{-1pc} \label{NematicLandauNext3}   \\
          &&\hspace{-.75pc} \!  \int \! d{\bf r}   \,      \delta \eta    \left[   2  \delta \gamma   \frac{m}{m_c}   \nabla_\parallel    + 8   m_c    +     \left(      \frac{m}{m_c}   \nabla_\parallel  +  i   \nabla_\perp  \right)   \delta \gamma  \right]    \delta \eta  \nonumber \\
            &&\hspace{-.75pc}+   8  \!  \int \! d{\bf r}     \left[   (m + m_c )^{1/2}  |m - m_c |^{1/2}        \rho   -   \frac{m}{m_c}  |m - m_c|  \right]  \delta \gamma^2        ,    \nonumber 
\end{eqnarray}
which, by virtue of the steps leading to Eq.~(\ref{Ginverseint}), gives the inverse propagator  
  \begin{eqnarray}
    &&\hspace{-1pc}G^{-1}_\gamma \left({\bf p}\right)  = \frac{ 1}{16 m_c^2}  \,  \left( \frac{m^2}{m_c^2}  \vert {\bf p}_\parallel   \vert^2   + \vert {\bf p}_\perp  \vert^2  \right)  \nonumber \\
      &&\hspace{-1pc}+   \,   8    \left[   (m + m_c )^{1/2}  |m - m_c |^{1/2}     \,    \rho   -   \frac{m}{m_c}  |m - m_c| \,  \right] .  \nonumber 
 \end{eqnarray}
Computing the correlations as we did in Sec.~\ref{MadForm} gives 
\begin{eqnarray}
 n({\bf r}, { \bf r}^\prime) \simeq  C     \vert {\bf  r} - {\bf r}^\prime \vert^{ - \nu  }      \, ,                 &   \hspace{2pc}    \tilde{m}  \ll   1/\lambda    \, ,                \label{correlation5} 
\end{eqnarray} 
with 
\begin{eqnarray}
&&\hspace{-1pc} \tilde{m}  \, \equiv  \,  128 \,  \lambda \,  m_c^2 \\
 &&\hspace{-1pc} \times   \left[    (m + m_c )^{1/2}  |m - m_c |^{1/2}     \,    \rho   -    \left( m / m_c \right)  |m - m_c| \,  \right] . \nonumber 
\end{eqnarray}
 Long-distance correlations decay algebraically as in our previous results, but here with critical exponent $\nu  = 1/2$ and coefficient 
\begin{eqnarray}
   &&\hspace{-1pc}C =  ( \lambda  g^2 )^{1/2} (m_c/m)^{1/2}   \\
   &&\hspace{-1pc}\times \left[   (m + m_c )^{1/2}  |m - m_c |^{1/2}       \rho   -    \left(m / m_c \right)  |m - m_c|  \right]^{-1/2}. \nonumber  
\end{eqnarray}
The single-particle density matrix computed for the meson phase, Eq.~(\ref{correlation3}), and that of the BCS phase, Eq.~(\ref{correlation5}), both exhibit discontinuities in the critical exponent at the critical point $m = m_c$. An important point to note here is that correlations for chiral fluctuations diverge where the symmetry group of the ground state changes, signaling a quantum phase transition.

\section{Quantum Berezinskii-Kosterlitz-Thouless Phase Transition}
\label{QBKTPT}

The physical mechanism underlying conventional BKT phase transitions involves the unbinding of paired bosonic vortices. The system is said to undergo a transition from a vortex insulating phase into a conduction phase characterized by the proliferation of free vortices. The identifying signature is exponential rather than algebraic decay of long-range order. Topologically, vortices are finite energy solutions of classical equations in two spatial dimensions that wrap around the $U(1)$ circle in the symmetry group for thermodynamic potentials in systems with infinitely degenerate ground states homeomorphic to $S^1$. Vortices of opposite rotation are paired at low temperatures since a relatively large and unavailable amount of thermal energy is required to produce a single isolated vortex. Notably, the energy increases logarithmically with the distance of separation between bound vortices, which is why such systems can be mapped to the two-dimensional Coulomb gas. Topological unwinding that characterizes the BKT transition occurs at finite temperature, where the energy peak at the center of the effective potential begins to flatten out. This allows for tunneling between $U(1)$ minima in real space. Indeed, isolated vortices are in fact spatial tunneling events between different ground-state minima analogous to kinks in one-dimensional systems.

\subsection{Large $N_c$ Topological Baryonic Solitons}

As we have seen, the single-quark ground state manifold (Fig.~\ref{ContSym}) is composed of right and left chiral circles associated with the baryon mean-field phase $\bar{\phi}_{0, R(L)}$, on the left hand side of Eq.~(\ref{HartreeSoln}) with zero baryon momentum $\mathrm{\bf k} =0$. This is built up of contributions from all $N_c$ quarks, displayed on the right hand side of Eq.~(\ref{HartreeSoln}). Up to this point, we have addressed quantum corrections to the large $N_c$ limit by expanding the baryon mean field effective potential around these minimal circles and integrating out the massive modes. This approach stresses the view of mesons and diquarks as bound states of excitations in the internal collective currents of a baryon or antibaryon.

Such current excitations can be either elementary or themselves topological collective modes that wrap around the phase directions $\bar{\phi}_{0, R(L)}$. These are baryonic solitons that can be viewed as a single quark excitation plus a vortex in the collective background of the other $N_c-1$ quarks that form a boson. There are two ways to map either the right or left internal phase circle to the spatial circle at infinity. One way tunnels through the central peak. The other circumnavigates the central peak by tunneling from $R \to - L \to -R$, for the right chiral mode, say, as shown in the diagram Fig. \ref{Minima}(b). The former describe thermal vortices, i.e., ones that require a finite amount of thermal energy to excite, whereas the latter are quantum activated vortices that may be excited at zero temperature when the quark mass equals the effective baryon chemical potential $m = \tilde{\mu}_B$.

Expanding on this, consider that a vortex which circumnavigates the central peak exhibits chiral switching from right (left) to left (right) while maintaining a constant overall particle density. The key point to note here is that such tunneling involves an intermediate meson layer that happens right where the magnitudes for the gradients of the right and left chiral wavefunctions are greatest. This is the essential point of the discussion in Sec.~\ref{Dissociation}. There we saw that chiral fluctuations $\delta \gamma$ induced quantum tunneling between right and left handed states passing through a virtual meson cloud. Thus, as quantum fluctuations, such vortices use virtual ``sea'' quarks to tunnel between right and left chiral states when passing from the outside into the interior core region. This spatial pattern of fluctuating chirality mediated by mesons remains at the quantum, or virtual, level as long as the quark mass does not exceed baryon chemical potential. Another way to say this is that such fluctuating patterns occur within bound vortices, only becoming real at the critical point where the mean vortex separation diverges. 

\begin{figure}[h]
\begin{center}
 \subfigure{
\label{fig:ex3-a}
\hspace{-.1in} \includegraphics[width=.48\textwidth]{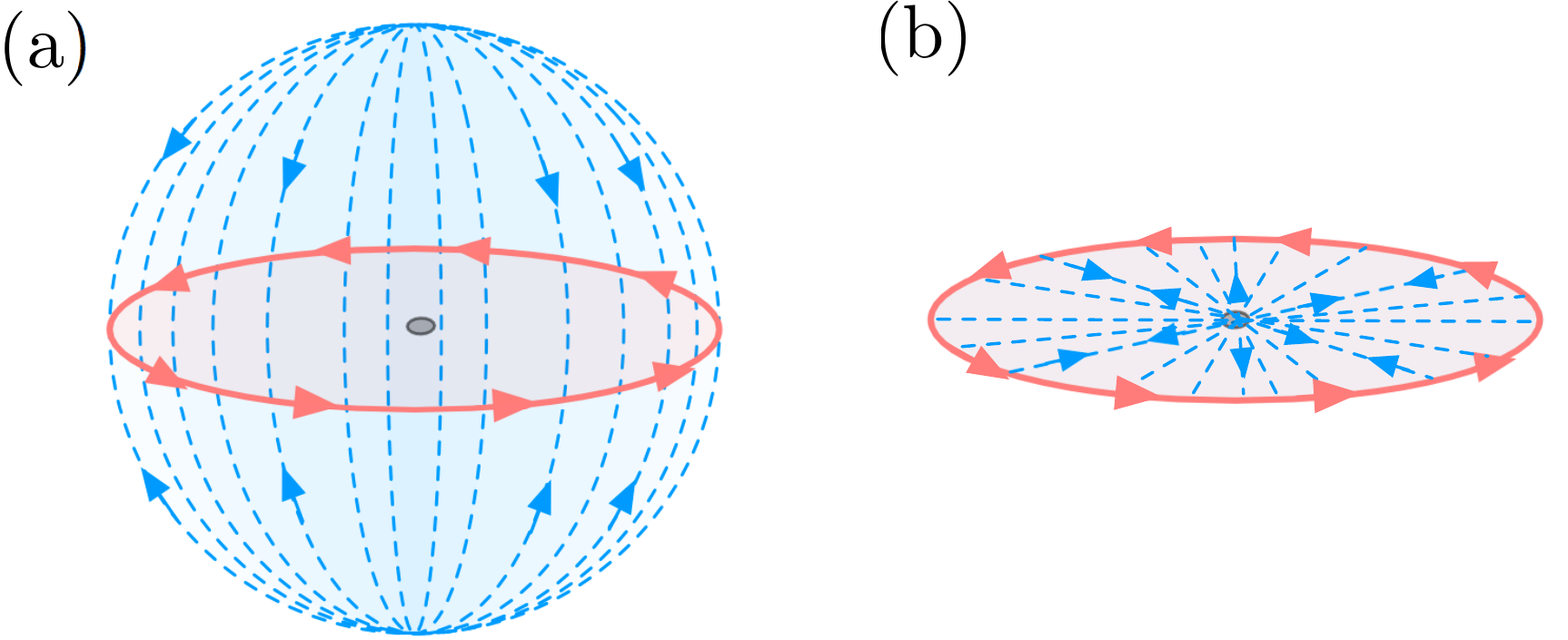}  } 
\vspace{0in}
\end{center}
\caption[]{(color online) \emph{Core structure of the baryonic soliton. } The sphere of left-chiral circles (a) is non-coherently mapped to the planar interior of the vortex (b). }
\label{BaryonVortex}
\end{figure}

From the substance of the discussion in Sec.~\ref{Dissociation}, we conclude that these quantum activated vortices can only exist at zero temperature when the $\delta \gamma$ fluctuations diverge, i.e., at the critical point $m = m_c$. Otherwise, vortices of opposite rotation must be bound with some finite distance of separation, with the $\delta \gamma$ fluctuations providing the effective ``thermal'' energy needed to unbind them. For small mass, $m \ll m_c$, there is a significant and clear distinction between the core and outer regions of a baryonic vortex. Specifically, the outer region has macroscopic rotation with well defined right (left) chirality. The inner core region has no rotation but corresponds to a classical radial s-wave state, so must be a Fourier superposition of left (right) chiral radial momentum states. This picture comes from mapping the two hemispheres in Fig.~\ref{BaryonVortex} onto the plane that contains the right (left) chiral circle, which forms the core region of a right (left) handed vortex. The key distinction between the outer rotating and core regions is that the outer vortex in real space is a direct one-to-one mapping of the right (left) chiral circle onto the circle at infinity, whereas the peak inner core is a quantum superposition of motion along the left (right) chiral circle superimposed onto every angular orientation inside the core. Hence, a vortex of opposite rotation is enfolded into the core region. 

\begin{figure}[h]
\begin{center}
 \subfigure{
\label{fig:ex3-a}
\hspace{-.3in} \includegraphics[width=.55\textwidth]{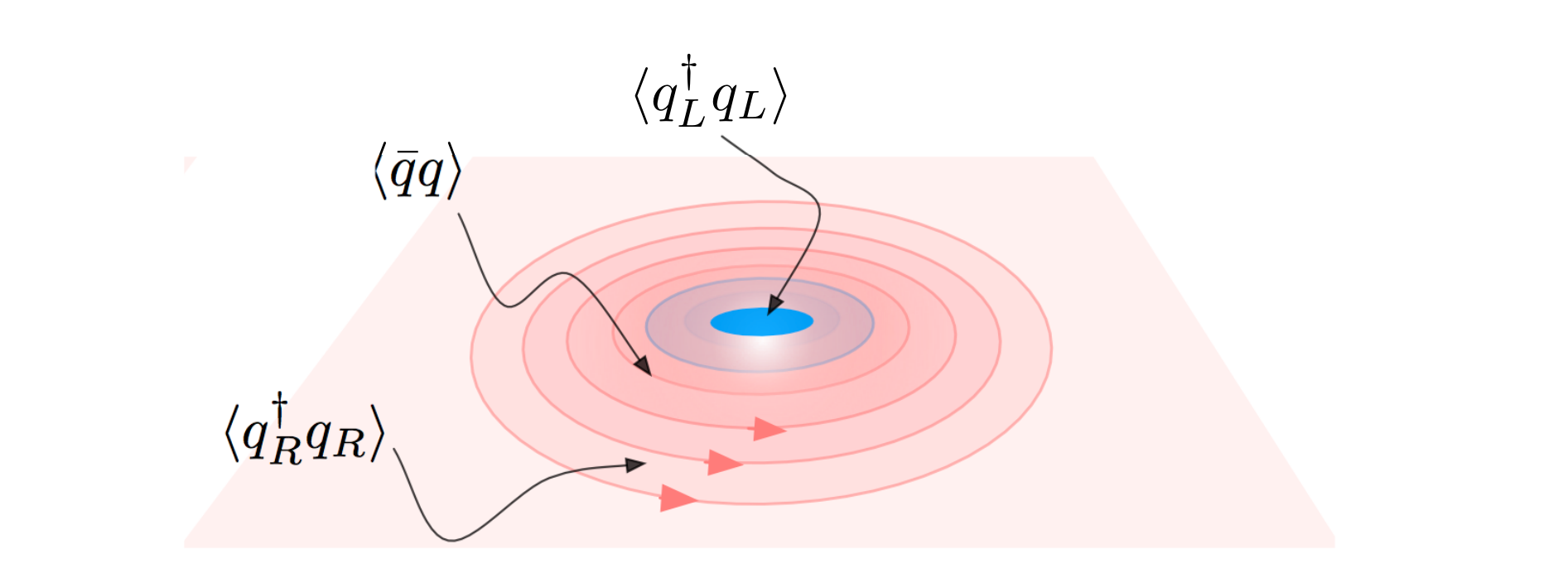}  } 
\vspace{0in}
\end{center}
\caption[]{(color online) \emph{A baryonic soliton. } The structure of a single baryonic soliton is shown with outer vortex region (red arrows), inner s-wave soliton core (solid blue), and intermediate transition region (no rotation). The outer and inner regions are composed of well-defined, but orthogonal, chiral diquarks. The intermediate transition region is a virtual meson cloud, alternatively described as a region of strong chiral quantum fluctuations of a real diquark field.   }
\label{BaryonVortex_2}
\end{figure}

Note that the vortex core in real space does not correspond to a macroscopic (classical) potential minimum, but instead highly spatially entangled radial motion (s-wave) with, however, well-defined chirality, opposite of that of the outer vortex. The endpoint of our analysis reveals a single baryonic vortex with three distinct regions: outer right(left)-chiral diquark, intermediate mesonic, and inner left(right)-chiral diquark regions, depicted in Fig.~\ref{BaryonVortex_2}. 

\subsection{Logarithmic Binding and Quantum Dissociation of Baryonic Solitons}

In this section we discuss vortex energetics and mechanics at the quantum critical point $m = m_c$, with the complete technical derivation of the zero-temperature BKT transition left for Sec.~\ref{QBKT}.  At ultra-low temperatures around a chosen ground state, massive fluctuations can be treated as non-dynamical fields with each propagator contributing an overall factor of the inverse mass. Near a quantum phase transition, however, the chiral mode $\delta \gamma$ becomes massless, with $m_\gamma \sim \sqrt{(1 -m/m_c)}$ near the critical point. There, the corresponding propagator contributes a logarithmically divergent energy, essentially coming from the Coulombic part of the gluon field. This structure is precisely that of a 2D Coulomb gas with an additional infinite repulsion at the defect cores. Consequently, implementing the standard renormalization group analysis reveals the BKT signature exponential decay of quasi-long-range order for melting (when $T \ne 0$) or dissolving (for $T=0$) of the diquark condensate. The crucial point to emphasize here is that, in order to work, the precise mathematics that maps baryonic vortices to Coulomb gas degrees of freedom requires a relativistic Dirac structure, as the latter is endowed with the chiral structure absent in ordinary BCS theory.

\subsubsection{Energetics of Vortex Binding and Unbinding}

As we have seen, the source of the breakdown of long-range order at the quantum critical point resides in the quantum fluctuations of the fermionic chiral angle $\delta \gamma$, whose vanishing mass signals the enhanced symmetry. It is these fluctuations that drive the transition by coupling diquarks to mesons as explained in Sec.~\ref{Dissociation}. The extreme fluctuations in $\delta \gamma$ reflect the system's instability at the critical point. In essence, the $\delta \gamma$ field acts as a convection for the diquark dissociation energy supplied by the energetic advantage offered by ``converting'' a diquark into a meson coming from the increased scalar-to-diquark condensate ratio, or equivalently that between the running quark mass and the baryon chemical potential. 

To investigate vortex unbinding we require an effective energy for bound vortices, which we can obtain by performing the Gaussian integration over the massive modes $\delta \eta$ and $\delta \gamma$. The resulting expression may then be tuned towards criticality allowing us to track the behavior of the vortex binding terms. The integration is done in two steps, integrating first over the quadratic contribution from the $\delta \eta$ modes in Eq.~(\ref{NematicLandauNext}), then collecting the $\delta \gamma$ quadratic terms in the resulting expression and integrating over these. At each step the functional integral is performed by Fourier transforming to momentum space, then applying the standard Gaussian prescription (saddle-point approximation), re-exponentiating the resulting determinant, expanding about the extremum of the action, and finally re-expressing the result in real space. This procedure yields an effective theory for the spin-wave (smooth) component of the quark current and a quark-spin (defect) contribution from the vortical term. The defect contribution to the effective energy near criticality ($m \lesssim m_c$) yields
\begin{eqnarray}
  \hspace{-1pc}E    = \! \! \int \! \! d{\bf r}\,  d{\bf r}^\prime   \psi^*_d({\bf r}^\prime) \!   \left[  \rho_\eta   \delta({\bf r} - {\bf r}^\prime )   +  \rho_\gamma    \mathrm{ln}\left(  {\bf r} - {\bf r}^\prime    \right)                  \right]  \! \psi_d({\bf r})   ,\label{DefectEnergy} 
  \end{eqnarray}
  with
  \begin{eqnarray}
  \hspace{-2pc} \hspace{1pc} \rho_\eta \equiv   \frac{  1}{16 g^4}m_c (m + m_c ) (m_c -m)\,   , \;\;\; \rho_\gamma \equiv     \frac{m}{4 \pi m_c} \rho_\eta \, , \label{coefficients}
\end{eqnarray}
 where the defect field is given by $\psi_d = \hat{\bf \mathrm{z}} \cdot ( \nabla \times {\bf n}_\varphi)$. The logarithmic contribution in the second term of Eq.~(\ref{DefectEnergy}) is key, and comes from the long distance behavior in the momentum integration of the $\delta \gamma$ Green's function. Defect fields of the same chirality (opposite spin and momentum) are now coupled through a repulsive contact term with coefficient $\rho_\eta$, which is just an ordinary mass gap from integrating out the amplitude field, and an attractive logarithmic term with coefficient $\rho_\gamma$ coming from the chiral field. Note that both coefficients in Eq.~(\ref{coefficients}) vanish at $m=m_c$, but at different rates.

Contrasting this with the result far from criticality ($m \ll m_c$), we find that the second term in Eq.~(\ref{DefectEnergy}) is an attractive delta function in the separation $\delta(|{\bf r} - {\bf r}^\prime|)$, which reduces the strength of the hard-core repulsion coming from the first term. This happens since the mass of $\delta \gamma$ fluctuations in this regime is large compared to the characteristic low-temperature momentum. Thus, in addition to the unbinding that occurs for $m \ge m_c$, a cross-over occurs between two regimes: logarithmic binding around $m \lesssim m_c$ to unbound defects for $m \ll m_c$. 

Equation~(\ref{DefectEnergy}) can be mapped to the 2D Coulomb gas as follows. Note first that for a closed path encircling delta function defect sources we have 
\begin{eqnarray}
&&\oint     {\bf n}_\varphi \cdot d {\bf s}  =   \int \!  \left( d{\bf r}  \, \hat{ \bf z} \right)  \cdot \nabla \times    {\bf n}_\varphi \\
 \Rightarrow    &&\nabla \times    {\bf n}_\varphi({\bf r})   =   2 \pi \, \hat{\bf z} \sum_i n_i^v \, \delta^2\! \left({\bf r} - {\bf r}_i  \right)\, , 
\end{eqnarray}
where $d {\bf s}$ is a tangent differential vector along the path. ``Magnetic'' charges for delta function defects at positions ${\bf r}_i$, ${\bf r}_j$, are denoted as $n_i^v$. Thus, in terms of an analog magnetic field notation $\nabla \times    {\bf n}_\varphi \equiv   \nabla \times {\bf \mathcal{B}}_d = \hat{ \bf z } \, \nabla^2 \mathcal{A}_d$, we then obtain $\nabla^2  \!  \mathcal{A}_d({\bf r})   =    \pi  \sum_i n_i^v \, \delta^2\! \left({\bf r} - {\bf r}_i  \right)$, which has the solution ${\bf \mathcal{A}}_d({\bf r})   =    \pi  \, \hat{\bf z}  \sum_i n_i^v \, \mathrm{ln} \left( {\bf r} - {\bf r}_i    \right)$. Incorporating these results into Eq.~(\ref{DefectEnergy}) and performing the integration over ${\bf r}$ and ${\bf r}^\prime$ gives 
\begin{eqnarray}
E=\pi^2  \sum_{i , j}   n_i^v n_j^v  \left[  \rho_\eta   \,  \delta^2 \! \left(   {\bf r}_i - {\bf r}_j    \right)      +      \rho_\gamma     \mathrm{ln} \! \left(  {\bf r}_i  - {\bf r}_j   \right) \right] \, , \label{CGE}
\end{eqnarray}
which leads to the diquark condensate correlation length obtained through standard renormalization techniques $\xi \sim  \exp \! \left[  c/ \sqrt{|m - m_c(T)|} \right]$, where $c$ is a constant and we have included finite temperature effects. The correlation length diverges exponentially near the critical point $m_c(T)$, consistent with the BKT theory: at $T=0$, a QBKT transition occurs at $m = m_c(0)$; for $T \ne 0$, $m_c(T)$ gives the critical point for a standard BKT transition, modified by quantum mechanical corrections.

 The zero-temperature transition displayed in Eq.~(\ref{CGE}) is driven by the quantity $\tilde{\mu}_B  - m$, which multiplies the logarithmic term and relates to the mass of chiral fluctuations $\delta \gamma$. The remarkable similarity of the second term in Eq.~(\ref{CGE}) to the free energy in conventional BKT for a system of size $R$ comprised of vortices with radius $a$ given by
 \begin{eqnarray}
 F = E - TS = (\kappa - 2 k_B T) \, \mathrm{ln}(R/a) \label{stndrdfree}
 \end{eqnarray}
 is evident by noting that the chemical potential in our system measures the energy of an isolated defect, analogous to $\kappa$ in Eq.~(\ref{stndrdfree}), with the scalar mass analogous to the entropy factor $2 k_BT$. Hence, the ``thermal'' energy required to overcome vortex binding in the QBKT transition is evidently supplied by quantum fluctuations as $m$ approaches criticality.

\subsubsection{Vortex Structure at Quantum Critical Point}

Baryonic vortices undergo dramatic structural changes near the quantum critical point. This is fundamentally linked to promotion of the discrete chiral symmetry to a continuous one at the critical point which introduces a flat direction in the central peak of the effective potential along the diagonal in Fig.~\ref{Minima}(b). Bound states that lie along the diagonal are baryonic, hence flattening along this direction allows for proliferation of free baryonic vortices and dissociation of diquarks. This is the general net effect but we would like to understand the dissociation process in finer detail.

\begin{figure}[]
\begin{center}
 \subfigure{
\label{fig:ex3-a}
\hspace{-.2in} \includegraphics[width=.525\textwidth]{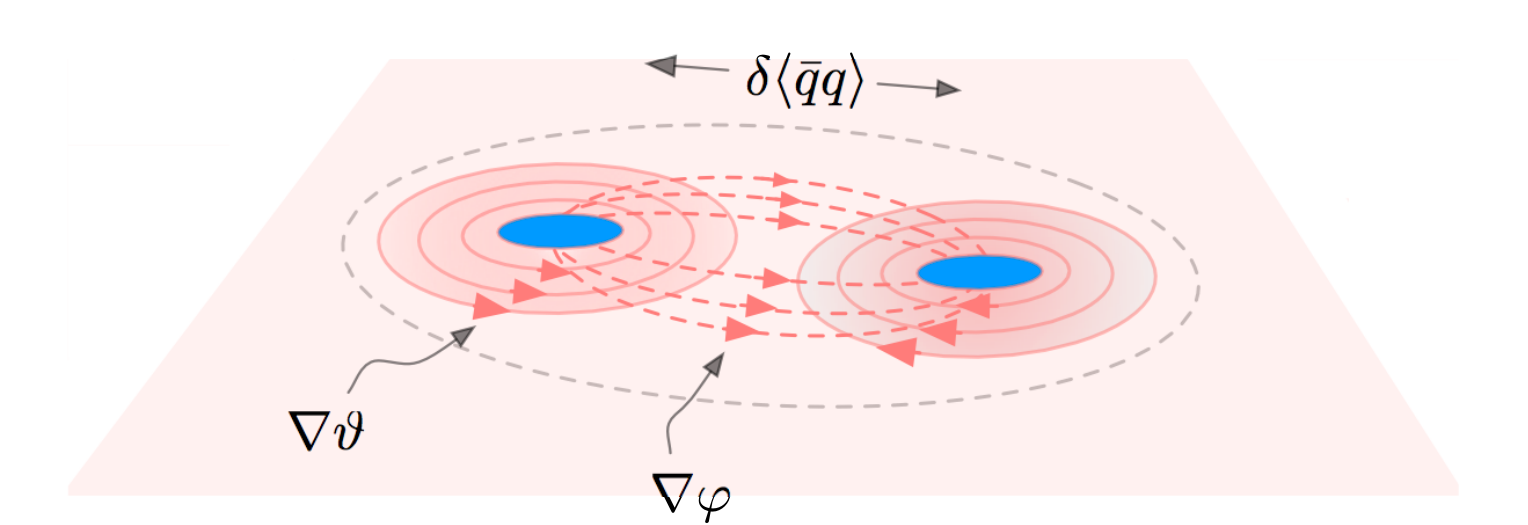}  } 
\vspace{0in}
\end{center}
\caption[]{(color online) \emph{Vortex dissociation mechanism near the quantum BKT transition. } The region between the outer rotation and inner soliton of each vortex overlaps and expands, filling in the intermediate space as the core separation tends to infinity. Intermediate meson fluctuations and spin and orbital current patterns are shown. Orthogonality of spin and orbital currents in the intermediate region simply reflects the extreme fluctuations in quark chirality near criticality.   }
\label{BaryonVortex_3}
\end{figure}

Let us examine the approach into the critical point from below, $m \to m_c^-$. When $m \ll m_c$, vortex rotation is mainly along the diagonal shown in Fig.~\ref{Minima}(b). Thus, each paired vortex has well defined chirality: spin and momentum are locked into either right or left chiral configurations. As the quark mass increases, the chiral angle of vortex rotation and core peak rotate away from the diagonal in spin space. In the case of the right chiral vortex with left handed core peak, this translates to an increase in the vortex angle and decrease in the core peak angle in spin space. How do we interpret this? This twist in $\gamma$, i.e., as the chiral angle for the outer vortex and inner core approach one another $\gamma_{-L} \to (\pi/2)^+$ and $\gamma_{R} \to (\pi/2)^-$, several changes in the structure of the vortex occur:

\begin{enumerate}

\item The rotating outer vortex and the s-wave soliton in the core become semi-classical chiral mixed states, i.e., states with a well specified superposition of chiral modes. 

\item The potential peak separating right and left chiral states becomes narrower and lower, thus increasing the amplitude of $\delta \gamma$ quantum fluctuations. 

\item The chiral character of the vortex and core regions give way to an expanding meson region that lies in the transition region the vortex and core.  

\item The expanding meson region is gradually converted from virtual meson/anti-meson pairs into real meson states.

\item Chiral mixing provides two orthogonal channels for vortex dissociation defined by the direction of baryonic orbital current: hedgehog or vortical decay, for radial directed orbital or spin currents, respectively.

\end{enumerate}

All of these features together result in a picture of vortex dissociation whereby the space between two bound vortices is gradually filled in by real scalar mesons, effectively pushing oppositely rotating vortices in opposite directions towards infinity. The details are depicted in Fig.~\ref{BaryonVortex_3}. The growing region begins as virtual fluctuations in the meson field $\delta \langle \bar{q} q \rangle = \langle q_R^\dagger  q_L \rangle \delta \gamma +  \langle q_L^\dagger  q_R \rangle \delta \gamma$, eventually becoming real scalar-meson density once the quantum fluctuations $\delta \gamma$ reach the dissociation threshold at the critical point $m = m_c$. From the perspective of spin-charge separation, initiated through the ansatz Eq.~(\ref{Madelung}) and implemented in the discussion that followed, a fluid dynamic picture of the strong chiral fluctuations in this intermediate region emerges. Fermionic spin current $\nabla \varphi$ flows radially outward leaving bosonic orbital current $\nabla \vartheta$ to form the inner vortex. This is the bosonic \emph{vortical} decay mechanism responsible for the zero-temperature BKT transition. The outward flow of spin current describes strong spin entanglement between the two vortices. A second, alternative decay channel occurs when the flow patterns are interchanged: radial flow of $\nabla \vartheta$ and vortical flow of $\nabla \varphi$. Here the spin part is a Pauli vortex and the radial orbital flow describes a bosonic \emph{hedgehog} decay mechanism.

\subsection{Derivation of the Quantum BKT Transition}
\label{QBKT}

In Sec.~\ref{MadForm}, we examined the quasi-long-range order associated with the spin-wave fluctuations $\delta \gamma$. In this section, we study the contribution to chiral symmetry breaking of the ground state coming from unbinding of paired baryonic topological defects, the mechanism that underlies the quantum BKT transition in our system. Introductions to many of the techniques in this section can be found in Refs.~\cite{Chaikin1995,Kardar2007}. An important step in our analysis is integration of the partition function over both the $\delta \eta$ and $\delta \gamma$ fluctuations in order to isolate the baryon phase degrees of freedom, which split into spin-wave and topological contributions. We find that fluctuations in the overall density $\delta \eta$ provide a hard contact repulsion, whereas fluctuations in the $\delta \gamma$ field counter the effective logarithmic diquark attraction between baryonic vortices. At zero temperature, the ``heat'' required to dissociate paired vortices comes from long-wavelength quantum fluctuations in $\delta \gamma$ field, when the corresponding mass vanishes at the critical point $m_c$.

\begin{figure}[]
\centering
\subfigure{
\hspace{-.25pc}\includegraphics[width=.48\textwidth]{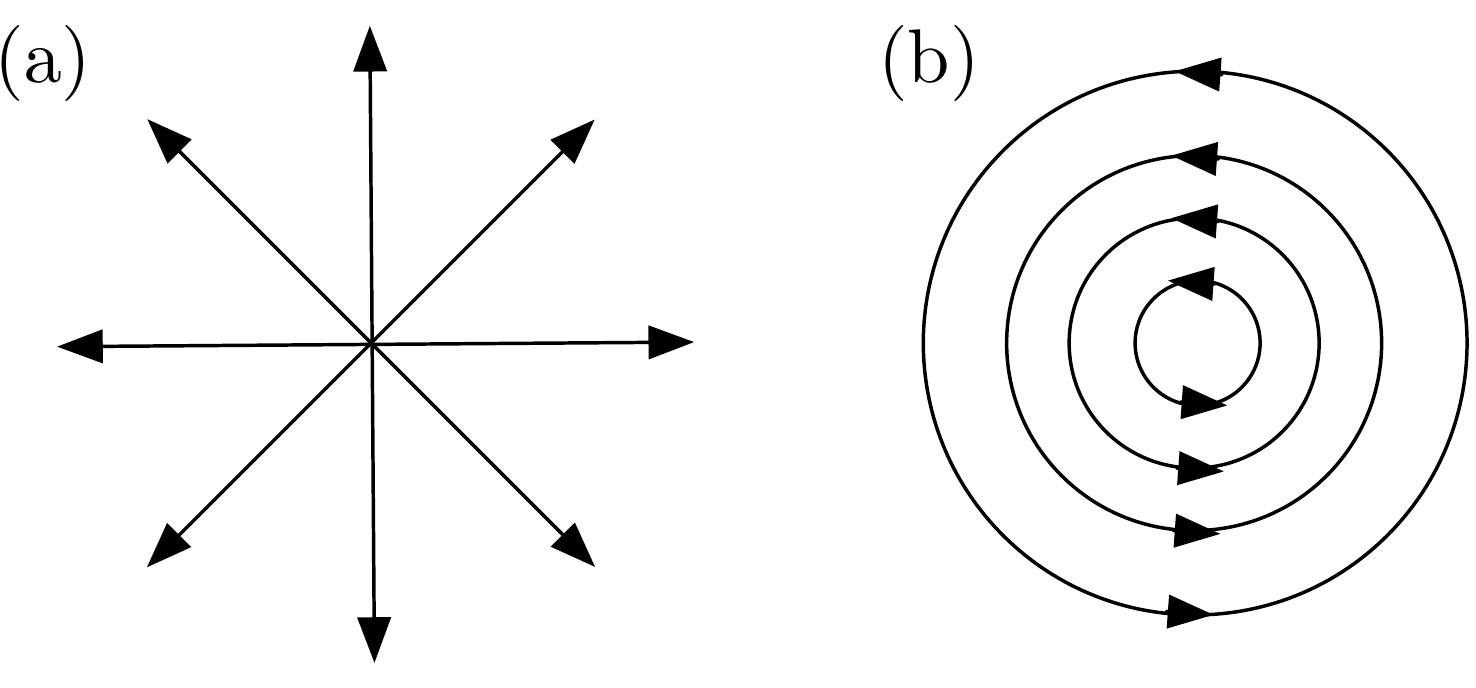}} \\
\caption[]{\emph{Basis for spin-charge separated spinon and chargon currents}. (a) Hedgehog flow patterns describe sources for either orbital or spin currents through a divergence term for ${\bf n}_\varphi$ in Eq.~(\ref{NematicLandauVortex}). (b) Vortical flow patterns describe rotational orbital or spin currents through the curl of the ${\bf n}_\varphi$ field in Eq.~(\ref{NematicLandauVortex}). For a ground state defined by a chiral symmetry group, orbital and spin currents remain aligned (spin-orbit locking), and the field ${\bf n}_\varphi$ will describe the flow direction for quark current. At a chiral phase transition, spin-orbit alignment is broken with spin and orbital currents described by one of the two basis modes (a) and (b). Baryons experience extreme fluctuations between left and right chirality at a critical point through two possible decay channels: 1) bosonic vortical current (chargon) with radial spin flow (spinon), depicted in Fig.~\ref{BaryonVortex_3}; or 2) fermionic vortical current with bosonic radial flow.  
}
\label{VortexHedge}
\end{figure}

We begin our analysis from Eq.~(\ref{NematicLandauNext}) and the associated expressions for $c_m$ and $s_m$ in Eqs.~(\ref{cm})-(\ref{sm}). Retaining zeroth and first-order $\delta \gamma$ terms in $c_m$ and $s_m$ followed by functional integration over second-order $\delta \eta$ fluctuations, leads to 
\begin{eqnarray}
      && \hspace{-1pc}H_\mathrm{eff}   \, =   \nonumber  \\
              && \hspace{-1pc}   \frac{   1  }{8 m_c}    \!  \int \! d{\bf r}                    \left\{   \left[  \left(  c_0- 2 c_1 \delta \gamma  \right)     \nabla_\parallel   \delta \gamma   - \frac{1}{2} \left(  s_0 + 2 s_1  \delta \gamma  \right)      \nabla \cdot {\bf n}_\varphi \right]^2   \right. \nonumber  \\
       && \hspace{-1pc}+ \left.                                \left[  \left(   s_0 + 2 s_1  \delta \gamma  \right)  \vert \nabla \vartheta  \vert   + \frac{1}{2} \left(   s_0 + 2 s_1  \delta \gamma \right)    \vert   \nabla \times  {\bf n}_\varphi  \vert \right]^2  \right\} \nonumber \\
       && \hspace{-1pc}+  \,  \, 4   \int \! d{\bf r}   \,  | m - m_c|  \,   \delta \gamma^2          \, ,   \label{NematicLandauVortex}  
\end{eqnarray}
where the zeroth and first-order coefficients are $c_0 = s_1 =  m/m_c$ and $c_1 = - s_0 =   - (m + m_c )^{1/2}  |m - m_c |^{1/2}/m_c$. We have inserted the divergence and curl of the direction vector field ${\bf n}_\varphi$. These forms arise from ${\bf n}_\varphi \cdot \nabla \varphi_\sigma= |\nabla_\parallel \varphi_\sigma | \to  |\nabla \times {\bf n}_\varphi|$ and $ \vert {\bf n}_\varphi  \times  \nabla \varphi \vert  = |\nabla_\perp \varphi | \to  |\nabla \cdot {\bf n}_\varphi|$. To relate the $\varphi$ fluctuations to the divergence and curl of the direction field ${\bf n}_\varphi$ one expands the two orthogonal directional derivatives of $\varphi$ as follows 
\begin{eqnarray}
\hspace{-2pc} \nabla_\perp \varphi   &=&    \left( \mathcal{R}_{\pi/2}   {\bf n}_\varphi   \right)  \cdot \nabla \varphi  =  - \mathrm{sin} \varphi  \frac{\partial}{\partial x}  \varphi   +    \mathrm{cos} \varphi    \frac{\partial}{\partial y} \varphi    \\
   &=&  \nabla \cdot \left(  \mathrm{cos} \varphi  , \, \mathrm{sin} \varphi  \right)  =  \nabla \cdot {\bf n}_\varphi   \, , \label{divergence}
\end{eqnarray}
and 
\begin{eqnarray}
\hspace{-2pc} \nabla_\parallel \varphi     &=&    {\bf n}_\varphi   \cdot  \nabla \varphi =  \mathrm{cos} \varphi  \frac{\partial}{\partial x}  \varphi   +    \mathrm{sin} \varphi    \frac{\partial}{\partial y} \varphi    \\
   &=&   \vert  \nabla \times \left(  \mathrm{cos} \varphi , \, \mathrm{sin} \varphi  \right)   \vert  =  \vert  \nabla \times  {\bf n}_\varphi  \vert \, ,  \label{curl}
\end{eqnarray}
where we have used the 2D rotation matrix $\mathcal{R}_{\pi/2}$ to rotate the direction field ${\bf n}_\varphi$ by $90$ degrees. Invoking the substitution $|\nabla_\perp   \varphi|    \;  \to \;  |\nabla \cdot {\bf n}_\varphi|$,  $|\nabla_\parallel    \varphi|   \;  \to \; \vert  \nabla \times {\bf n}_\varphi   \vert$ provides a more physically intuitive perspective, as the field ${\bf n}_\varphi$ describes the direction of orbital current flow, away from a chiral phase transition, and either orbital or spin current at a chiral critical point, as shown in Fig.~\ref{VortexHedge}. The paradigm depicted in Fig.~\ref{VortexHedge} is precisely that of spin-charge separation. Within the topological sector, chiral phase transitions occur through two possible ``slave-fermion'' channels involving spinon and chargon fields, ${\bf n}_\varphi({\bf r}, t )$ and $\vartheta({\bf r}, t )$: 1) decay of bound bosonic chargon vortex pairs mediated by fermionic spinon exchange; and 2) decay of bound fermionic spinon vortices (Pauli vortices) mediated by bosonic chargon exchange. In both cases, the beginning states are the same with the end products in channels 1) and 2) distinguished by the character of the free vortices and exchange modes in the final state. The quantum BKT transition occurs through channel 1).

Separating zeroth, first, and second-order terms in $\delta \gamma$, and applying integration by parts with total derivatives integrating to zero at large distances, Eq.~(\ref{NematicLandauVortex}) reduces to  
\begin{eqnarray}
        &&\hspace{-1pc}H_\mathrm{eff}   =     \label{NematicLandauVortex2}    \\
          &&\hspace{-1pc} \frac{ 1 }{2^4  m_c}    \!  \int \! d{\bf r}   \,          \delta \gamma \left(    c_0       \nabla_\parallel^2     +       \nabla_\perp^2  +   m_c  | m - m_c|  \right)  \delta \gamma   \nonumber \\
       &&\hspace{-1pc}+ \,    \frac{ 1}{2^4 m_c }    \!  \int \! d{\bf r}   \,  \left[ 4 s_1 s_0  \left\vert \nabla \vartheta  \right\vert + s_1 s_0 \left(  \nabla \cdot {\bf n}_\varphi   +  \left\vert \nabla \times  {\bf n}_\varphi  \right\vert    \right) \right]                    \delta \gamma      \nonumber \\
        &&\hspace{-1pc}+ \,    \frac{ 1}{2^4  m_c } s_0^2  \int \! d{\bf r} \, \left[  \left\vert \nabla \vartheta  \right\vert^2  -  \frac{1}{4} \left(   \left\vert  \nabla \cdot {\bf n}_\varphi   \right\vert^2  +  \left\vert   \nabla \times  {\bf n}_\varphi   \right\vert^2 \right)  \right]   \, .   \nonumber 
\end{eqnarray}
The first line on the right of Eq.~(\ref{NematicLandauVortex2}) contains second-order terms in $\delta \gamma$, with the non-interacting inverse Green's function appearing inside the brackets. The second line consists of interactions for the smooth part of the overall phase and the defect fields, mediated to first order by $\delta \gamma$ fluctuations. The third line contains interactions for the smooth phase and defects implicitly mediated by the overall amplitude field $\delta \eta$. Notice that if $\delta \gamma = \mathrm{constant}$ (trivial solution) only the third line in Eq.~(\ref{NematicLandauVortex2}) is nonzero leaving a spin wave and bound defects. Hence, individual defects can only be excited in the presence of a spatially varying background $\delta \gamma$. Integrating out the $\delta \gamma$ fluctuations in Eq.~(\ref{NematicLandauVortex2}) as we have previously done in Sec.~\ref{MadForm} gives 
\begin{widetext}
\begin{eqnarray}
        H_\mathrm{eff}  =   \frac{ 1}{16 \,m_c }  s_0^2       \int \! d{\bf r}\,  d{\bf r}^\prime d{\bf p}_\gamma \left\{     \nabla \vartheta({\bf r}^\prime )    \cdot  \left[   \delta({\bf r} - {\bf r}^\prime ) \delta( {\bf p}_\gamma ) -     \frac{m_c}{ 4     }         s_1^2    \,     G_{0 , \gamma}({\bf p}_\gamma)     e^{ - i  {\bf p}_\gamma \cdot ( {\bf r} - {\bf r}^\prime )   }                   \right]    \nabla \vartheta({\bf r})             \right. \nonumber \\
            \left. +   \, \frac{ 1}{ 4}  \sum_{i \in \{ h, v\}}   \psi^*_i({\bf r}^\prime)  \left[   \delta({\bf r} - {\bf r}^\prime ) \delta( {\bf p}_\gamma ) -     \frac{m_c }{ 4     }         s_1^2    \,     G_{0 , \gamma}({\bf p}_\gamma)     e^{ - i  {\bf p}_\gamma \cdot ( {\bf r} - {\bf r}^\prime )   }                   \right]  \psi_i({\bf r})          \right\} \, ,    \label{NematicLandauVortex3} 
\end{eqnarray}
\end{widetext}
where the subscripts $h$ and $v$ denote contributions from hedgehog and vortex defects with respective fields $\psi_{h}({\bf r})$$=\nabla \cdot {\bf n}_\varphi({\bf r})$, $\psi_{v}({\bf r})$$=$$\nabla \times {\bf n}_\varphi({\bf r})$, where $\nabla_h \equiv \nabla_\parallel$ and $\nabla_v \equiv \nabla_\perp$. Complex conjugation of the fields $\psi_i$ comes from including parity reversed defects. We do not include defect-pairing terms of the same rotation and radial flux since such contributions have divergent overall energy. The first line on the right side of Eq.~(\ref{NematicLandauVortex3}) accounts for spatial twists in the smooth part of the condensate phase. The second term sums over the two types of bound defects. For each defect type, the delta function terms result from integrating over $\delta \eta$, whereas the second term comes from $\delta \gamma$ contributions. Individual defects are correlated over a separation length $d = |{\bf r} - {\bf r}^\prime|$, where we integrate over all values of $d$ weighed by the Green's function $G_{0 , \gamma}({\bf p}_\gamma)$, to obtain the free energy. Here, the Green's function $G_{0 , \gamma}({\bf p}_\gamma)$ is explicitly given by
\begin{eqnarray}
G_{0 , \gamma}({\bf p}_\gamma) \equiv  \frac{1}{   c_0      |{\bf  p}_{\gamma , \parallel}|^2     + |{\bf p}_{\gamma, \perp}|^2  +   m_c  | m - m_c| }   \, . \label{GammaGreenFunction}
\end{eqnarray}

\subsubsection{BEC-BCS Transition: Dissolving the Diquark Condensate Through Vortex Dissociation}
\label{TopologicalSuperfluid}

At zero temperature, there are two important regimes to consider in Eq.~(\ref{NematicLandauVortex3}). Resolving these regimes reveals a crossover into phase coherence coexisting with a vortex-pair superfluid. First, far from criticality the characteristic momentum of $\delta \gamma$ fluctuations is small compared to the $\delta \gamma$ mass, i.e., $|{\bf p}_\gamma|^2   \ll  m_c  | m - m_c|$ in Eq.~(\ref{GammaGreenFunction}). In this case the zeroth-order approximation of the Green's function with respect to $|{\bf p}_\gamma|^2$ is valid, which leads to the approximate form of the Green's function 
\begin{eqnarray}
G_{0 , \gamma}({\bf p}_\gamma)  \simeq  \frac{1}{     m_c  | m - m_c| } \, . \label{limit1}
\end{eqnarray}
Consequently, the nonlocal correlations in Eq.~(\ref{NematicLandauVortex3}) reduce to local quantities upon integration over ${\bf p}_\gamma$ and ${\bf r}^\prime$, leading to the reduced effective Hamiltonian
\begin{eqnarray}
  \hspace{0pc} H_\mathrm{eff}     =        \int \! d{\bf r}     \,   \left(   \rho_\mathrm{sw}    \left\vert \nabla \vartheta \right\vert^2  +   \rho_d  \! \!  \sum_{i \in \{ h, v\}}        \left\vert  \psi_i  \right\vert^2     \right)   \, ,    \label{NematicLandauVortex4}  
\end{eqnarray}
 where the mass-dependent spin-wave stiffness $\rho_\mathrm{sw}$ and defect-binding coefficient $\rho_d$ are defined by
\begin{eqnarray}
\hspace{-1pc}\rho_\mathrm{sw} &\equiv&  \frac{ |m -m_c| (m+ m_c)}{16\,  m_c^3}    \left(  1   -     \frac{m^2 }{ 4 m_c | m- m_c| }          \right)   \, ,  \label{spinstiff} \\
\hspace{-1pc}\rho_d  &\equiv&   \frac{ |m -m_c| (m+ m_c)}{64\, m_c^3}  \left(  1   -     \frac{m^2 }{ 4 m_c | m- m_c| }          \right)   \label{hedgemass}   \,. 
\end{eqnarray}
In Eqs.~(\ref{spinstiff})-(\ref{hedgemass}), the positive term in parentheses comes from integrating over $\delta \eta$, and the negative term is the renormalized correction from the $\delta \gamma$ fluctuations. In particular, Eq.~(\ref{hedgemass}) decomposes as 
\begin{eqnarray}
\rho_d = \rho_d^{(\eta)} - \rho_d^{(\gamma)} \, , \label{rhodecomp}
\end{eqnarray}
where $\rho_d^{(\eta)}$ and $\rho_d^{(\gamma)}$ are the contributions to defect binding from fluctuations in the overall amplitude $\delta \eta$ and the relative amplitude $\delta \gamma$, respectively. Note, that in Eq.~(\ref{rhodecomp}) fluctuations in $\delta \gamma$ counter $\delta \eta$ effects due to the negative sign in the second term. To more accurately display the physics near the critical point, i.e., $m \lesssim  m_c$, we must take a different limit in Eqs.~(\ref{NematicLandauVortex3})-(\ref{GammaGreenFunction}), namely $m_c  | m - m_c| \ll |{\bf p}_\gamma|^2$. In this case, $c_0 = 1$ and the Green's function reduces to
\begin{eqnarray}
G_{0 , \gamma}({\bf p}_\gamma) \simeq \frac{1}{  |{\bf  p}_{\gamma}|^2  }   \, . \label{limit2}
\end{eqnarray}
Integrating over ${\bf p}_\gamma$ in Eq.~(\ref{NematicLandauVortex3}) gives 
\begin{widetext}
\begin{eqnarray}
   H_\mathrm{eff}  =  \frac{ (m + m_c ) |m -m_c|  }{16 \, m_c^3}   && \int \! d{\bf r}\,  d{\bf r}^\prime  \left\{   \nabla \vartheta({\bf r}') \cdot  \left[    \delta({\bf r} - {\bf r}^\prime )      +     \frac{4 m^2 }{ \pi  m_c     }              \mathrm{ln} \left(  \vert  {\bf r} - {\bf r}^\prime \vert \right)               \right]  \nabla \vartheta({\bf r})   \right. \nonumber  \\
   &&   \left. + \;\;   \frac{1}{4}  \sum_{i \in \{ h, v\}}   \psi^*_i({\bf r}^\prime)  \left[    \delta({\bf r} - {\bf r}^\prime )      +     \frac{4  m^2 }{\pi m_c   }           \mathrm{ln}\left( \vert  {\bf r} - {\bf r}^\prime \vert   \right)                  \right]  \psi_i({\bf r})         \,  \right\} \, .   \label{RenormalizedStiff}
\end{eqnarray}
\end{widetext}
Equation~(\ref{RenormalizedStiff}) implies the nonlocal renormalized $\rho_\mathrm{sw}$ and $\rho_d$
\begin{eqnarray}
   &&\hspace{-3pc}\rho_\mathrm{sw}( {\bf r})  \equiv \nonumber  \\
   &&\hspace{-3pc}\frac{|m -m_c| (m+ m_c) }{16 \, m_c^3}   \left[  1   +      \frac{4  m^2 }{\pi m_c  }          \;      \int \!  d{\bf r}^\prime \,  \mathrm{ln}\left( \vert  {\bf r} - {\bf r}^\prime \vert   \right)          \right]    ,  \label{renspinstiff} \\
   &&\hspace{-3pc}\rho_d( {\bf r}) \equiv  \nonumber \\
    &&\hspace{-3pc}\frac{  |m -m_c| (m+ m_c) }{64 \, m_c^3}     \left[  1   +      \frac{4 m^2 }{\pi m_c    }          \;    \int \!  d{\bf r}^\prime \,   \mathrm{ln}\left( \vert  {\bf r} - {\bf r}^\prime \vert   \right)          \right]  .  \label{renhedgemass}   
\end{eqnarray}

\subsubsection{Mapping to the 2D Coulomb Gas}
\label{Coulomb}

To get a more accurate detailed picture of the pairing and BKT transitions at $m = m_\mathrm{pair}$ and $m = m_\mathrm{BKT}$, we must elaborate on the defect functions $\psi_v({\bf r})$ and $\psi_h({\bf r})$ by expressing them as collections of point-like defects, i.e., as gases of defects. The two types of defects here are the hedgehog and the vortex. The former corresponds to local flow into or out of a particular chiral state, hence only present under non-equilibrium conditions brought on by a mass quench into the quantum critical point. In the vortex case, particularly where orbital angular momentum is zero, only the geometric phase contributes to the curl of the $SU(2)$ phase. In this case, the vortical defect is actually a Pauli vortex, which encodes the half-integer spin of the quark. In our analysis, we will find the 2D Coulomb gas picture to be an indispensable tool, given its ubiquity in statistical systems. The utility of this picture cannot be understated, as for example applications based on the relationship of the Coulomb gas to the Sine-Gordon and XY models~\cite{Jose1976,Amit1976}. For the purposes of mapping to the Coulomb gas, we can treat vortices and hedgehogs distinctly. This leads to a straightforward mapping to two copies of the Coulomb gas in two spatial dimensions. Having successfully mapped the energy to that of a Coulomb gas, we then proceed towards obtaining the corresponding BKT scaling law through standard renormalization group methods. Significantly, we will find that the quark mass $m$ appears in place of temperature in standard finite-temperature BKT theory.

To map our system to the 2D Coulomb gas we return to the definitions of the hedgehog and vortex fields, i.e., $\psi_{h}({\bf r}) =\nabla \cdot {\bf n}_\varphi({\bf r})$ and $\psi_{v}({\bf r}) = \hat{\bf z} \cdot \left[  \nabla \times {\bf n}_\varphi({\bf r})\right]$. We then note that for a closed path encircling delta function defect sources we have 
\begin{eqnarray}
     \oint     {\bf n}_\varphi \cdot d {\bf s}   &=&   \int \!  \left( d{\bf r}  \, \hat{ \bf z} \right)  \cdot \nabla \times    {\bf n}_\varphi    \\
   \Rightarrow  \;\;\;    \nabla \times    {\bf n}_\varphi({\bf r})   &=&   2 \pi \, \hat{\bf z} \sum_i n_i^v \, \delta^2\! \left({\bf r} - {\bf r}_i  \right) \, ,  \label{magsource}
\end{eqnarray} 
and 
\begin{eqnarray}
   \oint     {\bf n}_\varphi \cdot d {\bf n}  &=&    \int \! d{\bf r} \,   \nabla \cdot     {\bf n}_\varphi    \\
   \Rightarrow  \;\;\;    \nabla \cdot     {\bf n}_\varphi({\bf r})   &=&     \sum_j  n_j^h \, \delta^2\! \left({\bf r} - {\bf r}_j   \right) \, , \label{elecsource}
\end{eqnarray} 
where $d {\bf s}$ and $d {\bf n}$ are tangent and normal differential vectors along the path. Magnetic and electric charges for delta function vortex and hedgehog sources at positions ${\bf r}_i$, ${\bf r}_j$, are denoted as $n_i^v$ and $n_j^h$, respectively. Thus, in terms of electric and magnetic field notation $\nabla \times    {\bf n}_\varphi \equiv   \nabla \times {\bf B} = \hat{ \bf z } \, \nabla^2 A$, and $\nabla \cdot     {\bf n}_\varphi \equiv   \nabla \cdot  {\bf E} = \nabla^2 \phi $, Eq.~(\ref{magsource}) and Eq.~(\ref{elecsource}) become
\begin{eqnarray}
     \nabla^2  \!  A({\bf r})   &=&   \pi  \sum_i n_i^v \, \delta^2\! \left({\bf r} - {\bf r}_i  \right) \, ,  \\
     \mathrm{and} \;\;\; \;\;   \nabla^2  \phi({\bf r})  &=&   \sum_j  n_j^h \, \delta^2\! \left({\bf r} - {\bf r}_j   \right) \, , 
\end{eqnarray} 
which have solutions
\begin{eqnarray}
      {\bf A}({\bf r})   &=&    \pi  \, \hat{\bf z}  \sum_i n_i^v \, \mathrm{ln} \left( \vert {\bf r} - {\bf r}_i \vert   \right) \, , \\
       \;  \mathrm{and} \;\;\; \;\;    \phi({\bf r})  &=&  \sum_j  n_j^h \, \mathrm{ln} \left( \vert {\bf r} - {\bf r}_j \vert   \right) \, , \label{logvecscal}
\end{eqnarray} 
respectively. Near criticality, we require the forms Eq.~(\ref{renspinstiff})-(\ref{renhedgemass}), so that 
\begin{eqnarray}
 \rho_d({\bf r}) &=&  \rho_d^{(\eta)}  +  \rho_d^{(\gamma )}    \!  \int \!  d{\bf r}^\prime \,   \mathrm{ln}\left( \vert  {\bf r} - {\bf r}^\prime \vert   \right)  \, , \\
  \rho_{\mathrm{sw}}({\bf r}) &=& 4  \rho_d({\bf r})     \, , 
\end{eqnarray}
where
\begin{eqnarray}
\rho_d^{(\eta)}  &=&    \frac{1}{64 m_c^3} |m -m_c| (m+ m_c) \, , \\
 \rho_d^{(\gamma )}   &=&    \frac{ m^2}{16 \pi  m_c^4} |m -m_c| (m+ m_c) \, . 
\end{eqnarray}
We first compute the spin-wave term (first line) in Eq.~(\ref{RenormalizedStiff})
\begin{eqnarray}
 H_\mathrm{eff}[ \vartheta] &=&   4 \rho_d^{(\eta)}   \int \!  d{\bf r} \,  \vert \nabla \vartheta({\bf r}) \vert^2     \label{spinwave2} \\
 &+&  4   \rho_d^{(\gamma )}    \int \! d{\bf r}  \, d{\bf r}^\prime \,    \vert \nabla \vartheta({\bf r}') \vert \,  \mathrm{ln}\! \left( \vert  {\bf r} - {\bf r}^\prime \vert   \right)    \vert \nabla \vartheta({\bf r}) \vert  \, . \nonumber
\end{eqnarray} 
Using integration by parts twice converts the second term to $\int \!  d{\bf r}  \,d{\bf r}^\prime \,     \vartheta({\bf r}')  \delta^2\left( \vert  {\bf r} - {\bf r}^\prime \vert   \right)  \vartheta({\bf r})$, after dropping total derivatives, which further simplifies to $\int \!  d{\bf r}  \,    [  \vartheta({\bf r}) ]^2 $, producing a mass gap for the spin wave. Using Eq.~(\ref{logvecscal}), the defect contribution (second line) to Eq.~(\ref{RenormalizedStiff}), i.e., the vortex contribution, is 
\begin{widetext}
\begin{eqnarray} 
H_\mathrm{eff}[ \psi_v] &=& \rho_d^{(\eta)} \pi^2   \sum_{i , j}  n_i^v n_j^v  \int \!  d{\bf r} \, d{\bf r}'  \delta^2 \! \left( \vert  {\bf r} - {\bf r}^\prime \vert   \right)  \delta^2 \! \left( \vert  {\bf r} - {\bf r}_i  \vert   \right) \delta^2 \! \left( \vert  {\bf r}' - {\bf r}_j \vert   \right) \nonumber \\                 &+& \rho_d^{(\gamma )}  \pi^2   \sum_{i , j}   n_i^v n_j^v    \int \!  d{\bf r} \,  d{\bf r}^\prime \,    \delta^2 \! \left( \vert  {\bf r}' - {\bf r}_i  \vert   \right) \mathrm{ln}\!\left( \vert  {\bf r} - {\bf r}^\prime \vert   \right)   \delta^2 \! \left( \vert  {\bf r} - {\bf r}_j  \vert   \right) \\
&=&   \pi^2   \sum_{i , j}  n_i^v n_j^v  \left[  \rho_d^{(\eta)}   \,  \delta^2 \! \left( \vert  {\bf r}_i - {\bf r}_j  \vert   \right)      +      \rho_d^{(\gamma )}    \mathrm{ln} \! \left( \vert  {\bf r}_i  - {\bf r}_j  \vert   \right)                     \right]  \, . \label{finaldefect}
\end{eqnarray} 
\end{widetext}
The first term in Eq.~(\ref{finaldefect}) is the repulsive core energy between bound baryonic vortices. The second term is a logarithmic binding energy between vortices. A similar expression holds for hedgehog defects. Including all the terms, the effective Hamiltonian reads 
\begin{eqnarray}
     &&\hspace{-2.5pc}H_\mathrm{eff}[ \vartheta, \psi_v, \psi_h ]   = 4    \int \! d{\bf r}        \left[  \rho_d^{(\eta)}  \left\vert \nabla \vartheta({\bf r}) \right\vert^2   +   \rho_d^{(\gamma )}     \vartheta({\bf r})^2  \right]       \\
      &&\hspace{-2.5pc} +  \,   \pi^2   \sum_{i , j}  n_i^v n_j^v  \left[  \rho_d^{(\eta)}   \,  \delta^2 \! \left( \vert  {\bf r}_i - {\bf r}_j  \vert   \right)      +      \rho_d^{(\gamma )}    \mathrm{ln} \! \left( \vert  {\bf r}_i  - {\bf r}_j  \vert   \right)                     \right]  \nonumber  \\
      &&\hspace{-2.5pc} + \,    \pi^2  \sum_{i , j}   n_i^h n_j^h  \left[  \rho_d^{(\eta)}   \,  \delta^2 \! \left( \vert  {\bf r}_i - {\bf r}_j  \vert   \right)      +      \rho_d^{(\gamma )}    \mathrm{ln} \! \left( \vert  {\bf r}_i  - {\bf r}_j  \vert   \right)                     \right] \nonumber     \, .   \label{finalenergy}
\end{eqnarray}

Thus, the key logarithmic form of our analog vector and scalar potentials in Eq.~(\ref{logvecscal}) realizes two copies of the 2D Coulomb gas contained in the summed vortex and hedgehog terms in the free energy Eq.~(\ref{finalenergy}). Notice that an infinite contact binding energy exists in Eq.~(\ref{finalenergy}) exhibited by the diverging logarithmic functions for $\vert{\bf r}_i - {\bf r}_j\vert \to 0$. This condition simply reflects the fact that we treat the defect cores as point-like objects where one must take into account the limits of the continuum model as in standard treatments of the Coulomb gas~\cite{Kardar2007}. To regularize the contact binding energy one simply includes short distance correction terms $\mathcal{E}^0_{n_i }$ to account for the core energies in the Coulomb terms. The low-temperature partition function splits into distinct contributions where one finds
\begin{widetext}
\begin{eqnarray}
     &&Z   =  \int \mathcal{D}[ \vartheta({\bf r}) ]   e^{-   4    \int \! d{\bf r}        \left[  \rho_d^{(\eta)}  \left\vert \nabla \vartheta({\bf r}) \right\vert^2   +   \rho_d^{(\gamma )}     \vartheta({\bf r})^2  \right]    }   \;  \sum_{\{ n_i \}} \int \! d^2{\bf r}_i    \,  e^{-  2 \sum_i \beta  \mathcal{E}^0_{n_i }    \,  + \,   2 \pi^2       \sum_{i <  j}   n_i  n_j \;   \left[  \rho_d^{(\eta)}   \,  \delta^2  \left( \vert  {\bf r}_i - {\bf r}_j  \vert   \right)      +      \rho_d^{(\gamma )}    \mathrm{ln}  \left( \vert  {\bf r}_i  - {\bf r}_j  \vert   \right)                     \right]  }      \nonumber   \\
       && \equiv Z_\mathrm{sw} \,  Z_\mathrm{d} \, , 
\end{eqnarray}
\end{widetext}
where $Z_\mathrm{sw}$ and $Z_\mathrm{d}$ are the spin wave and topological defect contributions to the partition function. Note the factor of 2 which accounts for both defect types. If we consider  only the contribution from the lowest elementary charges $n_i = \pm 1$, the fugacity is $\mathrm{y}  \equiv   \mathrm{y}^0_{\pm 1}  = \exp[ \, - \beta  \mathcal{E}^0_{\pm 1}\,  ]$, we obtain
\begin{widetext}
\begin{eqnarray}
 Z_\mathrm{d}   =      \sum_{N =0}^\infty  \left( \mathrm{y}^0 \right)^N   \int   \prod_{i =1}^N d^2{\bf r}_i    \,  \exp \! \left\{    2 \pi^2       \sum_{i <  j}   n_i  n_j \;   \left[  \rho_d^{(\eta)}   \,  \delta^2  \left( \vert  {\bf r}_i - {\bf r}_j  \vert   \right)      +      \rho_d^{(\gamma )}    \mathrm{ln}  \left( \vert  {\bf r}_i  - {\bf r}_j  \vert   \right)                     \right]    \right\} \, , 
\end{eqnarray}
\end{widetext}
where $n_i = \pm 1$.

Let us return to the regime away from criticality described by Eqs.~(\ref{spinstiff})-(\ref{hedgemass}). In this regime, we have 
\begin{eqnarray}
 \rho_d    &=&  \rho_d^{(\eta)}  -  \rho_d^{(\gamma )}   \, , \\
  \rho_{\mathrm{sw}} &=& 4  \rho_d     \, , 
\end{eqnarray}
where
\begin{eqnarray}
\rho_d^{(\eta)}  &=&    \frac{1}{64 m_c^3} |m -m_c| (m+ m_c)  \, ,       \label{rhoetafinal}\\
 \rho_d^{(\gamma )}   &=&    \frac{ m^2}{256  m_c^4} (m+ m_c) \, , \label{rhogammafinal}
\end{eqnarray}
both constants. In this regime, the logarithms in Eq.~(\ref{spinwave2}) and  Eq.~(\ref{finaldefect}) are delta functions leading to a simpler form for the energy functional in Eq.~(\ref{finalenergy}):
\begin{eqnarray}
     &&E[ \psi_v, \psi_h ]   =    \label{finalenergy2}\\
         &&\pi^2   \rho_d \left[  \sum_{i , j}  n_i^v n_j^v  \,   \delta^2 \! \left( \vert  {\bf r}_i - {\bf r}_j  \vert   \right)  +    \sum_{i , j}   n_i^h n_j^h  \,  \delta^2 \! \left( \vert  {\bf r}_i - {\bf r}_j  \vert   \right)     \right]    \, , \nonumber  
\end{eqnarray}
where we have focused on the defect contribution. Equation~(\ref{finalenergy2}) exhibits a distinct phase transition when $\rho_d$ undergoes a sign change. For positive $\rho_d$, the delta functions act as repulsive cores in an otherwise noninteracting gas of particles. In this regime, Eq.~(\ref{finalenergy2}) describes an ideal Fermi gas, and Eqs.~(\ref{rhoetafinal})-(\ref{rhogammafinal}) satisfy $\rho_d^{(\eta)}  > \rho_d^{(\gamma)}$, where  
\begin{eqnarray}
  m^2  < 4  \, m_c  |m -m_c|    \, ,
\end{eqnarray}
or 
\begin{eqnarray}
  m   <  2 (\sqrt{2} -1)  \, m_c  \approx 0.828 \, m_c  \, . 
\end{eqnarray}
Here, the system is described by an asymptotically free quark gas with short range repulsive interactions. In contrast, for  $m   >  2 (\sqrt{2} -1) \, m_c$, we find that $\rho_d^{(\eta)} < \rho_d^{(\gamma)}$ where interactions are short range but infinitely attractive. Thus, here we find quarks tightly bound into diquarks. One must keep in mind that these results pertain to the approximate regime defined by Eq.~(\ref{limit1}) far from the critical mass $m_c$. Yet, here we examine phenomena near $m_\mathrm{pair} \equiv 2 (\sqrt{2} -1) \, m_c $. This means that the $m_\mathrm{pair}$ should be interpreted as a crossover, in contrast to a hard phase transition, from repulsive core (free) Fermi gas to a diquark Coulomb gas.

\subsubsection{Renormalization Group Analysis}
\label{Renormalization}

Before embarking on the technical aspects of the renormalization group we recall here the essential physical motivation which underlies the method. As we have seen up to this point, for low temperatures, i.e., $(T /T_\mathrm{ \, BKT} ) \ll 1 $, or for small values of the mass parameter, $(m /m_\mathrm{ \, BKT} )\ll 1 $, topological defects cannot exist as individual entities since the associated energy diverges logarithmically. On the other hand, a few tightly bound dipoles (defects) of opposite rotation or radial flux may exist since the effective quark momentum flux vanishes far from such configurations leading to a finite total energy. In contrast, for large $T$ or $m$, dipoles dissociate into a plasma. The way to understand precisely what happens between these two limits is to introduce a pair of ``external'' defects as test charges and then calculate their effect on the surrounding medium. In particular, the presence of an external charge polarizes the medium so that test charges are effectively screened. The precise details of this screening is determined by computing the modified effective interaction between the test charges as a function of their separation. Mathematically, this is done by integrating out the large momentum (divergent) contribution to the interaction energy. 

The renormalization group analysis proceeds by similar arguments as those found in standard texts on the subject such as Kardar~\cite{Kardar2007}. We compute the effective interaction between two defects located at positions ${\bf r}$ and ${\bf r}^\prime$ as a perturbative expansion in the fugacity $\mathrm{y}^0$. Since we treat shielding effects resulting only for the logarithmic interactions, we focus on these and take $\rho_d \equiv \rho^{(\gamma)}_{d}$, in order to simplify the notation. To seed the expansion we introduce a single ``internal'' dipole, i.e., two defects located at ${\bf x}$ and ${\bf x}^\prime$. The interaction $\mathcal{V}({\bf r} - {\bf r}^\prime)$ is then defined by 
\begin{widetext}
\begin{eqnarray}
   e^{ - \beta \mathcal{V}({\bf r} - {\bf r}^\prime)}  =  e^{ - 2 \pi^2   \rho_{d} \,       \mathrm{ln} \left( \vert {\bf r} - {\bf r}^\prime  \vert   \right)  }   \left\{  1 + \mathrm{y}^2 \int d{\bf x} \, d{\bf x}^\prime   e^{ - 2 \pi   \rho_{d}  \,       \mathrm{ln} \left( \vert {\bf x} - {\bf x}^\prime  \vert   \right)   }    \left[    e^{ - 2 \pi^2    \rho_{d}\,      D \left({\bf x}, {\bf x}^\prime; {\bf r}, {\bf r}^\prime \right)   }   -1              \right]  +  \mathcal{O}\left( \mathrm{y}^4 \right)   \right\}  \, , \nonumber 
\end{eqnarray}
\end{widetext}
where $D \left({\bf x}, {\bf x}^\prime; {\bf r}, {\bf r}^\prime \right)$ is the interaction between the internal and external defects. The steps that follow proceed identically to those in Kardar (see specifically Eqs.(8.60)-(8.66) in Ref.~\cite{Kardar2007}) which lead to the effective interaction $8 \pi^2  \,  \rho_{d}^\mathrm{eff} \,       \mathrm{ln} \left( \vert {\bf r} - {\bf r}^\prime  \vert   \right)   \equiv    \beta \mathcal{V}({\bf r} - {\bf r}^\prime)$, where 
\begin{eqnarray}
\rho_{d}^\mathrm{eff}  &=&  \rho_{d}   - 8  \pi^3 \,   {\rho^{(\gamma)}_{d}}^2 \,  \mathrm{y}^2  \, a^{4 \pi   \rho_{d} } \, \nonumber  \\
&&\times \int_a^\infty  \! dx_r  \, x_r^{3 - 4 \pi   \rho_{d}}  + \mathcal{O}\left(\mathrm{y}^4\right)  \, ,  \label{effM}
\end{eqnarray}
and the integration in Eq.~(\ref{effM}) is over the relative distance between the internal defects, $x_r  \equiv \vert {\bf x} - {\bf x}^\prime \vert$, from the cutoff $a$ to arbitrarily large separations. Note that the integral converges provided $\rho_{d}   \geq  1/\pi$. This condition differs from the usual value $2/\pi$ due to doubling of internal degrees of freedom for dyonic defects versus ordinary vortices. The renormalization group is based on fact that the core size, hence the minimum defect separation, is chosen arbitrarily. Increasing the core size changes the core energy as well as the interaction parameter since the minimum defect separation changes the polarization of the ambient medium. In particular, $a \to b a$ changes the fugacity by 
\begin{eqnarray}
\mathrm{y}(ba)  = b^{2 -  2 \pi \rho_d } \,  \mathrm{y}(a)  \, . 
\end{eqnarray}
The contribution from dipoles in the size range $a$ to $ba$ is (see Eq.~(8.69) in Kardar)
\begin{eqnarray}
&&\hspace{-2pc} \rho_{d}^\mathrm{eff} =  \nonumber \\
 &&\hspace{-2pc} \rho_{d} \left[  1   - 4  \pi^2\,  \rho_{d} \, \int_a^{ba}   \!   ( 2 \pi x_r dx_r )  \,  \mathrm{y}^2 e^{- 8 \pi^2 \rho_d  \,   \mathrm{ln} \left( x_r \right)} x_r^2      \right]  .   \label{effM2}
\end{eqnarray}
Taking the size change to be infinitesimal, i.e., $b = e^s \approx 1 + s $ ($s \ll 1$) in Eq.~(\ref{effM2}), the recursion relations for $\rho_d$ and the fugacity $\mathrm{y}$ are
\begin{eqnarray}
\frac{d  \rho_{d}^{-1}}{ds}  &=&  8 \pi^3  \,    a^4  \, \mathrm{y}^2 + \mathcal{O}\left(\mathrm{y}^4\right) \, ,  \label{recM} \\
 \frac{d \mathrm{y}}{ds}  &=&  \left( 2 - 2 \pi   \rho_d  \right) \mathrm{y} + \mathcal{O}\left(\mathrm{y}^3\right) \, . \label{recy} 
\end{eqnarray}
Equations~(\ref{recM})-(\ref{recy}) are the renormalization group equations, originally derived by Kosterlitz~\cite{Kosterlitz1974}, that describe the flow of the interaction parameter and fugacity as the defect separation is scaled up or down. Analysis of Eqs.~(\ref{recM})-(\ref{recy}) reveals two distinct phases. First, note that $\rho_{d}^{-1}$ varies directly with the mass parameter $m$ and temperature $T$ and that the recursion relation for $\mathrm{y}$ changes sign at $\rho_{d} = 1/\pi$. Thus, for small $\mathrm{y}$ and small $m$ or $T$ the system is in an insulating phase characterized by finite size dipoles. For $\rho_{d} > 1/\pi$ the perturbative formalism which leads to Eqs.~(\ref{recM})-(\ref{recy}) breaks down and the system enters a high temperature or large mass parameter phase with free defects.

To study the phase transition in detail we expand Eqs.~(\ref{recM})-(\ref{recy}) near the critical point by shifting and rescaling $X \equiv  \rho_{d}^{-1} -  \pi$, $Y  \equiv \mathrm{y} a^2$, which gives 
\begin{eqnarray}
\frac{d  X }{ds}  &=&  8 \pi^3  \,    Y^2 + \mathcal{O}\left(X Y^2 , Y^4\right) \, ,  \label{recM2} \\
 \frac{d Y }{ds}  &=&   \frac{1}{\pi} X  Y+ \mathcal{O}\left( X^2 Y,    Y^3\right) \, . \label{recy2} 
\end{eqnarray}
Equations~(\ref{recM2})-(\ref{recy2}) imply  
\begin{eqnarray}
\frac{d}{ds} \left(   X^2 - 8 \pi^4  Y^2 \right) = 0   \,\,\, \; \Rightarrow  \,\, \, \;  X^2 - 8 \pi^4  Y^2  = c  \, , \label{traj}
\end{eqnarray}
where  the constant $c$ characterizes a hyperbolic renormalization group flow with two critical trajectories for $c =0$ given by $Y = \pm  \, X/( 2 \sqrt{2} \pi^2)$. In particular, for $c > 0$ the foci are along the $X$-axis with two flow branches in the upper half plane $Y \ge 0$. The $X < 0$ branch corresponds to the \emph{insulating} phase of a Coulomb gas comprised of tightly bound defects. Here, flow trajectories terminate (for $s \to \infty$) at fixed points along the half line $X \in ( - \infty , 0 )$. with $c$ positive in the insulating phase, equivalently the low-mass or low-temperature phase, near criticality, i.e., $(X, \, Y) = (0, \, 0)$, linearization with respect to $m$ and $T$ is valid hence $c =  b_m^2 ( a_m  - m ) +  b_T^2 (a_T  - T) $. From Eq.~(\ref{traj}) we then have $X^2 - 8 \pi^4  Y^2 =   b_m^2 ( a_m - m ) +  b_T^2 ( a_T - T)$. In the long distance limit $s \to \infty$ the renormalization group gives $Y \to 0$, so that $\lim_{s \to \infty }X(s)  =   \sqrt{ b_m^2 ( a_m  - m ) +  b_T^2 ( a_T  - T)}$. Hence, 
\begin{eqnarray}
&&\rho_d^\mathrm{eff}(m, T)   =  \lim_{s \to \infty}   \frac{1}{ \pi + X(s) } \\
              &\approx& \lim_{s \to \infty}   \, \frac{1}{\pi} \left[ 1 - \frac{X(s)}{\pi} \right] \\
               &=&   \frac{1}{\pi}  - \frac{1}{\pi^2}  \sqrt{ b_m^2 ( a_m  - m ) +  b_T^2 ( a_T  - T)} \, . \label{finaleff}
\end{eqnarray}
From Eq.~(\ref{finaleff}) the critical mass can be expressed as a function of the temperature 
\begin{eqnarray}
m_c(T) = a_m + \frac{b_T^2}{b_m^2} \left( a_T - T\right) \, ,  \label{mofT}
\end{eqnarray}
from which we identify the zero-temperature effective interaction 
\begin{eqnarray}
\rho_d^\mathrm{eff}(m, 0)   =  \frac{1}{\pi}  - \frac{b_m}{\pi^2}  \sqrt{ m_c(0)   - m }  \, , 
\end{eqnarray}
with associated quantum critical point $ m_c(0) = a_m + b_T^2 a_T / b_m^2$.

The branch for $X > 0$ corresponds to the \emph{metallic} phase described by a plasma of free defects. In this case trajectories begin at points along the half line $X \in ( 0,  \infty  )$ and flow to $( X, \, Y) \to ( \infty, \,  \infty )$. To compute the decay of correlations in the metallic phase we note that now $c =  b_m^2 (  m - a_m   ) +  b_T^2 (T - a_T  ) > 0$, so that combining Eqs.~(\ref{recM2})-(\ref{recy2}) leads to 
\begin{eqnarray}
\frac{dX }{ X^2 +  b_m^2 (  m - a_m   ) +  b_T^2 (T - a_T  ) } = \frac{ds}{2 \pi} \, , 
\end{eqnarray}
which upon integration yields 
\begin{widetext}
\begin{eqnarray}
   \frac{1 }{ \sqrt{b_m^2 (  m - a_m   ) +  b_T^2 (T - a_T  )} } \; \;   \mathrm{tan}^{-1}  \! \left( \frac{X}{ \sqrt{b_m^2 (  m - a_m   ) +  b_T^2 (T - a_T  )}  }    \right)          = \frac{1}{2 \pi}  \, s \,  . \label{intX}
\end{eqnarray}
\end{widetext}
The limits of the integral Eq.~(\ref{intX}) are between $X_0 \propto  b_m^2 (  m - a_m   ) +  b_T^2 (T - a_T  )$ and $X_s \sim1$, where the latter condition must be imposed in order for our initial expansion Eqs.~(\ref{recM2})-(\ref{recy2}) to remain valid. The upper limit occurs for $s = s^*$
\begin{eqnarray}
s^*  \approx    \frac{2 \pi  }{ \sqrt{b_m^2 (  m - a_m   ) +  b_T^2 (T - a_T  )} } \;   \frac{\pi}{2}  \, , 
\end{eqnarray}
which leads to the correlation length 
\begin{eqnarray}
\xi &\approx&  a b \approx a e^{s^*} \\
    &\approx&   a \exp \! \left[  \frac{\pi^2  }{ b_m \sqrt{ m - m_c(T)} }  \right]  \, , \label{corleng}
\end{eqnarray}
where we have used the definition Eq.~(\ref{mofT}). Thus, the correlation length diverges exponentially near the critical point $m_c(T)$ consistent with standard BKT theory. In particular, at zero-temperature Eq.~(\ref{corleng}) gives the correlation length for a quantum BKT transition near the critical mass parameter $m = m_c(0)$. At finite temperature $m_c(T)$ gives the critical point for a standard BKT transition adjusted for quantum effects.

\section{Connection to Previous Work}
\label{Connection}

In the final part of our analysis we connect our results to established wisdom. A critical  step is to make connection to what is known, or supported through sold conjecture, about the QCD phase diagram. This requires that we express our results in terms of the baryon chemical potential $\mu_B$, which can be related directly to the quark density. The region of the QCD phase diagram that emerges is the transition region between hadronic matter, defined by $\langle \bar{q} q   \rangle > \langle q q \rangle$, and quark matter, $\langle \bar{q} q \rangle < \langle q q \rangle$, which is sometimes referred to as the coexistence phase of QCD~\cite{Fukushima2011,Weise2012}. This is intimately related to the idea of hadron-quark continuity, i.e., a smooth transition from superfluid/superconducting hadronic matter to superconducting quark matter.

Another way to view this region is in terms of what has been dubbed ``quarkyonic'' matter~\cite{McLerran2007}, wherein the fermionic sector is defined by a mixed system comprised of free quarks as well as baryons; a coexistence of quarks, mesons, gluons appearing when $\mu_B = M_B$, where $M_B$ is the baryon mass. Unique features of this phase include the possibility of chiral symmetry restoration in the presence of confinement. We have shown this to be true in our present model: chiral restoration occurs at low temperatures and for large chemical potential. Our model also exhibits the effect of a baryonic ``skin'' at the Fermi surface, another characteristic of the quarkyonic phase characterized by a layer of baryons of thickness $\sim \Lambda_\mathrm{QCD}$ measured from the Fermi surface that interpolates smoothly into the pure quark phase residing deep within the Fermi sea. We have shown that much of the established wisdom regarding the phases of QCD is well replicated within our model. 

Another topic which has direct relation to our results is the study of holographic BKT phase transitions. We will conclude by connecting our work to this fascinating topic. The original holographic BKT transition was discovered in the D3/D5 brane system which forms an $\mathrm{AdS}_5 \times \mathrm{S}^5$ bulk-theory background dual to $\mathcal{N}=4$ $SU(N)$ super Yang-Mills theory on the boundary~\cite{Jensen2010}. This seems like a very different kind of system than the one studied in this paper. Yet, models of QCD have been shown to exhibit features reminiscent of certain supersymmetric field theories. We leave the full connection of our present work to holographic or string theory based models for a follow up paper, but hear only comment on two points from the work presented in~\cite{Jensen2010}.

\subsection{QCD Phase Diagram}

To construct the temperature-chemical potential phase diagram, we recast the boundary conditions from Sec.~\ref{FiniteTPT} for the various phases of the ground state of our system by letting the quark mass vary with temperature through $m(T) = m_0 + \Delta_s(T)$, with $\Delta_s(T) = \Delta_s(0)\sqrt{1 - \left(T/T^*_s\right)^2}$ and $T^*_s$ the scalar meson dissociation temperature. The two critical conditions read
\begin{eqnarray}
\tilde{\mu}_B = \pm m(T) \, ,  \label{PhaseConst}
\end{eqnarray}
and yield the phase transition curves
\begin{eqnarray}
T^{\pm }_{c } =  T_s^*\,       \sqrt{  1  -  \frac{ \left[ m_0 \pm \left( \mu_B  - \bar{\Delta}_d \right) \right]^2}{ \Delta_s(0)^2  }   } \, .  \label{T1mu} 
\end{eqnarray}
with the independent variables given by $T$ and $\mu_B$. Thus, the diquark pairing field is an implicit function of both independent variables: $\bar{\Delta}_d \equiv  \bar{\Delta}_d(T, \mu_B)$. We must keep in mind that $\mu_B > \bar{\Delta}_d$ for $T^+_c$, and $\bar{\Delta}_d > \mu_B$ for $T^-_c$, with $\bar{\Delta}_d(T, \mu_B)$ evaluated locally along each curve. One should observe that $m$ runs inversely with $\mu_B$: introducing a scaling factor s through either $m \to s  m$ or $\mu_B \to \mu_B/s$, at the level of Eq.~(\ref{PhaseConst}), leads to the same scaling behavior for each of the critical curves $T^{\pm }_{c }$.

\begin{figure}[]
\centering
\subfigure{
\label{fig:ex3-a}
\includegraphics[width=.53\textwidth]{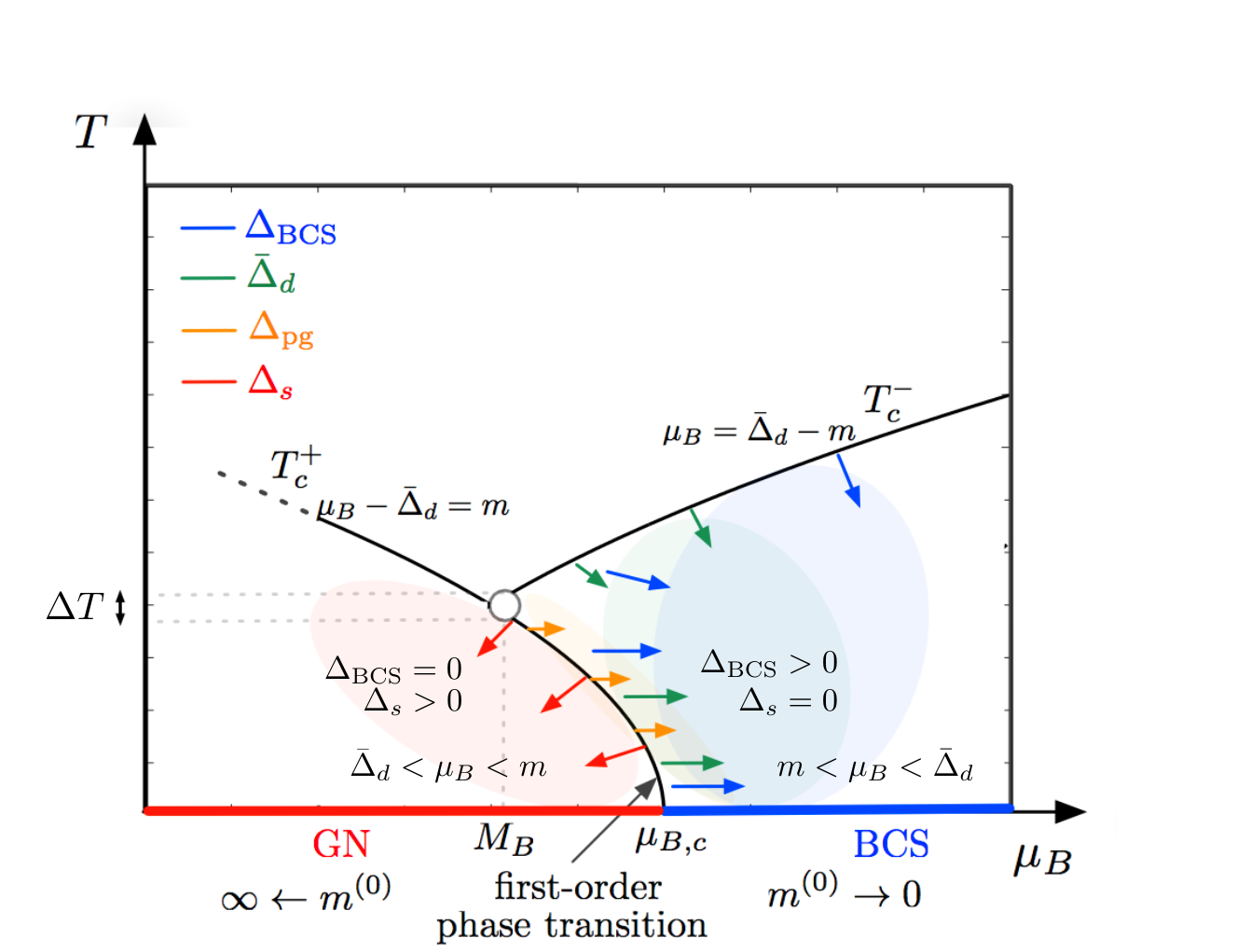}} \\
\caption[]{\emph{Mean-field features of the color superconducting region of the QCD phase diagram}. The arrows indicate directions of increasing values for the various pairings and order parameters. The blue and red regions indicate high concentration of BCS gap and scalar condensation, respectively. The two zero-temperature theories, Gross-Neveu (GN) and superconducting (BCS), are indicated in red and blue, respectively, and match the analysis in Fig.~\ref{BECBCS}.  }
\label{BECBCS3}
\end{figure}

We must now interpret the critical curves in Eq.~(\ref{T1mu}). If we consider temperatures well below that for scalar meson dissociation ($T \ll T_s^*$), which isolates the diquark phase transiton, we obtain the approximate curves 
\begin{eqnarray}
T^{+}_{c }    &\simeq&  \bar{T}  \, \sqrt{  1  -  \left( \frac{ \mu_B  - \bar{\Delta}_d }{ m^{(0)}  }  \right) } \, , \label{T2mu} \\
T^{-}_{c }   &\simeq& \bar{T} \, \sqrt{ 1  +  \left(  \frac{  \bar{\Delta}_d  - \mu_B}{ m^{(0)}  } \right) } \, , \label{T3mu}
\end{eqnarray}
where 
\begin{eqnarray}
\bar{T}  =   T_s^*  \sqrt{\frac{2   m^{(0)}}{\Delta_s(0)  }} \,  .  \label{Tbar}
\end{eqnarray}
Though the quark pairing function is evaluated locally in the $T-\mu_B$ plane, the running quark mass is evaluated at zero temperature: $m^{(0)} = m_0 + \Delta_s(0)$. For the discussion that follows it is helpful to consult Fig.~\ref{BECBCS3}.

Consider Eq.~(\ref{T2mu}). Recall the zero-temperature description of the curve $T^{+}_{c}$ in Eq.~(\ref{explicitdiquark2}), the quantum phase transition depicted in Fig.~\ref{BECBCS}. There, the difference between the chemical potential and diquark pairing was held fixed in order to simplify our task of understanding the critical region. In the full picture, however, we must now allow $\mu_B$, $\bar{\Delta}_d$, $m$, $\Delta_s$, and $\Delta_\mathrm{BCS}$ to vary through the critical curve.

Let us first examine the behavior of the positive part of the second term under the radical sign in Eq.~(\ref{T2mu}). With the pairing ansatz $\bar{\Delta}_d \sim  a \mu_B^r$ and mass scaling $m^{(0)} \sim c/\mu_B$, we have 
\begin{eqnarray}
\frac{d}{d \mu_B}\! \!  \left[  \frac{   \mu_B - \bar{\Delta}_d }{m^{(0)}    } \right] 
\sim  \frac{2}{c} \mu_B  - \frac{a}{c} \mu_B^r ( 1 + r)  \, , 
\end{eqnarray} 
which is guaranteed to be increasing if $r =1$ and $0 <a< 1$, consistent with increasing $\bar{\Delta}_d$. Another check is that along $T^+_c$ we have $\mu_B - \bar{\Delta}_d = m$, which gives the condition for the derivatives $1 - d \bar{\Delta}_d/d\mu_B = d m/d\mu_B$, or $d m/d\mu_B = 1 - a > 0$ (since $a < 1$). This is satisfied provided $T^+_c$ decays fast enough into the quantum critical point such that the increase in $m$ towards lower temperatures outweighs its decrease towards larger $\mu_B$.

Now let us look deeper into the phase transition across $T^+_c$. Taking $\bar{\Delta}_\mathrm{BCS}$ as the order parameter and $\mu_B$ as the independent tunable parameter, the phase transition is seen to be one of first order. To see that this is the case, we may examine the one-sided limits into $T^{+}_{c}$, which are found to be 
\begin{eqnarray}
\lim_{\mu_B \to \mu_{B, c}(T)^-}\bar{\Delta}_d  = \mu_B  - m  \, , \label{CritLimit1}
\end{eqnarray}
 and 
 \begin{eqnarray}
 \lim_{\mu_B \to \mu_{B, c}(T)^+}\bar{\Delta}_d =  \mu_B +m  \, , \label{CritLimit2}
 \end{eqnarray}
with the superscripts indicating the region to the right ($+$) or left ($-$) of the critical point. The limit from the right comes from extrapolating the condition at $T^{-}_c$ down to lower temperatures. Since $\Delta_\mathrm{BCS} \equiv \bar{\Delta}_d  - \mu_B$, the BCS gap is positive definite to the right of, and arbitrarily close to, the curve $T_c^+$, and zero to the left of it. In fact, to the right of $T_c^+$, $\Delta_\mathrm{BCS}$ approaches the quark mass $m$ at criticality, which is precisely the condition that drives the phase transition there. Expanding on this point, to the right of $T_c^+$ the hierarchy condition reads 
\begin{eqnarray}
m < \mu_B < \bar{\Delta}_d \, ,  \label{RightHier}
\end{eqnarray}
a signature indicating that the system favors strong diquark pairing over free quarks and free quarks over mesons. As $\mu_B$ is tuned downward towards criticality, the quark mass $m$ increases until $m = \Delta_\mathrm{BCS}$ at the phase transition. In alignment with the analysis depicted in Fig.~\ref{BECBCS}, a detailed dissection of the phase transition reveals a re-shuffling of the hierarchy in Eq.~(\ref{RightHier}) at criticality. The critical condition $m = \Delta_\mathrm{BCS}$ marks the onset of a decrease in energetic favorability for the BCS gap and a corresponding increase in that for the quark mass through a variable scalar condensate, according to the relation $m = \Delta_s + m_0$. But, to ``transform'' a BCS gap state into a scalar meson requires first dissociation of diquarks into free quarks, generated by the hierarchy switch $\bar{\Delta}_d > \mu_B \to \bar{\Delta}_d < \mu_B$, followed by binding of quarks with anti-quarks near the surface of the Fermi sea, generated by $\mu_B > m  \to \mu_B < m$.

We thus traverse the singular critical curve moving from larger to smaller values of the baryon chemical potential, emerging to the left of $T^+_c$ into a medium characterized by the hierarchy
\begin{eqnarray}
\bar{\Delta}_d < \mu_B < m  \, .  \label{LeftHier}
\end{eqnarray}
It is important to understand that our mean-field results show that the diquark pairing $\bar{\Delta}_d$ must be discontinuous at the critical curve, in contrast rather to a gradual decay as $\mu_B$ is reduced across the critical curve, as one would see in a standard BEC-BCS crossover. To prove this statement, recall that Eqs.~(\ref{CritLimit1})-(\ref{CritLimit2}) can be restated as 
\begin{eqnarray}
\lim_{\mu_B \to \mu_{B, c}(T)^-}m  = \mu_B  - \bar{\Delta}_d   \, , \label{CritLimit3}
\end{eqnarray}
 and 
 \begin{eqnarray}
 \lim_{\mu_B \to \mu_{B, c}(T)^+} m  =    \bar{\Delta}_d - \mu_B   \, .  \label{CritLimit4}
 \end{eqnarray}
Adding these equations with the assumption that both $\bar{\Delta}_d$ and $\mu_B$ are continuous across $T^+_c$, we get 
\begin{eqnarray}
m^+ + m^- = 0  \, ,  
\end{eqnarray}
with the $+$ and $-$ superscripts indicating, respectively, right and left limit values for the quark mass. Since $m$ is positive definite, our assumption must be false: the right and left sided limits for $\bar{\Delta}_d$ must not be equal, i.e., $\bar{\Delta}_d$ must be discontinuous across $T^+_c$. Thus, the correct relation reads $m^+ + m^- = \delta \bar{\Delta}_d$, where the discontinuity in diquark pairing is $\delta \bar{\Delta}_d \equiv  \bar{\Delta}_d^+ - \bar{\Delta}_d^-$. Expanding this result by expressing the mass in terms of the bare mass and scalar condensate gives $\delta \bar{\Delta}_d + \delta \Delta_s = 2 m_0$. Taking the difference of Eqs.~(\ref{CritLimit3})-(\ref{CritLimit4}) gives $\delta \bar{\Delta}_d + \delta \Delta_s = 2 \delta \Delta_\mathrm{BCS}$, hence the interesting result that the discontinuity in the BCS gap along the critical curve is equal to the bare quark mass along the curve
\begin{eqnarray}
\delta \Delta_\mathrm{BCS} = m_0 \, , 
\end{eqnarray}
and equal to the average of the diquark and scalar discontinuities 
\begin{eqnarray}
\delta \Delta_\mathrm{BCS} = \mathrm{Ave} \!\left( \delta \bar{\Delta}_d ,  \delta \Delta_s \right) \, . 
\end{eqnarray}
Finally, the diquark pairing function has largest gain across the critical curve, given by 
\begin{eqnarray}
\delta \bar{\Delta}_d  = 2 \delta \Delta_\mathrm{BCS} + |\delta \Delta_s | \, , 
\end{eqnarray}
where the fact that $\delta \Delta_s < 0$ is taken into account.

Let us now turn to the second boundary curve $T^-_c$ in Eq.~(\ref{T3mu}). Taking $\bar{\Delta}_d \sim a \mu_B^r + b$ and $m^{(0)} \sim c/\mu_B$, we find that 
\begin{eqnarray}
\frac{d}{d \mu_B} \left[  \frac{\bar{\Delta}_d - \mu_B }{m^{(0)}  }\right] \sim - \frac{2}{c} \mu_B + \frac{a}{c} \mu_B^r ( 1 + r) + \frac{b}{c}  \, . \label{DerCond}
\end{eqnarray}
Moreover, as before, we find that derivatives along this curve satisfy $d\bar{\Delta}/d \mu_B -1 = d m/d \mu_B$, so that $dm/d \mu_B = a - 1$. The mass $m$ must be decreasing along $T_c^-$ which requires that $0 < a<1$. We then find from Eq.~(\ref{DerCond}) that $b/\mu_B > 1 - a$, a condition which may be satisfied over some finite range of $\mu_B$, but ultimately causes $T_c^-$ to turn downward towards the $\mu_B$ axis. Along $T_c^-$, both $\Delta_s$ and $m$ are small and continue to decrease towards larger values of $\mu_B$. This means that $\Delta_\mathrm{BCS}$ is also small here, since $\Delta_\mathrm{BCS} = \bar{\Delta}_d - \mu_B = m $ along $T_c^-$. Also, $\Delta_\mathrm{BCS}$ continues to decrease towards larger temperatures, and increases towards lower ones consistent with the behavior of the pairing function $\bar{\Delta}_d$. Hence, one may safely identify $T_c^-$ as an ordinary thermal BEC-BCS crossover rather than an actual phase transition.

The interesting features of $T_c^-$ lie more towards the point where it seems to intersect $T_c^+$. But the two curves do not in fact intersect. Since we found $T_c^+$ to be the locus of a first-order phase transition, particularly with respect to the diquark pairing $\bar{\Delta}_d$, it appears in Eqs.~(\ref{T2mu})-(\ref{T3mu}) on different sides of the phase transition and so must have a different limit as $T_c^+$ and $T_c^-$ approach each other. We must take the point of nearest approach of the two critical curves to be located at the baryon mass $\mu_B = M_B$. The presence of $M_B$ sets the lower viable limit for our model: the Fermi surface and associated diquark condensate cannot form below the baryon mass. Taking $\bar{\Delta}_d$ to vary symmetrically across $T_c^+$, the temperature gap at the tricritical point where $\mu_B = M_M$ is 
\begin{eqnarray}
\Delta T^{(3)} = \sqrt{2} \left( \sqrt{ 1 + \frac{M_B}{c} \delta \Delta_\mathrm{BCS}^{(3)} } -  \sqrt{ 1 -  \frac{M_B}{c} \delta \Delta_\mathrm{BCS}^{(3)}    } \right) \, , 
\end{eqnarray}
where we have again inserted the mass scaling $m \sim c/\mu_B = c/M_B$, and the value of $\bar{T} = \sqrt{ 2 m^{(0)} / \Delta_s(0)} \approx \sqrt{2}$ varies smoothly along the $\mu_B$-axis with the predominant contribution to the mass coming from the scalar condensate. Here $c^{1/2}$ is just the limiting value of $\mu_B$ when approaching the quantum critical point from below, or the value of the gap discontinuity there $\delta \Delta_\mathrm{BCS}^{(qc)}$. At the tricritical point one would expect the discontinuity in the BCS gap to be small compared to the other scales involved: $\delta \Delta_\mathrm{BCS}^{(3)} \ll c^{1/2}$. Taking this into account we have the approximate form for the temperature gap
\begin{eqnarray}
\Delta T \approx \sqrt{2}  \, \frac{M_B}{c} \delta \Delta_\mathrm{BCS}^{(3)}   =   \sqrt{2}  \, M_B \frac{ \delta \Delta_\mathrm{BCS}^{(3)}}{  {\delta \Delta_\mathrm{BCS}^{(qc)}}^2} \, , 
\end{eqnarray}
where, in the denominator, we have introduced the square of the gap discontinuity at the quantum critical point. Hence, from our analysis of the Fermi surface we find that the tricritical point for deconfined, hadronic, and superconducting phases occurs at $\mu_B = M_B$.

\begin{figure}[]
\centering
\subfigure{
\label{fig:ex3-a}
\hspace{-.4pc}\includegraphics[width=.55\textwidth]{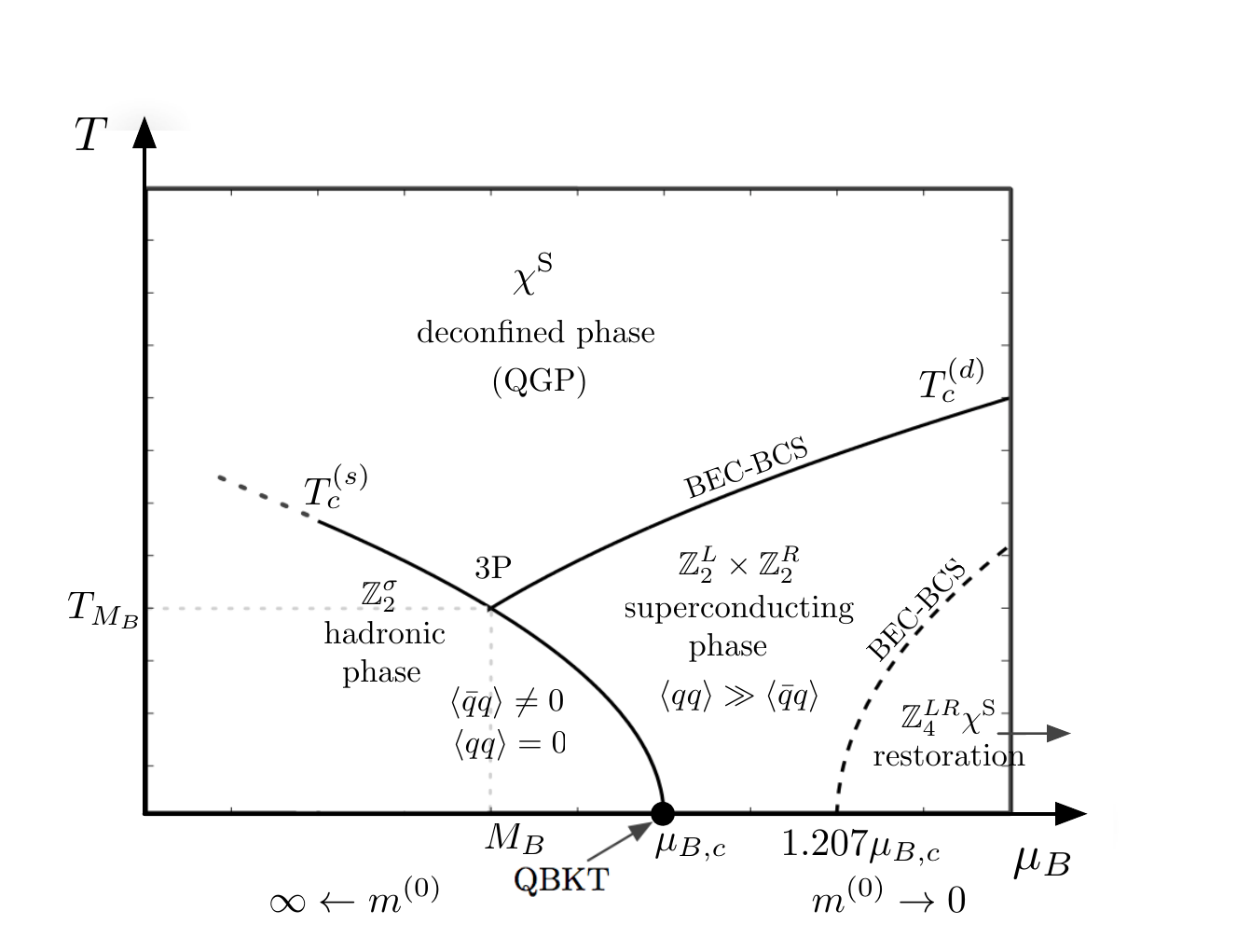}} \\
\caption[]{\emph{Full quantum treatment for the $T-\mu_B$ phase diagram for color superconductivity}. Chiral symmetry breaking and quark pairing are indicated based on quantum field calculations associated with formation of the Fermi surface. Symmetries of the ground state $\mathbb{Z}_2^\sigma$, $\mathbb{Z}_2^{LR} \times \mathbb{Z}_2^{LR}$, and $\mathbb{Z}_4^{LR}$ are indicated, connecting to the corresponding regions in Fig.~\ref{PseudospinDomain}. The $\mathbb{Z}_2^\sigma$-symmetric meson/hadron phase with vanishing diquark condensate, lies to the left of the quantum BKT transition (QBKT). To the right of the quantum BKT transition lies the superconducting phase described by a Coulomb gas of Cooper pairs with $\mathbb{Z}_2^{LR} \times \mathbb{Z}_2^{LR}$ symmetry. For very large $\mu_B$ the system crosses over to free quarks indicated by the dashed curve in the lower right corner. Directions of vanishing and increasing running (zero-temperature) quark mass $m^{(0)}$, $\mathbb{Z}_4^{LR}$-chiral symmetry restoration, and deconfined chiral symmetric ($\chi^\mathrm{S}$) quark-gluon plasma (QGP) are shown. The quark chemical potential is directly related to the quark density $\mu_B  \sim n_q$.}
\label{BECBCS2}
\end{figure}

So far, we have only discussed the mean field result for the QCD phase diagram. The full quantum calculation for the superconducting region of the QCD phase diagram is plotted in Fig.~\ref{BECBCS2}. Our results are remarkably consistent with ones obtained using more traditional approaches (see, for example, Fig.~1 and discussions in Ref.~\cite{Alford2008}). A key difference is the appearance in our graph of a quantum BKT transition that separates the meson dominant ($\langle q  q \rangle =  0$) and the color superconducting ($\langle q q \rangle \ne 0$) regimes. The BEC-BCS crossover region for large $\mu_B$ is indicated by a dashed curve towards the lower right corner of the plot. Interestingly, the crossover occurs where the color-flavor locking transition appears in more elaborate QCD models. We have identified the critical curves, $T_c^-$ and $T_c^+$, with the diquark and scalar meson superfluid transitions, $T_c^{(d)}$ and $T_c^{(s)}$.

\subsection{Holographic BKT Transition}

There are some striking similarities between the quantum BKT transition in our model and those associated with holographic systems. The first pertains to the parameter that drives the holographic BKT transition: the ratio of charge density to the applied magnetic field. In our system it is the ratio of the density to the mass. It has been known for quite some time that an applied magnetic field generates a dynamical fermion mass, even in the absence of attractive interactions. This was first observed in~\cite{Klimenko1991,Klimenko1992}, which lead to further investigations, as for instance~\cite{Gusynin1995}. The general argument can be stated briefly as follows. First, we consider a Lagrange density that includes only kinetic and mass terms 
\begin{eqnarray}
 \mathcal{L}  =   \bar{\Psi} \left( i \gamma^\mu {D}_\mu - m  \right)   \Psi \, .    \label{LB}
\end{eqnarray}
Equation~(\ref{LB}) is expressed in the four-spinor representation with the external applied magnetic field is incorporated through the covariant derivative $D_\mu = \partial_\mu - i e A_\mu^\mathrm{ext}$, with the vector potential given by $A_\mu^\mathrm{ext} = - B x_2 \delta_{\mu 1}$. The fundamental point to take away from Eq.~(\ref{LB}) is that the linear factor of $x_2$ in the vector potential reveals an infrared suppression of the dynamics, i.e., both momentum degrees of freedom are removed from the spectrum at long wavelengths. This is essentially what a mass term does. Hence, we expect to see the appearance of an effective mass at low energies.

 To see how this works, consider that condensation is given by the propagator
 \begin{eqnarray}
 S(x, y) = \langle 0 | T\left( \bar{\Psi}(x) \Psi(y)   \right) | 0 \rangle \, . 
 \end{eqnarray}
The Schwinger approach~\cite{Schwinger1951} yields
\begin{eqnarray}
 S(x, y) = \exp\left( i e \int_y^x A_\lambda^\mathrm{ext} dz^\lambda \right)   \bar{S}(x, y) \, , 
\end{eqnarray}
where the Fourier transform of the propagator is 
\begin{widetext}
\begin{eqnarray}
\bar{S}(k) &=& - i \int_0^\infty ds \exp\left[ - s \left( m^2 +k_3^2 + \mathrm{\bf k}^2 \frac{\mathrm{tanh}(eBs)}{eBs} \right) \right]     \nonumber  \\
      && \times   \left[ - k_\mu \gamma_\mu + m - i   \left( k_2 \gamma_1 - k_1 \gamma_2 \right)  \mathrm{tanh}(eBs) \right] \left[ 1 - i \gamma_1 \gamma_2 \mathrm{tanh}(eBs) \right]  \, . 
\end{eqnarray}
\end{widetext}
The condensate is then obtained through
\begin{widetext}
\begin{eqnarray}
 \langle 0 | \bar{\Psi}  \Psi  | 0 \rangle  &=&  - \lim_{x \to y} \mathrm{tr}S(x, y)  \nonumber \\
 &=& - \frac{i}{(2 \pi)^3} \mathrm{tr} \int d^3 k \bar{S}(k) = \lim_{\Lambda \to \infty}\frac{4m}{(2\pi)^3} \int d^3k \int_{1/\Lambda^2}^\infty ds \exp \left[ - s \left( m^2 + k_3^2 + \mathrm{\bf k}^2 \frac{\mathrm{tanh}( eBs )}{eBs} \right) \right] \, , 
\end{eqnarray}
\end{widetext}
where we take the limit as $m \to 0$ to obtain the explicit dependence on the magnetic field 
\begin{eqnarray}
\lim_{m \to 0} \langle 0 | \bar{\Psi}  \Psi  | 0 \rangle   = - \frac{|eB|}{2 \pi }\, , \label{Bmass}
\end{eqnarray}
 independent of the ultraviolet cutoff $\Lambda$. 

Thus, we see that spontaneous chiral symmetry breaking occurs even in the absence of an attractive scalar channel, due purely to the applied magnetic field. Although the exact result Eq.~(\ref{Bmass}) will be modified by the interactions in our model, the general dependence on $B$ will be retained. The simple result Eq.~(\ref{Bmass}) modifies our critical point by 
\begin{eqnarray}
\left(\frac{m}{|\tilde{\mu}_B|} \right)_c  \to \left(\frac{m + |e B|/2 \pi^3}{|\tilde{\mu}_B|} \right)_c  = 1 \, , 
\end{eqnarray}
so that in the absence of an explicit quark mass ($m \to 0$) the quantum BKT transition is driven by the ratio of the magnetic field to the density
\begin{eqnarray}
\frac{1}{2 \pi^3} \left| \frac{e B}{ \tilde{\mu}_B} \right| \, .  \label{HolRes}
\end{eqnarray}
A more accurate form for the critical behavior must include interactions, but even at this level Eq.~(\ref{HolRes}) essentially matches the holographic result.

Another point of interest is the question of Efimov states discussed in~\cite{Jensen2010} and expanded on in~\cite{Iqbal2010}. These states comprise an infinite discrete spectrum of possible AdS embeddings which count the number of oscillations between the bottom of the brane and the AdS boundary (see discussion around Eq.~(14) in \cite{Jensen2010}). Our results provide strong evidence and a natural mechanism by which such states are realized through the higher winding modes of the baryonic vortices we have studied.

\section{Conclusion}
\label{Conclusion}

We have presented the first example of a quantum BKT transition within a close model of QCD, in particular, and four-fermion models, more generally; a significant step towards deepening the connection between QCD and holography. We have elucidated the microscopic structure of BEC-BCS transitions in relativistic (2+1)-dimensional systems. At ordinary low densities quarks are tightly confined into the familiar nuclei from the standard table of elements. As densities increase, however, nuclear matter undergoes several phase transitions ultimately dissociating into constituent quarks and gluons forming a strongly interacting quantum liquid, or quark-gluon plasma, through a phenomenon known as asymptotic freedom. There are presently many open questions regarding the structure of matter between these extremes of low and high densities. However, the broader aim of the present work is to establish certain preliminary foundations from which investigations into the role that topological sectors of quantum field theories play in dual quantum gravity theories may proceed. Indeed, our present and future work is ultimately motivated by a singular fundamental interest: the connection between quantum field theories and theories of gravity known as holographic or gauge-gravity dualities. That is, the notion that certain quantum mechanical theories in flat space-time contain theories of gravity in one higher dimension hidden within the subtleties of their mathematical structure, and conversely that Einstein's field equations have something to say about quantum mechanics. This is undoubtedly one of the great discoveries of late twentieth century physics.

\begin{acknowledgments}
The author would like to thank the Department of Physics at Colorado School of Mines for support during the writing of this manuscript.
\end{acknowledgments}

\bibliography{QCD_Phases_Refs}

\end{document}